\documentclass[ reprint,  aps,nofootinbib,]{revtex4-2}
\usepackage{amssymb}
\usepackage{amsmath}
\usepackage{graphicx}
\usepackage{dcolumn}
\usepackage{bm}

\begin{document}

\title{Schwinger mechanism of magnon-antimagnon pair production on magnetic
field inhomogeneities and the bosonic Klein effect}

\author{T. C. Adorno}\email{Tiago.Adorno@xjtlu.edu.cn}
 \affiliation{Department of Physics, School of Mathematics and Physics, Xi'an
Jiaotong-Liverpool University, 111 Ren'ai Road, Suzhou Dushu Lake Science
and Education Innovation District, Suzhou Industrial Park, Suzhou 215123,
People's Republic of China.}

\author{S. P. Gavrilov}%
 \email{gavrilovsergeyp@yahoo.com;gavrilovsp@herzen.spb.ru}
\affiliation{Department of Physics, Tomsk State University, Lenin Ave. 36,
634050, Tomsk, Russia,}
\affiliation{Department of General and Experimental Physics, Herzen State
Pedagogical University of Russia, Moyka embankment 48, 191186, St.
Petersburg, Russia;}

\author{D. M. Gitman}\email{gitman@if.usp.br}
\affiliation{P.N. Lebedev Physical Institute, 53 Leninskiy prospect, 119991
Moscow, Russia,}
\affiliation{Institute of Physics, University of S\~{a}o Paulo, Rua do Mat%
\~{a}o, 1371, CEP 05508-090, S\~{a}o Paulo, SP, Brazil.}

\date{\today}

\begin{abstract}
Effective field theory of low-energy excitations (magnons) that describe
antiferromagnets is mapped into electrodynamics of a charged scalar field
interacting with an external magnetic background. In this theory magnons and
antimagnons are described by a corresponding scalar field. If the external
background is a constant inhomogeneous magnetic field in the quantum version
of the model, then there exists vacuum instability which can be analyzed by
an analogy with the scalar QED with electric potential steps. Here magnons
and antimagnons are treated as charged particles, whereas the magnetic
moment plays the role of the electric charge such that magnons and
antimagnons differ from each other in the sign of this moment. The vacuum
instability is related to the magnon-antimagnon production from the
corresponding vacuum by magnetic field inhomogeneities. Characteristics of
the vacuum instability can be calculated nonperturbatively using special
exact solutions of the Klein-Gordon equation. In particular, we consider
examples of the magnetic field that correspond to some regularizations of
the Klein step. In the case of smooth-gradient steps, we have derived an
universal behavior of the flux density of created magnon-antimagnon pairs.
It is noted that there exists an opportunity, for the first time, to observe
the Schwinger effect in the case of Bose particle creation. Moreover, it
turns out that in the case of the Bose statistics appears a new mechanism
for amplifying the effect of pair creation, which we call
statistically-assisted Schwinger effect. 
\end{abstract}







\maketitle

\section{Introduction\label{S1}}

Magnons, or quantized spin waves occur in various types of ordered magnets:
antiferromagnet, ferromagnet, and ferrimagnet. They present collective
magnetic excitations of the electron spin structure in a crystal lattice.
The emerging field of magnonics utilizes magnons for information processing;
see Ref. \cite{magnonics22} for a review. Using magnons as information
carriers has various advantages, in particular, the low power-consumption.
Although spin systems are originally described as lattice models, similar to
Dirac models of nanostructures, one can describe their low-energy dynamics
based on a continuum field theory at energy scales much lower than the
inverse lattice spacing; see Ref. \cite{HFMNS21,KamMoW05} and references
therein. Descriptions based on effective field theories (EFT) of spin
systems at low energies also allow including external fields in the model.
The magnon EFT can incorporate various symmetry-breaking terms. For example,
Zeeman-like interactions break the symmetry explicitly because of the
coupling with an external magnetic field. It turn out that the magnon EFT
that describes antiferromagnets is relativisticlike. Our special interest is
the case of an inhomogeneous magnetic field applied to an antiferromagnet in
the collinear (homogeneous) ground state. It was recently shown \cite%
{HFMNS21} that the magnon EFT with the easy-axis anisotropy can be mapped
into electrodynamics of a charged scalar field interacting with an external
electromagnetic potential. The mass of this field is determined by the sum
of the easy-axis potential and the ratio of magnetization and condensation
parameters. Magnetic moment here plays the role of the electric charge, and
magnons and antimagnons differ from each other by the sign of the magnetic
moment. In the framework of such a consideration, it is important to take
into account the vacuum instability (the Schwinger effect) under the
magnon-antimagnon production on magnetic field inhomogeneities. (an analog
of particle-antiparticle creation by constant inhomogeneous electriclike
fields). In this article, the leading imaginary part of the one-loop
effective action was calculated in the framework of the semiclassical
world-line formalism for the case of a linearly varying magnetic field and
agrees with the Schwinger's formula for a constant electric field \cite%
{Schw51}. In the presence of a constant inhomogeneous external magnetic
field, one can see that the latter problem is technically reduced to the
problem of charged-particle creation from the vacuum by an electric
potential step. In this context, it is important to mention recent works
addressed the problem of magnon-antimagnon pair creation by a sufficiently
high rectangular step (the Klein step) and barrier, formed by magnetic field
inhomogeneities. \cite{HarYD22,HarYD22b,ZhouWY19,Nature23}. In these cases
the world-line formalism does not work and the problem is considered in the
framework of the relativistic quantum mechanics.{\large \ }In relativistic
quantum mechanics, problems of this type were considered in relation to the
Klein paradox in the pioneer works \cite{Klein27,Sauter31a,Sauter-pot} (a
detailed historical review can be found in Refs. \cite{DomCal99,HansRavn81}%
). We note that to avoid possible confusion, the Klein paradox should be
distinguished from the Klein\ tunneling through the square barrier. This
tunneling without an exponential suppression occurs when a particle is
incident on a high barrier, even when it is not high enough to create
particles. It is known that attempts to consider overlapping amplitudes as
amplitudes of particle transmission and reflection by the Klein step in the
same manner as in the relativistic quantum mechanics often leads to
contradictions and paradoxes. As it is known, processes in which the number
of particles is not constant must be considered in the framework of quantum
field theory (QFT). Recently, a consistent nonperturbative treatment of the
vacuum instability with respect to charged particle creation was developed
in the framework of strong-field quantum electrodynamics (QED) with
time-independent external electric potential steps (we call them
conditionally $x$ steps) in Refs. \cite{GavGi16,GavGi20,BGavGi23}.\ In the
case of bosons, the latter nonperturbative treatment is based on the
existence of special exact solutions of the Klein-Gordon equation with the
corresponding $x$ step. This enables the consideration of pair creation by $%
x $ steps of arbitrary form, including the Klein step. We hope that the
present article will promote consistent application of strong field QED
methods in magnonics, avoiding contradictions and non-existent paradoxes in
interpreting the theoretical results.

In the present work we use the strong field QED to study the
magnon-antimagnon pair production on magnetic field inhomogeneities.%
\footnote{%
Here we are using the natural system of units $\hslash =1$.}. The article is
organized as follows: In Sec. \ref{S2} the{\large \ }EFT{\large \ }model
describing the low-energy dynamics of antiferromagnetic magnons is mapped
into scalar electrodynamics with $x$ steps. In Sec. \ref{S3}, we construct a
Fock space realization of the $EFT$\ model\ in the framework of strong-field
QED with $x$ steps.{\large \ }Initial and final one-particle states are
constructed with the help of stationary plane waves satisfying the
Klein-Gordon equation.{\large \ }Initial and final vacua are defined and
initial and final states of the Fock space are constructed.{\large \ }Mean
numbers of magnons and antimagnons created from the vacuum are expressed via
overlap amplitudes of the stationary plane waves.{\large \ }Observable
physical quantities specifying the vacuum instability are determined. We
calculate and analyze the fluxes of energy and magnetic moments of created
magnons.{\large \ }In Sec. \ref{S4}, we present characteristics of the
vacuum instability obtained for some magnetic steps that allows exact
solving the Klein-Gordon equation. In particular, we consider examples of
magnetic steps with very sharp field derivatives $\partial _{x}U$\ that
correspond to a regularization of the Klein step.{\large \ }In the case of
smooth-gradient steps, we describe an universal behavior of the flux density
of created pairs.{\large \ }In the last Sec. \ref{S5}, we summarize the main
results of the present work.{\large \ }Some details of the scalar field
quantization in the presence of critical potential steps are placed in
Appendix \ref{Ap}. Examples of some exact solutions with $x$ steps are given
in Appendix \ref{B}.

\section{EFT model describing low-energy dynamics of magnons\label{S2}}

The system under consideration consists of localized spins which live on
sites of a cubic-type lattice. These sites are numbered by the index $n$.
The corresponding spin vector operators are denoted by $\mathbf{\hat{s}}^{n}$%
. It is assumed that the spins are involved in the antiferromagnetic
interaction. Its original $SO\left( 3\right) $ spin-rotation symmetry is
explicitly (but softly) broken due to an external magnetic field $\mathbf{B}$
and an anisotropic interaction $C$ known as single-ion anisotropy. The
Hamiltonian describing such a system reads:%
\begin{eqnarray}
&&\ \hat{H}_{\mathrm{spin}}=\sum_{n}\sum_{i=1}^{d}J\delta ^{ab}\hat{s}%
_{a}^{n}\hat{s}_{b}^{n+\hat{\imath}}  \notag \\
&&-\sum_{n}\left[ \mu B^{a}\left( \mathbf{r}_{n}\right) \hat{s}%
_{a}^{n}+C^{ab}\hat{s}_{a}^{n}\hat{s}_{b}^{n}\right] \ .  \label{d1}
\end{eqnarray}%
Here $\hat{s}_{a}^{n}$ denote spin operator components on the site $n$ $%
\left( \left[ \hat{s}_{a}^{n},\hat{s}_{b}^{n}\right] =i\epsilon _{ab}^{c}\
\hat{s}_{c}^{n}\right) $;$\ J>0$ is the antiferromagnetic interaction
coupling constant. To describe the nearest-neighbor pairs, the direction $%
\hat{\imath}=1,2,\ldots ,d$ with a spatial dimension $d$ is introduced. The
sum over $\hat{\imath}$\ means summing over nearest neighboring spins and
the sum over $n$\ means summing over sites $n$\ of a cubic-type lattice. $%
B^{a}\left( \mathbf{r}_{n}\right) $ are external magnetic field components
on the site{\Large \ }$n$, they depend on the coordinates $\mathbf{r}%
_{n}=\left( x_{n},y_{n},z_{n}\right) $ of the site; $\mu >0$ is the modulus
of\emph{\ }the magnetic moment projection onto the direction of the magnetic
field, which is called magnetic moment in what follows; the single-ion
anisotropic interaction is presented by the term $C^{ab}\hat{s}_{a}^{n}\hat{s%
}_{b}^{n}$, where the $C^{ab}$ is a constant symmetric rank-two tensor. Here
$\hat{\imath}$ is a vector of the length $l$ ($l$ is the lattice spacing)
pointing in the $i$-direction. The Kronecker delta $\delta ^{ab}$ and the
Levi-Civita symbol $\epsilon _{abc}$ are used for the internal spin indices,
$a,b=1,2,3$, and the summation over the repeated indices is implied.

We only consider the simple cubic-type lattice and the G-type
antiferromagnet, in which the N\'{e}el order appears along all the spatial
directions. In the absence of explicit symmetry-breaking terms ($\mu B^{a}=0$
and $C^{ab}=0$) Hamiltonian (\ref{d1}) enjoys $SO\left( 3\right) $ symmetry.
We can study effects of symmetry-breaking terms using the background field
(spurion) method if these terms are small enough compared to the symmetric
interaction ($\mu B^{a},\;C^{ab}\ll J$).

A continuum field-theoretical description of magnons is given by a $O\left(
3\right) $ nonlinear sigma model, in which a three-component unit vector $%
\mathbf{n}=\left( n^{1},n^{2},n^{3}\right) $ with $n^{a}n_{a}=1$ plays a
role as a dynamic degree of freedom. This unit vector expresses the N\'{e}el
order parameter. Taking the continuum limit of the background field, $%
B^{a}\left( \mathbf{r}_{n}\right) \rightarrow B^{a}\left( \mathbf{r}\right) $%
, and following the way described in Ref. \cite{HFMNS21}, one can construct
a $SO(3)$ gauge invariant effective Lagrangian. The corresponding space-time
is parametrized by coordinates $X=\left( t,\mathbf{r}\right) $, $t=X^{0}$, $%
\mathbf{r}=X^{j}=\left( x,y,z\right) $, $j=1,2,3$. The local $SO(3)$
transformation simply acts on the vector field $\mathbf{n}$ as $\mathbf{n}%
\rightarrow g(X)\mathbf{n}$ with $g(X)\in SO(3)$, as in the lattice case.
One can identify the background field $B^{a}$ as the $SO(3)$ gauge field on
which the local $SO(3)$ transformation acts as follows:%
\begin{equation}
B^{a}\rightarrow g(X)B^{a}g^{-1}(X)+g(X)\partial _{0}g^{-1}(X)\ .
\label{add1}
\end{equation}

The smallness of the terms $\mu B^{a}/J$\emph{\ }and\emph{\ }$C^{ab}/J$
allows us to neglect their higher orders in the future.\emph{\ }The only
effect of the symmetry-breaking term $\mu B^{a}$ is that the constant piece
of $B^{a}\left( \mathbf{r}\right) $ is used to tune the collinear ground
state.\emph{\ }In this case,\emph{\ }as it will be seen further, the field $%
B^{a}\left( x\right) $ can be treated as zero component $A_{0}^{a}\left(
x\right) $ of the electromagnetic potential in the theory of the charged
scalar field. However, the constant part of $B^{a}\left( \mathbf{r}\right) $
is a physical quantity.

In the leading order of the derivative expansion at low-energies, namely,
preserving derivatives only up to the second order, the $SO(3)$ gauge
invariant effective Lagrangian can be written as:%
\begin{equation}
\mathcal{L}=\frac{f_{t}^{2}}{2}\left( D_{0}n^{a}\right) ^{2}-\frac{f_{s}^{2}%
}{2}\left( \partial _{i}n^{a}\right) ^{2}+rC^{ab}n_{a}n_{b}\ ,  \label{d2}
\end{equation}%
where the covariant derivative $D_{0}$ with the $SO(3)$ background gauge
field is defined as:%
\begin{equation}
D_{0}n^{a}=\partial _{0}n^{a}-\epsilon _{bc}^{a}\ n^{b}\mu B^{c},\;\partial
_{0}=\frac{\partial }{\partial t}\,,  \label{add2}
\end{equation}%
and low-energy parameters $f_{t}$, $f_{s}$, and $r$ can be determined from
the underlying lattice model by the matching condition.

Suppose that our spin system possesses a potential with an easy-axis
anisotropy and develops the collinear ground state. We apply an
inhomogeneous magnetic field along the spin direction of the ground state.
We assume that the magnetic field points to the direction of axis $z$ and
depends on the coordinate $x$, $B^{a}\left( x\right) =B\left( x\right)
\delta ^{a3}$, and the sign of the field is positive, $B\left( x\right) >0$.
This field gives the collinear ground state with the N\'{e}el vector
pointing to the direction of axis $z$ as $\left\langle \mathbf{n}%
\right\rangle =\left( 0,0,1\right) $. Then one can introduce magnon complex
scalar fields $\Phi \left( X\right) $ and $\Phi ^{\ast }\left( X\right) $ as
fluctuations on the top of the ground state, which parametrize the vector $%
\mathbf{n}$ as%
\begin{equation}
\mathbf{n}=\left( \frac{\Phi +\Phi ^{\ast }}{\sqrt{2}},\frac{\Phi -\Phi
^{\ast }}{\sqrt{2}i},\sqrt{1-\Phi ^{\ast }\Phi }\right) ,  \label{add3}
\end{equation}%
where the constraint $n^{a}n_{a}=1$ is explicitly solved. Substituting this
parametrization into Eq. (\ref{d2}), one obtains the effective Lagrangian of
magnons at the quadratic order of fluctuation fields around the ground state
in the following form:%
\begin{eqnarray}
&&\mathcal{L}^{\left( 2\right) }=f_{t}^{2}\left( D_{0}\Phi ^{\ast }D_{0}\Phi
-\Delta ^{2}\Phi ^{\ast }\Phi \right) -f_{s}^{2}\delta ^{ij}\partial
_{i}\Phi ^{\ast }\partial _{j}\Phi ,  \notag \\
&&D_{0}\Phi =\left( \partial _{0}+iU\right) \Phi ,\;D_{0}\Phi ^{\ast
}=\left( \partial _{0}-iU\right) \Phi ^{\ast },  \label{d4}
\end{eqnarray}%
where the notation $U=\mu B$ and $rC^{ab}=\frac{1}{2}f_{t}^{2}\Delta
^{2}\delta ^{a3}\delta ^{b3}$ are used. We see that the field-theoretical
description of antiferromagnetic magnons embedded into an external
inhomogeneous magnetic field $B$ can be realized in terms of scalar
electrodynamics where a charged complex field $\Phi \left( X\right) $ (with $%
\mu $ playing the role of an electric charge) coupled to the zero component
of the electromagnetic potential $A_{0}=B$. Here the constant $%
v_{s}=f_{s}/f_{t}$ plays the role of the speed of light and the energy gap $%
\Delta $ plays the role of a mass term. We note that the constant $v_{s}$ is
not related to the speed of the light $c$ and is relatively small, e.g. $%
\Delta \sim 1\mathrm{meV}$ and $v_{s}\sim 60\mathrm{m/s}$ for
antiferromagnetic $MnF_{2}$ \cite{expdata06}.

It follows from the effective Lagrangian (\ref{d4}) that the corresponding
wave equation for the field $\Phi \left( X\right) $ is a modification of the
Klein-Gordon equation,%
\begin{equation}
\left( D_{0}^{2}-v_{s}^{2}\delta ^{ij}\partial _{i}\partial _{j}+\Delta
^{2}\right) \Phi \left( X\right) =0.  \label{d5}
\end{equation}

Summarizing, we can say that in the example under consideration, the $EFT$
model describing low-energy dynamics of antiferromagnetic\emph{\ }magnons
can technically be identified with the theory of a charged scalar field
interacting with an external constant electric field, which in our
terminology is an $x$ step; see Refs. \cite{GavGi16,GavGi20,BGavGi23}. In
this theory, the wave equation describing the corresponding charged
particles has the form of Eq. (\ref{d5}). A nonperturbative study (with
respect to the interaction with the external field) of various quantum
effects in such a system, in particular, the study of the vacuum
instability, can use the technique (with the necessary modifications due to
the specifics of the wave equation (\ref{d5})) developed earlier by two
coauthors (S.P. Gavrilov and D.M. Gitman) for strong-field $QED$ with $x$%
-potential steps and presented in Refs. \cite{GavGi16,GavGi20}. In the next
section, we study the vacuum instability in the system of low-energy magnons
following the formulated idea.

\section{Consideration in the framework of strong-field QED \label{S3}}

\subsection{Solutions of the Klein-Gordon equation with critical $x$ steps}

Solutions of the Klein-Gordon equation with critical $x$ steps are known in
the form of stationary plane waves. A complete set of such solutions reads:%
\begin{eqnarray}
&&\phi _{m}\left( X\right) =\exp \left( -i\varepsilon t+i\mathbf{p}_{\bot }%
\mathbf{r}_{\bot }\right) \varphi _{m}\left( x\right) \,,  \notag \\
&&\mathbf{r}_{\bot }=\left( 0,y,z\right) ,\ \ m=(\varepsilon ,\mathbf{p}%
_{\bot }).  \label{d6}
\end{eqnarray}%
In fact these are stationary states with well defined the total energy of a
particle $\varepsilon $ and with definite momenta $\mathbf{p}_{\bot }$ in
the perpendicular to the axis $x$ directions. Substituting Eq. (\ref{d6})
into Eq. (\ref{d5}), we obtain a second-order differential equation for the
function $\varphi _{m}(x)$,%
\begin{eqnarray}
&&\left\{ v_{s}^{2}\partial _{x}^{2}+\left[ \varepsilon -U\left( x\right) %
\right] ^{2}-\pi _{\bot }^{2}\right\} \varphi _{m}\left( x\right) =0\,,
\notag \\
&&\pi _{\bot }=\sqrt{v_{s}^{2}\mathbf{p}_{\bot }^{2}+\Delta ^{2}},
\label{d7}
\end{eqnarray}

Before considering the case of inhomogeneous magnetic field, where it is
necessary to use the recently developed in Refs. \cite%
{GavGi16,GavGi20,BGavGi23}\ approach to strong field QED with $x$ steps, it
is useful to discuss a simple case of homogeneous magnetic field with $%
A_{0}=B=\emph{const}>0$\emph{\ }such that $U=\mu B<\Delta $\emph{. }In this
case, the field\emph{\ }$\Phi \left( X\right) $\emph{\ }is a free field
satisfying equation (\ref{d5}) with the constant $U$\emph{. }A complete set
of solutions of this equation reads:%
\begin{eqnarray}
\phi _{m}^{\left( \pm \right) }\left( X\right) &=&N^{\left( \pm \right)
}\exp \left( -i\varepsilon ^{\left( \pm \right) }t+i\mathbf{pr}\right) ,
\notag \\
\varepsilon ^{\left( \pm \right) }-U &=&\pm \sqrt{v_{s}^{2}\mathbf{p}%
^{2}+\Delta ^{2}},  \label{dop1}
\end{eqnarray}%
where $N^{\left( \pm \right) }$\ is a normalization factor.\emph{\ }A
time-independent inner product of solutions of the Klein-Gordon equation is
proportional to the matrix elements of a field charge given by the following
integral over a spatial volume:
\begin{equation}
\int \left\{ \Phi ^{\ast }\left( i\partial _{0}-U\right) \Phi ^{\prime
}+\Phi \left[ \left( i\partial _{0}-U\right) \Phi ^{\prime }\right] ^{\ast
}\right\} dV.  \label{dop2}
\end{equation}%
This integral is positive (negative) for any superpositions of $\phi
_{m}^{\left( +\right) }$\emph{\ }($\phi _{m}^{\left( -\right) }$).{\large \ }%
The collective excitation described by the classical fields\emph{\ }$\phi
_{m}^{\left( +\right) }$\emph{\ }and\emph{\ }$\phi _{m}^{\left( -\right) }$%
\emph{\ }are fluctuations of the N\'{e}el\emph{\ }vector on the top of the
ground state, that is, semiclassically speaking, these fields describe the
fluctuations of the spin vector with clockwise and counterclockwise rotation
seen from the north pole.\ The realization of the scalar field in a Fock%
\emph{\ }space implies that field quanta be boson particles (with the
positive frequency $\varepsilon ^{\left( +\right) }$\ and the effective
charge $\mu $) and antiparticles (with the negative frequency $\varepsilon
^{\left( -\right) }$\ and the effective charge $-\mu $).\emph{\ }In this
case, the Hamiltonian of the quantum scalar field is positive defined and%
\emph{\ }the corresponding vacuum is the uncharged Fock state with minimal
energy.\emph{\ }The normalized plane waves $\phi _{m}^{\left( +\right) }$\
and $\phi _{m}^{\left( -\right) }$\ describe\emph{\ }a single-particle state
with the energy $\varepsilon ^{\left( +\right) }>0$\ and a
single-antiparticle state with the energy $-\varepsilon ^{\left( -\right)
}>0 $\ as well as with Zeeman energy terms for positive and negative
projections of a single-magnon magnetic moment,%
\begin{equation}
\pm \varepsilon ^{\left( \pm \right) }=\sqrt{v_{s}^{2}\mathbf{p}^{2}+\Delta
^{2}}\pm \mu B.  \label{dop3}
\end{equation}%
This is a charge conjugation rule. It implies also the following
interpretation of momentum quantum number: the quantum number $\mathbf{p}$\
is the physical momentum of a particle while the physical momentum of an
antiparticle is $-\mathbf{p}$. One sees that $\pm \left( \varepsilon
^{\left( \pm \right) }-U\right) $\ are kinetic energies (the energy gap $%
\Delta $ is included in the definition) of a particle and an antiparticle,
respectively. In this way, we establish a relation between
particle-antiparticle and quantum magnon interpretations. This resembles the
electron-hole interpretation of states in semiconductors. In the framework
of the effective scalar QED the collinear ground state with the given N\'{e}%
el vector can be considered as a vacuum state. Then a single-particle
(antiparticle) is a quant of the quantum magnon field with positive
(negative)\ magnetic moment projection $\mu $\ ($-\mu $) onto the direction
of the magnetic field, respectively. In which follows, keeping in mind this
particle-antiparticle interpretation, we'll call these quanta with opposite
effective charges magnon and antimagnon, respectively.

In general case with inhomogeneous $x$-dependent $B$\ field of a step form,
we chose the potential energy $U\left( x\right) $\ in the form of a
monotonically decreasing function,\emph{\ }$\partial _{x}U\left( x\right) <0$%
\emph{. }If the field derivative $\partial _{x}U\left( x\right) $\ is not
very big then the terms $\pm \left[ \varepsilon -U\left( x\right) \right] $\
can be considered as kinetic energies of particle and antiparticle,
respectively.\emph{\ }One sees that the kinetic energy of the particle
(antiparticle) would grow monotonously along (in inverse direction of) the
axis $x$, that is, the magnon and antimagnon accelerate under the influence
of the field derivative $\partial _{x}U\left( x\right) $\ in opposite
directions and form a spin current.

If the field derivative $\partial _{x}U\left( x\right) $ (playing here the
role of the electric field in the scalar electrodynamics) acting on magnons
produces enough work, it may lead to the production of magnon-antimagnon
pairs from the corresponding vacuum. We assume that the action of the field
derivative contributes significantly to mean numbers of created pairs in the
restricted region{\large \ }$S_{\mathrm{int}}${\large \ }between two planes $%
x=x_{\mathrm{L}}$ and $x=x_{\mathrm{R}}$ during the sufficiently large
(macroscopic)\ time period $T$. It is either negligible or switches off in
the macroscopic regions $S_{\mathrm{L}}=\left( x_{\mathrm{FL}},x_{\mathrm{L}}%
\right] $\ on the left of the plane $x=x_{\mathrm{L}}$ and in $S_{\mathrm{R}%
}=\left[ x_{\mathrm{R}},x_{\mathrm{FR}}\right) $ on the right of\ the plane $%
x=x_{\mathrm{R}}$. We also assume that the points $x_{\mathrm{FL}}$\ and $x_{%
\mathrm{FR}}$\ are separated from the origin by macroscopic but finite
distances. In this way, the magnetic field $B$ plays the role of an electric
potential step, which we call shortly an $x$ step. Its magnitude is:%
\begin{equation}
\delta U=U_{\mathrm{L}}-U_{\mathrm{R}}>0\,,\;U_{\mathrm{L}}=U\left( -\infty
\right) \,,\;U_{\mathrm{R}}=U\left( \infty \right) \,.  \label{d8}
\end{equation}%
We distinguish two types of $x$ steps, noncritical $\delta U<2\Delta ,$ and
critical $\delta U>2\Delta $. The pair production from the vacuum occurs due
to the critical $x$ step.

Nonperturbative calculations of the vacuum instability effects in the
framework of strong-field QED with $x$ steps are possible if there are
special complete sets of exact solutions of the corresponding relativistic
wave equations (solutions of the Klein-Gordon equations in the case under
consideration) orthonormalized on a $t$ constant\ hyperplane; see Ref. \cite%
{GavGi16}.{\large \ }In the strong-field QED with $x$ step, which acts
during the macroscopic time{\large \ }$T$, one can construct such sets of
solutions, see below. In our work these solutions are chosen in the form of
stationary plane waves with given real longitudinal momenta{\large \ }$p^{%
\mathrm{L}}${\large \ }and{\large \ }$p^{\mathrm{R}}${\large \ }in the
regions $S_{\mathrm{L}}$\ and $S_{\mathrm{R}}$. Moreover, it is necessary to
construct two types of complete sets of solution in the form of Eq. (\ref{d6}%
). The first one $_{\;\zeta }\phi _{m}\left( X\right) $ is constructed with
the help of the functions $\varphi _{m}\left( x\right) $ denoted as $%
_{\;\zeta }\varphi _{m}\left( x\right) $ and the second one,$^{\;\zeta }\phi
_{m}\left( X\right) $, with the help of functions $\varphi _{m}\left(
x\right) $ denoted as $^{\;\zeta }\varphi _{m}\left( x\right) $.
Asymptotically, these functions have the following forms:
\begin{eqnarray}
&&_{\;\zeta }\varphi _{m}\left( x\right) =\ _{\zeta }\mathcal{N}\exp \left[
ip^{\mathrm{L}}\left( x-x_{\mathrm{L}}\right) \right] ,\ x\in S_{\mathrm{L}},
\notag \\
&&^{\;\zeta }\varphi _{m}\left( x\right) =\ ^{\zeta }\mathcal{N}\exp \left[
ip^{\mathrm{R}}\left( x-x_{\mathrm{R}}\right) \right] ,\ x\in S_{\mathrm{R}},
\notag \\
&&p^{\mathrm{L}}=\frac{\zeta }{v_{s}}\sqrt{\left[ \pi _{0}\left( \mathrm{L}%
\right) \right] ^{2}-\pi _{\bot }^{2}},\ \ p^{\mathrm{R}}=\frac{\zeta }{v_{s}%
}\sqrt{\left[ \pi _{0}\left( \mathrm{R}\right) \right] ^{2}-\pi _{\bot }^{2}}%
,  \label{d9}
\end{eqnarray}%
where $\zeta =\pm $, $\ _{\zeta }\mathcal{N}$ and $^{\zeta }\mathcal{N}$ are
normalization constants that shall be determined below. We introduce the
quantities $\pi _{0}\left( \mathrm{L/R}\right) =\varepsilon -U_{\mathrm{L/R}%
} $. Note that $\pi _{0}\left( \mathrm{R}\right) >\pi _{0}\left( \mathrm{L}%
\right) $.

By analogy with the way it is done in the time-independent potential
scattering due to noncritical steps, it is assumed that time-independent
observables exist in the presence of critical $x$ steps. For example, it
seems quite natural that the pair-production rate and the flux of created
particles are constant during the macroscopic time\ $T$.\ This means that a
leading contribution to the number density of created particle-antiparticle
pairs is assumed to be proportional to the large dimensionless parameter $%
\sqrt{v_{s}\left\vert \partial _{x}U\right\vert }T$\ and is independent from
switching-on and -off if this parameter satisfies inequality%
\begin{equation}
T\gg \left( v_{s}\left\vert \partial _{x}U\right\vert \right) ^{-1/2}\max
\left\{ 1,\Delta ^{2}/v_{s}\left\vert \partial _{x}U\right\vert \right\} .
\label{d12}
\end{equation}%
{\large \ }It is clear that the process of pair creation is transient.\
Nevertheless, the condition of the smallness of backreaction shows there is
a window in the parameter range of $\partial _{x}U$\ and $T$\ where the
constant field approximation is consistent\ \cite{GavG08}.

For any two solutions $\Phi \left( X\right) $ and $\Phi ^{\prime }\left(
X\right) $ of the Klein-Gordon equation, the inner product on the hyperplane
$x=\mathrm{const}$ has the form of a longitudinal flux:
\begin{equation}
\left( \Phi ,\Phi ^{\prime }\right) _{x}=i\int \left[ \Phi ^{\prime }\left(
X\right) \partial _{x}\Phi ^{\ast }\left( X\right) -\Phi ^{\ast }\left(
X\right) \partial _{x}\Phi ^{\prime }\left( X\right) \right] dtd\mathbf{r}%
_{\bot }.  \label{scip}
\end{equation}%
Note, that this flux is proportional to the effective charge current.{\large %
\ }We consider the system under consideration in a large space-time box that
has a spatial area $V_{\bot }=K_{y}K_{z}$\ and the time dimension $T$, where
all $K_{y}$, $K_{z}$,\ and $T$\ are macroscopically large. In general the
wave packets $\Phi (X)$\ and $\Phi ^{\prime }(X)$\ can be decomposed into
plane waves $\phi _{m}(X)$\ and $\phi _{m}^{\prime }(X)$. Along with the
introduced plane waves, it is assumed that all the solutions $\Phi (X)$ are
periodic under transitions from one box to another. Then the integration in
integral (\ref{scip}) over the transverse coordinates runs from $-K_{j}/2$\
to $+K_{j}/2$, $j=y,z,$ and over the time $t$\ from $-T/2$\ to $+T/2$. We
assume that the macroscopic time{\large \ }$T$ is the system surveillance
time. Under these suppositions, one can verify, integrating by parts, that
the inner product (\ref{scip}) does not depend on $x$.

Nontrivial solutions $_{\zeta }\phi _{m}\left( X\right) $ and $^{\;\zeta
}\phi _{m}\left( X\right) $ exist only for the quantum numbers $m$ that obey
the following relations:%
\begin{equation}
\left[ \pi _{0}\left( \mathrm{L/R}\right) \right] ^{2}>\pi _{\bot
}^{2}\Longleftrightarrow \left\{
\begin{array}{l}
\pi _{0}\left( \mathrm{L/R}\right) >\pi _{\bot } \\
\pi _{0}\left( \mathrm{L/R}\right) <-\pi _{\bot }%
\end{array}%
\right. .  \label{2.6c}
\end{equation}%
In the case of critical $x$ steps and for $2\pi _{\bot }\leq \delta U\ $%
there exist five ranges\ $\Omega _{k}$, $k=1,...,5$, of quantum numbers%
{\large \ }$m${\large ,}%
\begin{eqnarray}
&&\pi _{0}\left( \mathrm{L}\right) \geq \pi _{\bot }\ \mathrm{\;if\;}m\in
\Omega _{1},  \notag \\
&&\ \left\vert \pi _{0}\left( \mathrm{L}\right) \right\vert <\pi _{\bot
},\;\pi _{0}\left( \mathrm{R}\right) >\pi _{\bot }\mathrm{\;if\;}m\in \Omega
_{2},  \notag \\
&&\pi _{0}\left( \mathrm{L}\right) \leq -\pi _{\bot },\;\pi _{0}\left(
\mathrm{R}\right) \geq \pi _{\bot }\mathrm{\;if\;}m\in \Omega _{3},  \notag
\\
&&\pi _{0}\left( \mathrm{L}\right) <-\pi _{\bot },\;\left\vert \pi
_{0}\left( \mathrm{R}\right) \right\vert <\pi _{\bot }\ \mathrm{\;if\;}m\in
\Omega _{4},  \notag \\
&&\pi _{0}\left( \mathrm{R}\right) \leq -\pi _{\bot }\ \mathrm{\;if\;}m\in
\Omega _{5}\ ,  \label{ranges}
\end{eqnarray}%
where the solutions$\ _{\zeta }\phi _{m}\left( X\right) $ and $\ ^{\zeta
}\phi _{m}\left( X\right) $ have similar forms and properties for a given $%
\pi _{\bot }${\large . }The manifold of all the quantum numbers $m$ is
denoted by $\Omega ,$ so that $\Omega =\Omega _{1}\cup \cdots \cup \ \Omega
_{5}$. In the ranges{\large \ }$\Omega _{2}${\large \ }and{\large \ }$\Omega
_{4}${\large \ }we deal with standing waves completed by linear
superpositions of solutions $\ _{\pm }\phi _{m}\left( X\right) $ and $\
^{\pm }\phi _{m}\left( X\right) $ {\large \ }with corresponding longitudinal
fluxes that are equal in magnitude for a given $m$. In fact, $\phi
_{m}\left( X\right) $\ for $m\in \Omega _{2}$ are wave functions that
describe an unbounded motion in $x\rightarrow \infty $ direction while $\phi
_{m}\left( X\right) $ for $m\in \Omega _{4}$ are wave functions that
describe an unbounded motion toward $x=-\infty $. It was demonstrated in the
framework of strong-field QED with $x$ steps; see Secs. V and VII and
Appendices C1 and C2 in Ref. \cite{GavGi16}, by using one-particle mean
currents and the energy fluxes that, depending on the asymptotic behavior in
the regions $S_{\mathrm{L}}$\ and $S_{\mathrm{R}}$, the plane waves $\ _{\pm
}\phi _{m}\left( X\right) $ and $\ ^{\pm }\phi _{m}\left( X\right) $\ are
identified unambiguously as components of the initial and final wave packets
of particles and antiparticles.

The plane waves are subjected to the following orthonormality conditions:%
\begin{eqnarray}
&&\left( \ _{\zeta }\phi _{m},\ _{\zeta ^{\prime }}\phi _{m^{\prime
}}\right) _{x}=\zeta \delta _{\zeta ,\zeta ^{\prime }}\delta _{m,m^{\prime
}}\,,  \notag \\
&&\left( \ ^{\zeta }\phi _{m},\ ^{\zeta ^{\prime }}\phi _{m^{\prime
}}\right) _{x}=\zeta \delta _{\zeta ,\zeta ^{\prime }}\delta _{m,m^{\prime
}}.  \label{d10}
\end{eqnarray}%
In fact, integrals (\ref{d10}) represent the flux densities of\ particles
with given{\large \ }$m${\large .} The normalization factors with respect to
the inner product (\ref{scip}) are:%
\begin{eqnarray}
&&\ _{\zeta }\mathcal{N}=\ _{\zeta }CY,\;\ ^{\zeta }\mathcal{N}=\ ^{\zeta
}CY,\;Y=\left( V_{\bot }T\right) ^{-1/2},  \notag \\
&&\ _{\zeta }C=\left\vert 2p^{\mathrm{L}}\right\vert ^{-1/2},\;\ ^{\zeta
}C=\left\vert 2p^{\mathrm{R}}\right\vert ^{-1/2}.  \label{d11}
\end{eqnarray}%
In the $K_{j}\rightarrow \infty $ and $T\rightarrow \infty $ limits, one has
to replace $\delta _{m,m^{\prime }}$ by the factor $\delta \left(
\varepsilon -\varepsilon ^{\prime }\right) \delta \left( \mathbf{p}_{\bot }-%
\mathbf{p}_{\bot }^{\prime }\right) $ in the normalization conditions (\ref%
{d10})\ and then to set $Y=\left( 2\pi \right) ^{-\left( d-1\right) /2}$.

Stationary plane waves of type (\ref{d9}) are commonly used in potential
scattering theory. They represent one-particle states with corresponding
conserved longitudinal currents. It is clear that in the ranges $\Omega _{1}$%
{\large \ }and $\Omega _{2}$\ we have deal with states of a particle whereas
in the ranges $\Omega _{4}$\ and $\Omega _{5}$\ the plane waves describe
states of an antiparticle.{\large \ } In these ranges particles and
antiparticles are subjected to the scattering and the reflection only. Such
one-particle consideration does not work in the range $\Omega _{3}$, where
the many-particle quantum field theory consideration is essential. Note that
the range $\Omega _{3}$ is often referred to as the Klein zone. In contrast
to the case $\Omega _{1}$ (and $\Omega _{5}$), where signs of $\pi
_{0}\left( \mathrm{L}\right) $ and $\pi _{0}\left( \mathrm{R}\right) $
coincide, they are opposite in the Klein zone. This reflects the fact that
the interpretation of overlapping between amplitudes{\large \ }$_{\zeta
}\phi _{m}\left( X\right) ${\large \ }and{\large \ }$\ ^{\zeta }\phi
_{m}\left( X\right) ${\large \ }using the quantities $\pi _{0}\left( \mathrm{%
L/R}\right) $ by analogy with potential scattering theory can be erroneous.

Indeed, it is known that attempts to consider the overlapping amplitudes in
the range $\Omega _{3}$ as amplitudes of particle transmission and
reflection as its works in the relativistic quantum mechanics led (and often
still lead researchers) to contradictions and paradoxes, before the advent
of consistent consideration in the framework of QFT. Therefore, we will
discuss this issue again below.

It is assumed that each pair of solutions $\ _{\zeta }\phi _{m}\left(
X\right) $ and $\ ^{\zeta }\phi _{m}\left( X\right) $ with given quantum
numbers $m\in \Omega _{1}\cup \Omega _{3}\cup \Omega _{5}$ is complete in
the space of solutions with each given $m$. Due to Eq. (\ref{d10}) the
corresponding mutual decompositions of such solutions have the form:%
\begin{eqnarray}
\ ^{\zeta }\phi _{m}\left( X\right) &=&\ _{+}\phi _{m}\left( X\right)
g\left( _{+}\left\vert ^{\zeta }\right. \right) -\ _{-}\phi _{m}\left(
X\right) g\left( _{-}\left\vert ^{\zeta }\right. \right) ,  \notag \\
\;_{\zeta }\phi _{m}\left( X\right) &=&\ ^{+}\phi _{m}\left( X\right)
g\left( ^{+}\left\vert _{\zeta }\right. \right) -\ ^{-}\phi _{m}\left(
X\right) g\left( ^{-}\left\vert _{\zeta }\right. \right) ,  \label{rel1}
\end{eqnarray}%
where decomposition coefficients $g$ are given by the relations:
\begin{eqnarray}
&&\left( \ _{\zeta }\phi _{m},\ ^{\zeta ^{\prime }}\phi _{m^{\prime
}}\right) _{x}=\delta _{m,m^{\prime }}g\left( \ _{\zeta }\left\vert ^{\zeta
^{\prime }}\right. \right) ,  \notag \\
&&g\left( \ ^{\zeta ^{\prime }}\left\vert _{\zeta }\right. \right) =g\left(
\ _{\zeta }\left\vert ^{\zeta ^{\prime }}\right. \right) ^{\ast },\ \ m\in
\Omega _{1}\cup \Omega _{3}\cup \Omega _{5}\ .  \label{c12}
\end{eqnarray}

Substituting Eq. (\ref{rel1}) into the orthonormality conditions (\ref{d10}%
), we derive the following unitary relations for the decomposition
coefficients:%
\begin{eqnarray}
&&g\left( \ ^{\zeta ^{\prime }}\left\vert _{+}\right. \right) g\left( \
_{+}\left\vert ^{\zeta }\right. \right) -g\left( \ ^{\zeta ^{\prime
}}\left\vert _{-}\right. \right) g\left( \ _{-}\left\vert ^{\zeta }\right.
\right) =\zeta \delta _{\zeta ,\zeta ^{\prime }}\ ,  \notag \\
&&g\left( \ _{\zeta ^{\prime }}\left\vert ^{+}\right. \right) g\left( \
^{+}\left\vert _{\zeta }\right. \right) -g\left( \ _{\zeta ^{\prime
}}\left\vert ^{-}\right. \right) g\left( \ ^{-}\left\vert _{\zeta }\right. \
\right) =\zeta \delta _{\zeta ,\zeta ^{\prime }}\ .  \label{UR}
\end{eqnarray}%
In particular, these relations imply that%
\begin{eqnarray}
&&\left\vert g\left( _{-}\left\vert ^{+}\right. \right) \right\vert
^{2}=\left\vert g\left( _{+}\left\vert ^{-}\right. \right) \right\vert
^{2},\;\left\vert g\left( _{+}\left\vert ^{+}\right. \right) \right\vert
^{2}=\left\vert g\left( _{-}\left\vert ^{-}\right. \right) \right\vert ^{2},
\notag \\
&&\frac{g\left( _{+}\left\vert ^{-}\right. \right) }{g\left( _{-}\left\vert
^{-}\right. \right) }=\frac{g\left( ^{+}\left\vert _{-}\right. \right) }{%
g\left( ^{+}\left\vert _{+}\right. \right) }.  \label{UR2}
\end{eqnarray}%
One can see that all the coefficients $g$ can be expressed via only two of
them, e.g. via $g\left( _{+}\left\vert ^{+}\right. \right) $ and $g\left(
_{+}\left\vert ^{-}\right. \right) $. However, even the latter coefficients
are not completely independent, they are related as follows:%
\begin{equation}
\left\vert g\left( _{+}\left\vert ^{+}\right. \right) \right\vert
^{2}-\left\vert g\left( _{+}\left\vert ^{-}\right. \right) \right\vert
^{2}=1.  \label{UR1}
\end{equation}%
Nevertheless, in what follows, we will use both coefficients $g\left(
_{+}\left\vert ^{-}\right. \right) $ and $g\left( _{+}\left\vert ^{+}\right.
\right) $ in our consideration. This maintains a certain symmetry in
important relations.

In canonical quantization of field theory, state vectors in the Fock space
are global objects, that is, the definition of vacuum and
particle(antiparticle) states has to be realized in the whole space in a
given time instant. To this end one has to use a time-independent inner
product of solutions of the Klein-Gordon equation. In the case under
consideration, this inner product must be adopted for the stationary plane
waves (\ref{d6}). We recall that the inner product between two solutions of
the Klein-Gordon equation can be defined on $t$-const. hyperplane as a
charge (in the case under consideration, the role of an effective charge
plays the magnetic moment). Note that physical states are wave packets that
vanish on the remote boundaries, that is why the effective charge of the
scalar field is finite. It allows one to integrate by parts in the inner
product neglecting boundary terms. The latter property provides the inner
product to be time independent. However, considering the stationary plane
waves (\ref{d6}) that are generalized states, which do not vanish at the
spatial infinity, one should take some additional steps. In the case under
consideration, the motion of particles in the $x$\ direction is unlimited,
therefore the corresponding wave functions cannot be subjected to any
periodic boundary conditions in the $x$-direction without changing their
physical meaning. For this reason one has to use a special kind of the
volume regularization; see Sec. C2 and Appendix B in Ref. \cite{GavGi16} and
Sec. 2.1 in Ref. \cite{GavGi20} for details. Following Refs. \cite%
{GavGi16,GavGi20}, the strong field $\partial _{x}U$ under consideration is
located inside the region $S_{\mathrm{int}}$\ during the time $T$.
Consequently, causally related to the area $S_{\mathrm{int}}$ can there be
only such parts of the areas $S_{\mathrm{L}}=\left( x_{\mathrm{FL}},x_{%
\mathrm{L}}\right] $ and $S_{\mathrm{R}}=\left[ x_{\mathrm{R}},x_{\mathrm{FR}%
}\right) $, which are located from it at distances not exceeding $v_{s}T$.%
{\large \ }The field derivative $\partial _{x}U$ is either negligible or
switches off in these macroscopic regions. We assume that there exist some
macroscopic but finite{\large \ }parameters $K^{\left( \mathrm{L/R}\right) }$
of the volume regularization which are in spatial areas where{\large \ }the
\ contribution of the field $\partial _{x}U$ is negligible,{\large \ }$%
\left\vert x_{\mathrm{FL}}\right\vert >K^{\left( \mathrm{L}\right) }\gg
\left\vert x_{\mathrm{L}}\right\vert >0${\large \ }and $x_{\mathrm{FR}%
}>K^{\left( \mathrm{R}\right) }\gg x_{\mathrm{R}}>0${\large .}

Following Refs. \cite{GavGi16,GavGi20} we propose the time-independent inner
product between two solutions $\Phi (X)$\ and $\Phi ^{\prime }(X)$\ of the
Klein-Gordon equation on $t$-constant\ hyperplane as%
\begin{eqnarray}
&&\left( \Phi ,\Phi ^{\prime }\right) =\frac{1}{v_{s}^{2}}\int_{V_{\bot }}d%
\mathbf{r}_{\bot }\int\limits_{-K^{\left( \mathrm{L}\right) }}^{K^{\left(
\mathrm{R}\right) }}\Psi ^{\dag }\left( X\right) \sigma _{1}\Psi \left(
X\right) dx\mathbf{\ ,}  \notag \\
&&\Psi \left( X\right) =\left(
\begin{array}{c}
i\partial _{0}-U\left( x\right)  \\
1%
\end{array}%
\right) \Phi \left( X\right) ,  \label{d13}
\end{eqnarray}%
where the integral over the spatial volume $V_{\bot }${\large \ }is
completed by an integral\ over the interval $\left[ K^{\left( \mathrm{L}%
\right) },K^{\left( \mathrm{R}\right) }\right] $ in the $x$ direction and $%
\sigma _{1}$ is a Pauli matrix. The parameters $K^{\left( \mathrm{L}/\mathrm{%
R}\right) }$ are assumed sufficiently large in final expressions. First, we
note that{\large \ }states with different quantum numbers $m$\ are
independent, therefore decompositions of wave packets $\Phi $ into the plane
waves (\ref{d6}) in Eq. (\ref{d13})\ do not contain interference terms.%
{\large \ }That is why it is enough to consider Eq. (\ref{d13}) only for a
particular case of plane waves{\large \ }$^{\zeta }\phi _{m}$ and{\large \ }$%
_{\zeta }\phi _{m}${\large \ }with equal{\large \ }$m$. Assuming that the
areas $S_{\mathrm{L}}$\ and $S_{\mathrm{R}}$\ are much wider than the area%
{\large \ }$S_{\mathrm{int}}${\large , }%
\begin{equation}
K^{\left( \mathrm{L}\right) }-\left\vert x_{\mathrm{L}}\right\vert
,K^{\left( \mathrm{R}\right) }-x_{\mathrm{R}}\gg x_{\mathrm{R}}-x_{\mathrm{L}%
},  \label{m0}
\end{equation}%
and the potential energy $U\left( x\right) $\ is a continuous function,%
{\large \ }the principal value of integral (\ref{d13}) is determined by
integrals over the areas $x\in \left[ -K^{\left( \mathrm{L}\right) },x_{%
\mathrm{L}}\right] $ and $x\in \left[ x_{\mathrm{R}},K^{\left( \mathrm{R}%
\right) }\right] $, where the field derivative $\partial _{x}U$ is
negligible small. Thus, it is possible to evaluate integrals of the form of
Eq. (\ref{d13}) for any form of the external field, using only the
asymptotic behavior (\ref{d9}) of functions in the regions $S_{\mathrm{L}}$%
and $S_{\mathrm{R}}$ where particles are free. The form of the field $%
\partial _{x}U$ in the area{\large \ }$S_{\mathrm{int}}$ affects only
coefficients $g$ entering into the mutual decompositions of the solutions
given by Eq. (\ref{rel1}). One can see that the norms of the plane waves
{\large \ }$_{\zeta }\phi _{m}${\large \ }and{\large \ }$_{\zeta }\phi _{m}$%
{\large \ } with respect to the inner product (\ref{d13}) are proportional
to the macroscopically large parameters{\large \ }$\tau ^{\left( \mathrm{L}%
\right) }=K^{\left( \mathrm{L}\right) }/v^{\mathrm{L}}${\large \ }and{\large %
\ }$\tau ^{\left( \mathrm{R}\right) }=K^{\left( \mathrm{R}\right) }/v^{%
\mathrm{R}}${\large , }where $v^{\mathrm{L}}=v_{s}^{2}\left\vert p^{\mathrm{L%
}}/\pi _{0}\left( \mathrm{L}\right) \right\vert >0$ and $v^{\mathrm{R}%
}=v_{s}^{2}\left\vert p^{\mathrm{R}}/\pi _{0}\left( \mathrm{R}\right)
\right\vert >0$ are absolute values of the longitudinal velocities of
particles in the regions $S_{\mathrm{L}}${\large \ }and{\large \ }$S_{%
\mathrm{R}}$, respectively; see Sec. IIIC.2 and Appendix B in Ref. \cite%
{GavGi16} for details.

It was shown (see Appendix B in Ref. \cite{GavGi16}) that\ the following
couples of plane waves are orthogonal with \ respect to the inner product (%
\ref{d13})%
\begin{eqnarray}
&&\left( _{\zeta }\phi _{m},^{-\zeta }\phi _{m}\right) =0,\ \ m\in \Omega
_{1}\cup \Omega _{5}\,,  \notag \\
&&\left( _{\zeta }\phi _{m},^{\zeta }\phi _{m}\right) =0,\ \ m\in \Omega
_{3}\ ,  \label{i7}
\end{eqnarray}%
if the parameters of the volume regularization $\tau ^{\left( \mathrm{L/R}%
\right) }$\ satisfy the condition%
\begin{equation}
\tau ^{\left( \mathrm{L}\right) }-\tau ^{\left( \mathrm{R}\right) }=O\left(
1\right) ,  \label{i8}
\end{equation}%
where $O\left( 1\right) $ denotes terms that are negligibly small in
comparison with the macroscopic quantities $\tau ^{\left( \mathrm{L/R}%
\right) }${\large . }One can see that $\tau ^{\left( \mathrm{R}\right) }$
and $\tau ^{\left( \mathrm{L}\right) }$ are macroscopic times of motion of
particles and antiparticles in the areas $S_{\mathrm{R}}$ and $S_{\mathrm{L}%
} $, respectively and they are equal,{\large \ }%
\begin{equation}
\tau ^{\left( \mathrm{L}\right) }=\tau ^{\left( \mathrm{R}\right) }=\tau .
\label{m4}
\end{equation}%
It allows one to introduce an unique time of motion $\tau $\ for all the
particles in the system under consideration. This time can be interpreted as
a system monitoring time during its evolution. Under condition (\ref{i8})
the norms of the plane waves on the $t$-constant hyperplane{\large \ }are%
\begin{eqnarray}
&&\left( \ _{\zeta }\phi _{m},\ _{\zeta }\phi _{m}\right) =\left( \ ^{\zeta
}\phi _{m},\ ^{\zeta }\phi _{m}\right) =\kappa \mathcal{M}_{m}\,,\ \kappa
=\left\{
\begin{array}{c}
1,\ m\in \Omega _{1} \\
-1,\ m\in \Omega _{5}%
\end{array}%
\right. \,,  \notag \\
&&\left( \ \phi _{m},\ \phi _{m}\right) =\mathcal{M}_{m}\;,\ m\in \Omega
_{2}\,;\left( \ \phi _{m},\ \phi _{m}\right) =-\mathcal{M}_{m}\ ,\ m\in
\Omega _{4}\ ,  \notag \\
&&\left( \ _{\zeta }\phi _{m},\ _{\zeta }\phi _{m}\right) =-\left( \ ^{\zeta
}\phi _{m},\ ^{\zeta }\phi _{m}\right) =\mathcal{M}_{m}\ ,\ \ m\in \Omega
_{3}\ ,  \notag \\
&&\mathcal{M}_{m}=2\frac{\tau }{T}\left\{
\begin{array}{l}
\left\vert g\left( _{+}\left\vert ^{+}\right. \right) \right\vert ^{2}\ ,\
m\in \Omega _{1}\cup \Omega _{5} \\
1\ ,\ m\in \Omega _{2}\cup \Omega _{4} \\
\left\vert g\left( _{+}\left\vert ^{-}\right. \right) \right\vert ^{2}\ ,\
m\in \Omega _{3}%
\end{array}%
\right. ;  \label{d14}
\end{eqnarray}%
see Appendix B in Ref. \cite{GavGi16} for details. The $\pm $ signs of
integrals in Eq. (\ref{d14}) correspond to the signs of the effective charge
(magnetic moment) of a particle; see appendix \ref{Ap}) for details.

Thus, there are constructed two linearly independent couples of complete%
{\large \ }on the $t$-constant hyperplane states with a given $m${\large \ }%
that are either initial ("\textrm{in}") or final ("\textrm{out}") states,%
\begin{eqnarray}
&&\mathrm{in\ states:}\mathrm{\ }\phi _{m_{1}}^{\left( \mathrm{in},+\right)
}=\ _{+}\phi _{m_{1}},\ \phi _{m_{1}}^{\left( \mathrm{in},-\right) }=\
^{-}\phi _{m_{1}};  \notag \\
&&\phi _{m_{5}}^{\left( \mathrm{in},+\right) }=\ ^{+}\phi _{m_{5}},\mathrm{\
}\phi _{m_{5}}^{\left( \mathrm{in},-\right) }=\ _{-}\phi _{m_{5}};  \notag \\
&&\phi _{m_{3}}^{\left( \mathrm{in},+\right) }=\ _{-}\phi _{m_{3}},\;\phi
_{m_{3}}^{\left( \mathrm{in},-\right) }=\ ^{-}\phi _{m_{3}};\ \   \notag \\
&&\mathrm{out\ states:}\mathrm{\ }\phi _{m_{1}}^{\left( \mathrm{out}%
,-\right) }=\ _{-}\phi _{m_{1}},\ \phi _{m_{1}}^{\left( \mathrm{out}%
,+\right) }=\ ^{+}\phi _{m_{1}};  \notag \\
&&\phi _{m_{5}}^{\left( \mathrm{out},-\right) }=\ ^{-}\phi _{m_{5}},\mathrm{%
\ }\phi _{m_{5}}^{\left( \mathrm{out},+\right) }=\ _{+}\phi _{m_{5}};  \notag
\\
&&\phi _{m_{3}}^{\left( \mathrm{out},+\right) }=\ _{+}\phi _{m_{3}},\ \ \phi
_{m_{3}}^{\left( \mathrm{out},-\right) }=\ ^{+}\phi _{m_{3}},\ \;m_{k}\in
\Omega _{k}.  \label{in-out}
\end{eqnarray}

Note that standing waves{\large \ }$\phi _{m}\left( X\right) ${\large \ }for%
{\large \ }$m\in \Omega _{2}\cup \Omega _{4}${\large \ }are the same for
initial and final sets.{\large \ }In the ranges $\Omega _{1}$\ and $\Omega
_{2}$\ we have deal with states of a particle whereas in the ranges $\Omega
_{4}$\ and $\Omega _{5}$\ the plane waves describe states of an
antiparticle. In these cases, description of the one-particle scattering and
reflection in the framework of strong-field QED is the same as in the
framework of the potential scattering. In particular, the sign of the flux
density, given by Eq. (\ref{d10}), allows one to determine initial and final
states, this fact was already used above. In doing so we take into account
the charge conjugation, which implies that the physical longitudinal
momentum of an antiparticle differs in sign from the quantum numbers $p^{%
\mathrm{L}}$ and $p^{\mathrm{R}}$; see Appendix \ref{Ap} for more details.%
{\large \ }We also note that in the range $\Omega _{3}$ the choice of
initial and final states is not so obvious and will be justified below.

\subsection{Quantization in terms of particles}

In the absence of an explicit time evolution of physical quantities
presented by the stationary plane waves (\ref{d6}), an interpretation of
these plane waves as states of initial and final particle or antiparticle in
the Klein zone $\Omega _{3}$ demands a consideration of possible violation
of vacuum stability due to effects of switching on and off the external
field. It is clear that the effect of pair creation is transient and
therefore must be limited in time. However, it can be assumed that moderate
intensity of pair creation one can neglect backreaction and that the
external field remains unchanged for some macroscopic period of time $T$.
Then physically it make sense to believe that the field of the $x$ step, $%
\partial _{x}B\left( x\right) $, should be considered as a part of a
time-dependent inhomogeneous field $E_{\mathrm{pristine}}=\partial _{x}B(x)$
directed along the $x$-direction, which was switched on at the time instant $%
t_{1}$ sufficiently fast before the time instant $t_{\mathrm{in}}$, by this
time it had time to spread to the whole area $S_{\mathrm{int}}$ and
disappear in the macroscopic regions $S_{\mathrm{L}}$\ and $S_{\mathrm{R}}$.
Thus, in the region $S_{\mathrm{int}}$ a time-independent field
configuration $E_{\mathrm{pristine}}$ is formed whereas in the two regions$%
S_{\mathrm{L}}$\emph{\ }and\emph{\ }$S_{\mathrm{R}}$\emph{\ }the field\emph{%
\ }$\partial _{x}B\left( x\right) $\emph{\ }is zero.\emph{\ }However, there
are two different uniform fields\emph{\ }$B\left( x_{\mathrm{L}}\right) \neq
0$\emph{\ }in\emph{\ }$S_{\mathrm{L}}$\emph{\ }and\emph{\ }$B\left( x_{%
\mathrm{R}}\right) \neq 0$\emph{\ }in\emph{\ }$S_{\mathrm{R}}$;\emph{\ }$\mu %
\left[ B\left( x_{\mathrm{L}}\right) -B\left( x_{\mathrm{R}}\right) \right]
=\delta U$. Such a field configuration remains unchanged for quite a long
time\emph{\ }$T$\emph{. }Then at time instant\emph{\ }$t_{2}$\emph{\ }the
field\emph{\ \ }$E_{\mathrm{pristine}}$\emph{\ }is switched off sufficiently
fast\ just after the time instant\emph{\ }$t_{\mathrm{out}}=t_{\mathrm{in}%
}+T $\emph{. }The main effect of the pair creation occurs in the region\emph{%
\ }$S_{\mathrm{int}}$ during the time period $T$. However, the switching on
and off of an external field\emph{\ }$E_{\mathrm{pristine}}$ may also lead
to a pair creation. We believe that its contribution is much less then the
one in the region $S_{\mathrm{int}}$. \emph{\ }The created magnons and
antimagnons enter in the regions $S_{\mathrm{L}}$\emph{\ }and\emph{\ }$S_{%
\mathrm{R}}$\emph{\ }respectively already as free particles and remain there
separately after switching off the field $E_{\mathrm{pristine}}$\emph{. }%
Thus, by observing over a long period of time $T$\emph{\ }fluxes of created
particles crossing the boundaries of the field at the planes\emph{\ }$x=x_{%
\mathrm{L}}$\emph{\ }and\emph{\ }$x=x_{\mathrm{R}}$\emph{\ }respectively, it
is possible to determine the parameters of these fluxes without waiting for
switching off the field\emph{\ }$E_{\mathrm{pristine}}$\emph{. }This can be
done using stationary plane wave solutions of the Klein-Gordon equation in
the framework of approach presented in Ref. \cite{GavGi16}. (We note that
some relevant details of scalar field quantization in the presence of
critical potential steps is given in Appendix \ref{Ap}).

Now you can set a quantitative criterion that allows one to evaluate the
accuracy of the approximation in which the effects of switching on and off
are neglected. Let $N^{\mathrm{true}}$ be the total number of pairs\ created
from the vacuum\ due to field\ $E_{\mathrm{pristine}}$ from the time it is
turned on $t_{1} $ to the time it is turned off $t_{2} $ , and $N^{\mathrm{cr%
}}${\large \ }be the total number of pairs\ created from the vacuum\ by the
field $\partial _{x}B\left( x\right) $ from a moment $t_{\mathrm{in}}>t_{1}$
to a moment $t_{\mathrm{out}}=t_{\mathrm{in}}+T$.

According to a widely accepted formulation of $QED$ with a strong background
there exist initial $\left\vert 0,\mathrm{true\;in}\right\rangle $ and final
$\left\vert 0,\mathrm{true\;out}\right\rangle $ vacua of free particles
related by an unitary transformation under the condition that $N^{\mathrm{%
true}}$ is finite; see, e.g., Refs.\emph{\ }\cite{FGS91,GavGT06}\emph{. }In
the Heisenberg representation these vacua are null vectors for particle
number operators $\hat{N}\left( t_{1}\right) $ and $\hat{N}\left(
t_{2}\right) $,
\begin{equation}
\hat{N}\left( t_{1}\right) \left\vert 0,\mathrm{true\;in}\right\rangle =0,\;%
\hat{N}\left( t_{2}\right) \left\vert 0,\mathrm{true\;out}\right\rangle =0,
\label{q1a}
\end{equation}%
where $\hat{N}\left( t_{1}\right) $ is free particle number operator at the
time $t_{1}$ and $\hat{N}\left( t_{2}\right) $ is free particle number
operator at the time $t_{2}$. The difference of the true final vacuum from
the initial one can be specified by the total number of\ pairs created from
the vacuum,%
\begin{equation}
N^{\mathrm{true}}=\sum_{m}N_{m}^{\mathrm{true}}=\left\langle 0,\mathrm{%
true\;in}\left\vert \hat{N}\left( t_{2}\right) \right\vert 0,\mathrm{true\;in%
}\right\rangle ,  \label{q1b}
\end{equation}%
and\emph{\ }$N_{m}^{\mathrm{true}}$\emph{\ }are differential mean numbers of
created from the vacuum pairs with given quantum numbers $m$. If there
exists a complete set of solutions of the Klein-Gordon equation with the
background under consideration, then one can find all the numbers $N_{m}^{%
\mathrm{true}}$. \ From general considerations, it is possible to establish
some properties of these numbers without knowing their form explicitly. It
seems quite natural that the pair-production rate and the flux of created
particles are constant during the macroscopic time\ $T$.\ It means that a
leading contribution to the number density $N^{\mathrm{true}}$ is assumed to
be proportional to the large dimensionless parameter\emph{\ }$\sqrt{%
v_{s}\left\vert \partial _{x}U\right\vert }T$\emph{\ }and is independent
from switching-on and -off if this parameter satisfies inequality (\ref{d12}%
). In this case, the numbers $N^{\mathrm{true}}$\ can be approximated by $N^{%
\mathrm{cr}}$,%
\begin{eqnarray}
&&N^{\mathrm{true}}=N^{\mathrm{cr}}\left\{ 1+O\left( \left[ \sqrt{%
v_{s}\left\vert \partial _{x}U\right\vert }T\right] ^{-1}\right) \right\} ,
\notag \\
&&N^{\mathrm{cr}}=\sum_{m\in \Omega _{3}}N_{m}^{\mathrm{cr}}\ ,  \label{q5}
\end{eqnarray}%
\emph{\ }where the terms\emph{\ }$N_{m}^{\mathrm{cr}}=N_{m}^{\mathrm{true}}$%
\emph{\ }if $m\in \Omega _{3}$\emph{\ }appear due to the time-independent
part of the field $E_{\mathrm{pristine}}$\emph{\ }and do not depend on the
oscillations related to fast switching-on and -off of the field $E_{\mathrm{%
pristine}}$. This possibility does exist due to the fact that\ both fast
switching-on and -off produce particle-antiparticle pairs with quantum
numbers in a tiny range of the kinetic energy, such that one can neglect the
corresponding contributions to total characteristics of vacuum instability
that are determined by sums over all the kinetic energies; see Ref. \cite%
{GavGi20} for more details.

The numbers $N_{m}^{\mathrm{cr}}$ can be found by using the decomposition
coefficients $g$\ , given by Eq. (\ref{c12}). Here one interprets stationary
plane wave solutions of the Klein-Gordon equation as states of initial and
final particle or antiparticle in the Klein zone $\Omega _{3}$.

Neglecting effects of fast switching-on and -off,\emph{\ }one can use
instead of the true vacua $\left\vert 0,\mathrm{true\;in}\right\rangle $\
and $\left\vert 0,\mathrm{true\;out}\right\rangle $\ some states $\left\vert
0,\mathrm{in}\right\rangle $\ and $\left\vert 0,\mathrm{out}\right\rangle $\
respectively. Namely, we choose these states as ones with minimum kinetic
energies (kinetic energies of the states $\left\vert 0,\mathrm{in}%
\right\rangle $\ and $\left\vert 0,\mathrm{out}\right\rangle $\ are the
same) and such that\ the leading contribution to the quantity $N^{\mathrm{%
true}}$\ is determined by the number $N^{\mathrm{cr}}$.\emph{\ }In the
Heisenberg representation these vacua are null vectors of the particle
number operators $\hat{N}($\textrm{in}$)$\ and $\hat{N}\left( \mathrm{out}%
\right) $,
\begin{equation}
\hat{N}(\mathrm{in})\left\vert 0,\mathrm{in}\right\rangle =0,\;\hat{N}\left(
\mathrm{out}\right) \left\vert 0,\mathrm{out}\right\rangle =0.  \label{q2a}
\end{equation}%
The difference between these vectors is determined by the total number of\
created pairs,%
\begin{equation}
N^{\mathrm{cr}}=\left\langle 0,\mathrm{in}\left\vert \hat{N}\left( \mathrm{%
out}\right) \right\vert 0,\mathrm{in}\right\rangle .  \label{q2b}
\end{equation}

In what follows, the number operators\emph{\ }$\hat{N}(\mathrm{in})$ and$\;%
\hat{N}\left( \mathrm{out}\right) $ will be expressed via corresponding
annihilation and creation operators\emph{\ }and the states $\left\vert 0,%
\mathrm{in}\right\rangle $\ and $\left\vert 0,\mathrm{out}\right\rangle $\
are called the initial and the final vacua, respectively.\emph{\ }%
Accordingly, magnon and antimagnon excitations over these vacua are called
initial and final particles respectively. Mean fluxes of the effective
charge and the energies of the vacuum states\ $\left\vert 0,\mathrm{in}%
\right\rangle $\ and\ $\left\vert 0,\mathrm{out}\right\rangle $\ are quite
distinct.\emph{\ }That allows us to define unambiguously these vacua\ and to
construct initial and final states in a Fock space, using the plane waves (%
\ref{d6}); see Appendix \ref{Ap} for details.

One can decompose the Heisenberg operators of the scalar fields $\hat{\Phi}%
\left( X\right) $\ and $\hat{\Phi}\left( X\right) ^{\dagger }$ into
solutions of either the initial or final complete sets (\ref{in-out}). It is
useful to represent $\hat{\Phi}\left( X\right) $ (and $\hat{\Phi}\left(
X\right) ^{\dagger }$ respectively) as sums of five operators, $\hat{\Phi}%
\left( X\right) =\sum_{k=1}^{5}\hat{\Phi}_{k}\left( X\right) $, where the
operators $\hat{\Phi}_{k}\left( X\right) $ are defined in the ranges $\Omega
_{k}$; see Eqs. (\ref{dec}) and (\ref{ab}) in Appendix \ref{Ap} for details.
The operators $\hat{\Phi}_{k}\left( X\right) $, $k=1,2$, are decomposed via
the creation and annihilation operators $a$ and $a^{\dag }$ of the magnons,
while the operators $\hat{\Phi}_{k}\left( X\right) $, $k=4,5$, are
decomposed via the creation and annihilation operators $b\ $and $b^{\dag }$
of antimagnons. In the range $\Omega _{3}$ the operators $\hat{\Phi}%
_{3}\left( X\right) $ and $\hat{\Phi}_{3}^{\dagger }\left( X\right) $ are
decomposed via creation and annihilation operators of both magnons and
antimagnons,

\begin{widetext}
\begin{eqnarray}
\hat{\Phi}_{3}\left( X\right) &=& \sum_{m\in \Omega _{3}}\mathcal{M}%
_{m}^{-1/2}\left[ \ _{-}a_{m}(\mathrm{in})\ _{-}\phi _{m}\left( X\right) +\
^{-}b_{m}^{\dagger }(\mathrm{in})\ ^{-}\phi _{m}\left( X\right) \right]\,,
\notag \\
&=&\sum_{m\in \Omega _{3}}\mathcal{M}_{m}^{-1/2}\left[ \ _{+}a_{m}(\mathrm{%
out})\ _{+}\phi _{m}\left( X\right) +\ ^{+}b_{m}^{\dagger }(\mathrm{out})\
^{+}\phi _{m}\left( X\right) \right]\,.  \label{d15}
\end{eqnarray}%
\end{widetext}

Here $\mathcal{M}_{m}$ are normalization factors given by Eq. (\ref{d14}).
All the operators labeled by the argument "in" are interpreted\ as the
operators of the initial particles, whereas all the operators labeled by the
argument "out" are interpreted as the operators of the final particles. This
identification can be confirmed as follows. The equal-time commutation
relations given by Eq. (\ref{a2}) in Appendix \ref{Ap} yield the standard
commutation rules for the creation and annihilation \textrm{in}- or \textrm{%
out}-operators introduced. The vacuum vectors $\left\vert 0,\mathrm{in}%
\right\rangle $ and $\left\vert 0,\mathrm{out}\right\rangle $ are null
vectors for all annihilation operators $a$ and $b\ $given by Eq. (\ref{ab})
in Appendix \ref{Ap}. In particular, in the range $\Omega _{3}$ nonzero
commutators are:%
\begin{equation}
\left[ \ _{\mp }a_{m}(\mathrm{in/out}),\ _{\mp }a_{m^{\prime }}^{\dagger }(%
\mathrm{in/out})\right] =\left[ \ ^{\mp }b_{m}(\mathrm{in/out}),\ ^{\mp
}b_{m}^{\dagger }(\mathrm{in/out})\right] =\delta _{m,m^{\prime }},
\label{d16}
\end{equation}%
{\large \ }and one-particle states of initial(final) magnon and antimagnon
have the form:%
\begin{equation}
\ _{\mp }a_{m^{\prime }}^{\dagger }(\mathrm{in/out})\left\vert 0,\mathrm{%
in/out}\right\rangle ,\;\ ^{\mp }b_{m}^{\dagger }(\mathrm{in/out})\left\vert
0,\mathrm{in/out}\right\rangle .  \label{add4}
\end{equation}

Using the both alternative decompositions (\ref{d15}) for $\hat{\Phi}%
_{3}\left( X\right) $ and the orthonormality conditions (\ref{d10}), one can
find the following linear canonical transformation between the introduced
\textrm{in}- and \textrm{out}- creation and annihilation operators%
\begin{eqnarray}
_{+}a_{m}(\mathrm{out}) &=&g\left( ^{-}\left\vert _{+}\right. \right)
^{-1}g\left( ^{-}\left\vert _{-}\right. \right) \;_{-}a_{m}(\mathrm{in})
\notag \\
&-&g\left( ^{-}\left\vert _{+}\right. \right) ^{-1}\;^{-}b_{m}^{\dagger }(%
\mathrm{in})\,,  \notag \\
^{+}b_{m}^{\dagger }(\mathrm{out}) &=&-g\left( _{-}\left\vert ^{+}\right.
\right) ^{-1}\;_{-}a_{m}(\mathrm{in})  \notag \\
&+&g\left( _{-}\left\vert ^{+}\right. \right) ^{-1}g\left( _{-}\left\vert
^{-}\right. \right) \;^{-}b_{m}^{\dagger }(\mathrm{in})\,.  \label{d17}
\end{eqnarray}%
One can verify that the transformation is unitary; see section VIID in Ref.
\cite{GavGi16}. The inverse transformation reads%
\begin{eqnarray}
_{-}a_{m}(\mathrm{in}) &=&g\left( ^{+}\left\vert _{-}\right. \right)
^{-1}g\left( ^{+}\left\vert _{+}\right. \right) \ _{+}a_{m}(\mathrm{out})
\notag \\
&+&g\left( ^{+}\left\vert _{-}\right. \right) ^{-1}\ ^{+}b_{m}^{\dagger }(%
\mathrm{out})\,,  \notag \\
^{-}b_{m}^{\dagger }(\mathrm{in}) &=&g\left( _{+}\left\vert ^{-}\right.
\right) ^{-1}\ _{+}a_{m}(\mathrm{out})  \notag \\
&+&g\left( _{+}\left\vert ^{-}\right. \right) ^{-1}g\left( _{+}\left\vert
^{+}\right. \right) \ ^{+}b_{m}^{\dagger }(\mathrm{out})\,.  \label{d18}
\end{eqnarray}

The $\pm $ sign of integrals in Eq. (\ref{d14}) corresponds to the sign of
the effective charge (magnetic moment) of a particle, which justifies the
above interpretation of states as magnons and antimagnons in the ranges $%
\Omega _{k},$ $k=1,2,4,5$. In the ranges $\Omega _{1}$ and $\Omega _{5}$ the
flux densities of\ particles through the surfaces\ $x=x_{\mathrm{L}}$\ and\ $%
x=x_{\mathrm{R}}$ given by Eq. (\ref{d10}) allow one to define initial and
final states of particles. In particular, it can be seen that the sign of
the flux densities of\ magnons with a given $m$\ is equal to $\zeta $\ in
the range $\Omega _{1},$\ while the sign of the flux densities of\
antimagnons with a given $m$ \ is equal to $-\zeta $ in the range $\Omega
_{5}$. The corresponding reasons in the framework of QFT are presented in
Appendix \ref{Ap}. One can see that the partial vacua in the Fock subspaces
with a given{\large \ }$m${\large \ }are stable in $\Omega _{k},$ $k=1,2,4,5$%
. In the ranges $\Omega _{1}$\ and $\Omega _{5}$\ we meet a realization of
rules of the potential scattering theory in the framework of QFT and see
that relative probabilities of reflection and transmission (under the
condition that the vacuum remain the vacuum),{\large \ }%
\begin{equation}
\left\vert R_{m}\right\vert ^{2}=\left\vert g\left( _{+}\left\vert
^{+}\right. \right) ^{-1}g\left( _{-}\left\vert ^{+}\right. \right)
\right\vert ^{2},\;\left\vert T_{m}\right\vert ^{2}=\left\vert g\left(
_{+}\left\vert ^{+}\right. \right) \right\vert ^{-2},  \label{RT}
\end{equation}%
coincide with mean currents of reflected particles $J_{R}=\left\vert
R_{m}\right\vert ^{2}$\ and transmitted particles $J_{T}=\left\vert
T_{m}\right\vert ^{2}$. The correct result $J_{R}+J_{T}=1$\ follows from the
unitary relation (\ref{UR1}).

In contrast to the ranges $\Omega _{1}$ and $\Omega _{5},$ one can see from
Eq. (\ref{d14}) that in the range $\Omega _{3}$ the magnetic moment of the
particle states $_{\zeta }\phi _{m}$ is positive while the magnetic moment
of the particle states $\ ^{\zeta }\phi _{m}$ is negative. This allows us to
use the magnon/antimagnon classification for the creation and annihilation
operators in Eq. (\ref{d15}). Considering diagonalized forms of the kinetic
energy and magnetic moment operators given by Eq. (\ref{a12}) in Appendix %
\ref{Ap}, we see that such identification is also confirmed in the framework
of QFT. In particular, we see that the spectrum of the kinetic energy
operator is positively defined and one-particle states $_{\zeta }\phi _{m}$
represent magnons with the kinetic energy $_{\zeta }E_{m}>0$\ and the
magnetic moment $\mu ,$\ whereas one-particle states $^{\zeta }\phi _{m}$\
are antimagnons with the kinetic energy $-\;^{\zeta }E_{m}>0$.

Since transformations (\ref{d17}) and (\ref{d18}) entangle annihilation and
creation operators, the vacua $\left\vert 0,\mathrm{in}\right\rangle $ and $%
\left\vert 0,\mathrm{out}\right\rangle $ are essentially different. The
total vacuum-to-vacuum transition amplitude $c_{v}$ is formed due to the
vacuum instability in the range $\Omega _{3}$. Differential mean numbers $%
N_{m}^{a}\left( \mathrm{out}\right) $ and $N_{m}^{b}\left( \mathrm{out}%
\right) $, $m\in \Omega _{3}$, of the magnons and antimagnons respectively,\
created from the vacuum are equal, $N_{m}^{b}\left( \mathrm{out}\right)
=N_{m}^{a}\left( \mathrm{out}\right) =N_{m}^{\mathrm{cr}}$, and have the
forms:%
\begin{eqnarray}
N_{m}^{a}\left( \mathrm{out}\right) &=&\left\langle 0,\mathrm{in}\left\vert
\ _{+}a_{m}^{\dagger }(\mathrm{out})\ _{+}a_{m}(\mathrm{out})\right\vert 0,%
\mathrm{in}\right\rangle  \notag \\
&=&\left\vert g\left( _{+}\left\vert ^{-}\right. \right) \right\vert ^{-2},
\notag \\
N_{m}^{b}\left( \mathrm{out}\right) &=&\left\langle 0,\mathrm{in}\left\vert
\ ^{+}b_{m}^{\dagger }(\mathrm{out})\ ^{+}b_{m}(\mathrm{out})\right\vert 0,%
\mathrm{in}\right\rangle  \notag \\
&=&\left\vert g\left( _{-}\left\vert ^{+}\right. \right) \right\vert ^{-2},
\label{Na}
\end{eqnarray}%
where the coefficients $g$\ are given by Eq. (\ref{c12}).{\large \ }

To distinguish initial and final states in the range $\Omega _{3}$, one
needs to consider one-particle mean values of the operators of the fluxes,
of the energy and the effective charge (that is, the magnetic moment
current) through the surfaces{\large \ }$x=x_{\mathrm{L}}${\large \ }and%
{\large \ }$x=x_{\mathrm{R}}$, given by Eqs. (\ref{a7}) and (\ref{a8}) in
Appendix \ref{Ap}. In the beginning we note that in the range $\Omega _{3}$
the spatial distribution of physical states, presented by wave packets of
plane waves, is the same as in the ranges $\Omega _{2}$ and $\Omega _{4}$.
Therefore, it can be shown that particles (magnons) can be situated only in
the region $S_{\mathrm{R}}$, whereas antiparticles (antimagnons) can be
situated only in the region $S_{\mathrm{L}}$. The field $\partial
_{x}U\left( x\right) $ does not allow particles to penetrate through the
region $S_{\mathrm{int}}$, and turns them in the opposite direction. For the
plane waves such a behavior can be easily seen in the case of weak external
fields (but still strong enough, $\delta U>2\pi _{\bot }$, to provide the
existence of the $\Omega _{3}$-range) using a semiclassical approximation.
If $N_{m}^{\mathrm{cr}}$ tends to zero, then $\left\vert g\left(
_{+}\left\vert ^{-}\right. \right) \right\vert ^{2}\rightarrow \infty $ and,
at the same time, $\left\vert g\left( _{+}\left\vert ^{+}\right. \right)
\right\vert ^{2}\rightarrow \infty $ in accordance to relation (\ref{UR1}).
Relations (\ref{rel1}) imply that for an arbitrary $m\in \Omega _{3}$ the
magnon densities $\left\vert \ _{\zeta }\phi _{m}\left( X\right) \right\vert
^{2}$ are concentrated in the region $S_{\mathrm{R}},$ whereas the
antimagnon densities $\left\vert \ ^{\zeta }\phi _{m}\right\vert ^{2}$ \ are
concentrated in the region $S_{\mathrm{L}}$.

In the general case when the quantities $N_{m}^{\mathrm{cr}}$\ are not
small, it is natural to expect a\ similar behavior, namely: the region $S_{%
\mathrm{L}}$\ is not available for magnons, and the region $S_{\mathrm{R}}$
is not available for antimagnons. However, when the quantities $N_{m}^{%
\mathrm{cr}}$\ are not small, the latter property may hold only for the
corresponding wave packets, but not for the separate plane waves. That means
that these plane waves may be different from zero in the whole space. Namely
this fact leads often to a misinterpretation, since the behavior of these
plane waves looks like the one in the ranges $\Omega _{1}$ and $\Omega _{5}$%
, where they represent one-particle densities both in the region $S_{\mathrm{%
L}}$ and $S_{\mathrm{R}}$. However, this similarity is misleading. Indeed,
within our context it is assumed \ that the magnons and antimagnons in one
of corresponding asymptotic regions may occupy quasistationary states, i.e.,
they should be described by wave packets that pertain their form a
sufficiently long time in these regions.\ Note, that in the ranges $\Omega
_{1}$ and $\Omega _{5}$, the sign of the longitudinal momentum $p^{\mathrm{%
L/R}}$ is related to the sign of the mean energy flux in the region $S_{%
\mathrm{L/R}}$. In the range $\Omega _{3}$ the magnon states $_{\zeta }\phi
_{m}$ are states with a$\ $definite quantum number $p^{\mathrm{L}}$, whereas
the antimagnon states $^{\zeta }\phi _{m}$ are states with a\ definite
quantum number $p^{\mathrm{R}}$. This fact together with relation (\ref{rel1}%
)\textrm{\ }implies, for example, that a partial wave $_{+}\phi _{m}$ of a
magnon, in the region where this particle can really be observed, i.e., in
the region $S_{\mathrm{R}}$, is always a superposition of two waves $%
^{+}\phi _{m}$ and $^{-}\phi _{m}$ with opposite signs of the quantum number
$p^{\mathrm{R}}$. Thus, the sign of the mean energy flux in the region $S_{%
\mathrm{R}}$ cannot be related to the sign of an asymptotic momentum in this
region. Similarly, one can see, for example, that the partial wave $^{+}\phi
_{m}$ of an antimagnon, in the region $S_{\mathrm{L}}$, is always a
superposition of two waves with quantum number $p^{\mathrm{L}}$ of opposite
signs\ and, therefore, the sign of the mean energy flux cannot be related to
the sign of an asymptotic momentum in the region where this particle can
really be observed. However, as it will be demonstrated, these are states
with well-defined asymptotic energy flux and, therefore, with a
corresponding well-defined asymptotic field momentum. Namely, these
properties of the constituent plane waves are responsible for the fact that
stable magnon wave packets can exist only in the region $S_{\mathrm{R}}$,
whereas stable antimagnon wave packets can exist only in the region $S_{%
\mathrm{L}}$; see appendix D in Ref. \cite{GavGi16} for details.

Taking into account such a space separation of the magnons and antimagnons
one can use the one-particle mean values of fluxes, of the kinetic energy,
and the magnetic moment, given by Eq. (\ref{a15}) in appendix \ref{Ap}, to
differ initial and final states in the range $\Omega _{3}$. So if the flux
of the magnetic moment and the kinetic energy in the region $S_{\mathrm{R}}$
coincides with the acceleration direction of a magnon in the region $S_{%
\mathrm{int}}$, then the state under consideration is a final state of the
magnon, since such a particle can only move away from the region $S_{\mathrm{%
int}}$ ($x\rightarrow \infty $). And vice versa, if these fluxes are
opposite to the acceleration direction of a magnon in the region $S_{\mathrm{%
int}}$, then the state under consideration is an initial state of the
magnon, since such a particle can only move to the region $S_{\mathrm{int}}$%
. In the case of antimagnons, the direction of the flux of the kinetic
energy coincides with the direction of the flux density, but is opposite to
the direction of the flux of the magnetic moment. The antimagnons do exist
in the region $S_{\mathrm{L}}$ only. Therefore, if the direction of the flux
of the kinetic energy in the region $S_{\mathrm{L}}$ coincides with the
acceleration direction of the antimagnon in the region $S_{\mathrm{int}}$,
then the state under consideration is the final state of an antimagnon. And
vice versa, if the direction of the flux of the kinetic energy in the region
$S_{\mathrm{L}}$ is opposite to the acceleration direction of an antimagnon
in the region $S_{\mathrm{int}}$, then the state under consideration is an
initial state of an antimagnon. Namely in such a manner initial and final
states in Eq. (\ref{in-out}) are defined.

\subsection{Observable physical quantities specifying the vacuum instability}

In the preceding section, we used representations (\ref{Na}) to calculate
the differential mean number of magnon-antimagnon pairs created from the
vacuum only. However, one can obtain additional characteristics of the
vacuum instability. This section is devoted to their study.

The probability of the transition from the vacuum\emph{\ }$\left\vert 0,%
\mathrm{in}\right\rangle $\emph{\ }to the vacuum\emph{\ }$\left\vert 0,%
\mathrm{out}\right\rangle $\emph{,\ }%
\begin{equation}
P_{v}=\left\vert c_{v}\right\vert ^{2},\;c_{v}=\left\langle 0,\mathrm{out}%
\right. \left\vert 0,\mathrm{in}\right\rangle ,  \label{q3}
\end{equation}%
is related to the mean numbers\emph{\ }$N_{m}^{\mathrm{cr}}$\emph{\ }as%
\begin{equation}
\ln P_{v}=\sum_{m\in \Omega _{3}}\ln p_{m},\;p_{m}=\left( 1+N_{m}^{\mathrm{cr%
}}\right) ^{-1}\ ;  \label{q6}
\end{equation}%
see appendix A in Ref. \cite{GavGi16} for details. However, this probability
can be represented via the imaginary part of a one-loop effective action $S$%
\ by the seminal Schwinger formula \cite{Schw51},%
\begin{equation}
P_{\mathrm{v}}=\exp \left( -2\mathrm{Im}S\right) .  \label{np1a}
\end{equation}%
A relation of this representation with the one that follows from the locally
constant field approximation for the Schwinger's effective action was found
in Ref. \cite{GGSh19}. The probabilities of the magnon reflection and the
magnon-antimagnon pair creation can be expressed via the mean numbers $%
N_{m}^{\mathrm{cr}}$ as follows:
\begin{eqnarray}
P(+|+)_{m,m^{\prime }} &=&|\langle 0,\mathrm{out}|\ _{+}a_{m}(\mathrm{out}%
)\;_{-}a_{m}^{\dagger }(\mathrm{in})|0,\mathrm{in}\rangle |^{2}  \notag \\
&=&\delta _{m,m^{\prime }}\left( 1+N_{m}^{\mathrm{cr}}\right) ^{-1}P_{v}\,,
\notag \\
P(+-|0)_{m,m^{\prime }} &=&|\langle 0,\mathrm{out}|\ _{+}a_{m}(\mathrm{out}%
)\ ^{+}b_{m}(\mathrm{out})|0,\mathrm{in}\rangle |^{2}  \notag \\
&=&\delta _{m,m^{\prime }}N_{m}^{\mathrm{cr}}\left( 1+N_{m}^{\mathrm{cr}%
}\right) ^{-1}P_{v}\,,  \label{d21}
\end{eqnarray}%
The probabilities of the antimagnon reflection and the magnon-antimagnon
pair annihilation coincide with the quantities $P(+|+)$ and $P(+-|0)$,
respectively. In the case of bosons in a given state $m$\ any number of
pairs can be created from the vacuum and from the one particle state.{\large %
\ }By this reason probabilities (\ref{d21}) are not representative if the
mean numbers $N_{m}^{\mathrm{cr}}$\ are large.{\large \ }In the partial
state with a given $m$\ the probability of the creation of any pairs with
given $m$\ is $1-p_{m}$\ where $p_{m}$\ is the probability that the partial
vacuum state remains a vacuum, given by Eq. (\ref{q6}). If all the mean
numbers $N_{m}^{\mathrm{cr}}$ are sufficiently small, $N_{m}^{\mathrm{cr}%
}\ll 1$, then\ the simple relations{\large \ }$p_{m}\approx 1-N_{m}^{\mathrm{%
cr}}${\large \ }and $1-P_{v}\approx N^{\mathrm{cr}}\ll 1$\ hold true in the
leading approximation. In this case $P(+|+)_{m,m}\approx 1$ and $%
P(+-|0)_{m,m}\approx N_{m}^{\mathrm{cr}}$. Therefore, information about the
quantity $P_{v}$ allows one to estimate the total number $N^{\mathrm{cr}}$.
It is in this case that the Schwinger's effective action approach \cite%
{Schw51} to calculating $P_{v}$ turns out to be useful. We note that this%
{\Huge \ }approach is a base of a number of approximation methods; see,
e.g., Ref. \cite{Adv-QED22} for a review. In this relation, it should be
noted that the probability $P_{v}$ by itself is not very useful in the case
of strong fields when $P_{v}\ll 1$.

Taking into account Eq. (\ref{Na}), the total number of pairs\ created from
the vacuum{\large \ }reads:{\large \ }
\begin{equation}
N^{\mathrm{cr}}=\sum_{m\in \Omega _{3}}N_{m}^{\mathrm{cr}}=\sum_{m\in \Omega
_{3}}\left\vert g\left( _{+}\left\vert ^{-}\right. \right) \right\vert ^{-2}.
\label{TN}
\end{equation}%
Magnons and antimagnons created with quantum numbers $m$ leaving the area $%
S_{\mathrm{int}}$ enter the areas $S_{\mathrm{L}}$ and $S_{\mathrm{R}}$,
respectively. At the same time, the magnons continue to move in the $x$
direction with a constant velocity $v^{\mathrm{R}}$. The motion of the
magnons forms the flux density%
\begin{equation}
\left\langle j_{x}\right\rangle _{m}=N_{m}^{\mathrm{cr}}(TV_{\perp })^{-1}
\label{d19a}
\end{equation}%
in the area $S_{\mathrm{R}}$, while the antimagnon motion in the opposite
direction with the constant velocities $-v^{\mathrm{L}}$ forms the flux
density $-\left( j_{x}\right) _{m}$ in the area $S_{\mathrm{L}}$. Here it is
taken into account that differential mean numbers of created magnons and
antimagnons with a given $m$ are equal. The total flux densities of the
magnons and the antimagnons are%
\begin{equation}
\left\langle j_{x}\right\rangle =\sum_{m\in \Omega _{3}}\left\langle
j_{x}\right\rangle _{m}=N^{\mathrm{cr}}(TV_{\perp })^{-1}  \label{d19b}
\end{equation}%
and $-\left\langle j_{x}\right\rangle $, respectively. The effective charge
(the magnetic moment) current density of both created {\large \ }magnons and
antimagnons is $J_{x}^{\mathrm{cr}}=\mu \left\langle j_{x}\right\rangle $.
This corresponds to the spin current{\large \ }$\left\langle
j_{x}\right\rangle $\emph{. }It is conserved in the $x$-direction.

During the time $T,$ the created magnons carry the magnetic moment $\mu
\left\langle j_{x}\right\rangle _{m}T$ over the unit area $V_{\bot }$ of the
surface $x=x_{\mathrm{R}}$. This magnetic moment is evenly distributed over
the cylindrical volume of the length $v^{\mathrm{R}}T$. Thus, the magnetic
moment density of the magnons created with a given $m$ is $\mu
j_{m}^{0}\left( \mathrm{R}\right) $, where $j_{m}^{0}\left( \mathrm{R}%
\right) =\left\langle j_{x}\right\rangle _{m}/v^{\mathrm{R}}$ is the number
density of the magnons. During the time $T,$ the created antimagnons carry
the magnetic moment $\mu \left\langle j_{x}\right\rangle _{m}T$ over the
unit area $V_{\bot }$ of the surface $x=x_{\mathrm{L}}$. Taking into account
that this magnetic moment is evenly distributed over the cylindrical volume
of the length $v^{\mathrm{L}}T$, we can see that the magnetic moment density
of the antimagnons created with a given $m$ is $-\mu j_{m}^{0}\left( \mathrm{%
L}\right) $, where $j_{m}^{0}\left( \mathrm{L}\right) =\left\langle
j_{x}\right\rangle _{m}/v^{\mathrm{L}}$ is the number density of the
magnons. The total magnetic moment density of the created particles reads:%
\begin{equation}
\rho ^{\mathrm{cr}}\left( x\right) =\mu \left\{
\begin{array}{c}
-\sum_{m\in \Omega _{3}}j_{m}^{0}\left( \mathrm{L}\right) ,\ \ x\in S_{%
\mathrm{L}} \\
\sum_{m\in \Omega _{3}}j_{m}^{0}\left( \mathrm{R}\right) ,\ x\in S_{\mathrm{R%
}}%
\end{array}%
\right. .\   \label{d22}
\end{equation}%
Due to a relation between the velocities $v^{\mathrm{L}}$ and $v^{\mathrm{R}%
} $, the total number densities of the created magnons and antimagnons are
the same,%
\begin{equation}
\sum_{m\in \Omega _{3}}j_{m}^{0}\left( \mathrm{L}\right) =\sum_{m\in \Omega
_{3}}j_{m}^{0}\left( \mathrm{R}\right) \ .  \label{add5}
\end{equation}%
We also note that \ the created magnons and antimagnons are spatially
separated and carry magnetic moments that tend to smooth out the
inhomogeneity of the external magnetic field.

In the same manner, one can derive some representation for the nonzero
components of EMT of the created particles:%
\begin{eqnarray}
&&T_{\mathrm{cr}}^{00}(x)=\left\{
\begin{array}{c}
\sum_{m\in \Omega _{3}}j_{m}^{0}\left( \mathrm{L}\right) \left\vert \pi
_{0}\left( \mathrm{L}\right) \right\vert ,\ x\in S_{\mathrm{L}} \\
\sum_{m\in \Omega _{3}}j_{m}^{0}\left( \mathrm{R}\right) \pi _{0}\left(
\mathrm{R}\right) ,\mathrm{\ }x\in S_{\mathrm{R}}%
\end{array}%
\right. ,  \notag \\
&&T_{\mathrm{cr}}^{11}(x)=\left\{
\begin{array}{c}
\sum_{m\in \Omega _{3}}\left\langle j_{x}\right\rangle _{m}\left\vert p^{%
\mathrm{L}}\right\vert ,\mathrm{\ }x\in S_{\mathrm{L}} \\
\sum_{m\in \Omega _{3}}\left\langle j_{x}\right\rangle _{m}\left\vert p^{%
\mathrm{R}}\right\vert ,\mathrm{\ }x\in S_{\mathrm{R}}%
\end{array}%
\right. ,  \notag \\
&&T_{\mathrm{cr}}^{kk}(x)=\left\{
\begin{array}{c}
\sum_{m\in \Omega _{3}}\left\langle j_{x}\right\rangle _{m}\left(
p_{k}\right) ^{2}/\left\vert p^{\mathrm{L}}\right\vert ,\mathrm{\ }x\in S_{%
\mathrm{L}} \\
\sum_{m\in \Omega _{3}}\left\langle j_{x}\right\rangle _{m}\left(
p_{k}\right) ^{2}/\left\vert p^{\mathrm{R}}\right\vert ,\mathrm{\ }x\in S_{%
\mathrm{R}}%
\end{array}%
\right. ,\ k=2,3,  \notag \\
&&T_{\mathrm{cr}}^{10}(x)=\left\{
\begin{array}{c}
-\frac{1}{v_{s}}\sum_{m\in \Omega _{3}}\left\langle j_{x}\right\rangle
_{m}\left\vert \pi _{0}\left( \mathrm{L}\right) \right\vert ,\mathrm{\ }x\in
S_{\mathrm{L}} \\
\frac{1}{v_{s}}\sum_{m\in \Omega _{3}}\left\langle j_{x}\right\rangle
_{m}\pi _{0}\left( \mathrm{R}\right) ,\mathrm{\ }x\in S_{\mathrm{R}}%
\end{array}%
\right. .  \label{d23}
\end{eqnarray}%
Here $T_{\mathrm{cr}}^{00}(x)$ and $T_{\mathrm{cr}}^{kk}(x)$, $k=1,2,3$, are
energy density and components of the pressure of the particles created in
the areas $S_{\mathrm{L}}$ and $S_{\mathrm{R}}$ respectively, whereas $T_{%
\mathrm{cr}}^{10}(x)v_{s}$, for $x\in S_{\mathrm{L}}$ or $x\in S_{\mathrm{R}%
} $, is the energy flux density of the created particles\ through the
surfaces $x=x_{\mathrm{L}}$ or $x=x_{\mathrm{R}}$ respectively. In a strong
field, or in a field with the sufficiently large potential step $\delta U$ ,
the energy density and the pressure{\large \ }along the direction of the
axis $x$ are near equal.

Let us consider effects of the backreaction on the external field due to the
vacuum instability, to establish the so-called consistency conditions. We
assume that the volume $V=V_{\bot }\left( x_{\mathrm{R}}-x_{\mathrm{L}%
}\right) $ contains the area $S_{\mathrm{int}}=\left( x_{\mathrm{L}},x_{%
\mathrm{R}}\right) .$ The total energy of the created particles in the
volume $V$ is given by the corresponding volume integral of the energy
density $T_{\mathrm{cr}}^{00}\left( t,x\right) .$ The corresponding energy
conservation low reads:%
\begin{equation}
\frac{\partial }{\partial t}\int_{V_{\bot }}d\mathbf{r}_{\bot }\int_{x_{%
\mathrm{L}}}^{x_{\mathrm{R}}}T_{\mathrm{cr}}^{00}\left( t,x\right)
dx=-\oint_{\Sigma }v_{s}T_{\mathrm{cr}}^{k0}(x)df_{k}\ ,  \label{m24}
\end{equation}%
where $\Sigma $ is a surface surrounding the volume $V$ and $df_{k}$, $%
k=1,2,3$, are the components of the surface element $d\mathbf{f}$. Taking
into account that $T_{\mathrm{cr}}^{00}\left( t,x\right) $ does not depend
on the transversal coordinates, and $T_{\mathrm{cr}}^{k0}(x)=0$ for $k\neq
1, $ we find using Eq. (\ref{d23}) that the rate of the energy density
change of the created particles in the region $S_{\mathrm{int}}$ per unit of
the spatial area $V_{\bot }$ is:
\begin{eqnarray}
\frac{\partial }{\partial t}\int_{x_{\mathrm{L}}}^{x_{\mathrm{R}}}T_{\mathrm{%
cr}}^{00}\left( t,x\right) dx &=& v_{s}\left[ \left. T_{\mathrm{cr}%
}^{10}(x)\right\vert _{x\in S_{\mathrm{L}}}-\left. T_{\mathrm{cr}%
}^{10}(x)\right\vert _{x\in S_{\mathrm{R}}}\right]  \notag \\
&=&-\delta Uj_{x}\ .  \label{m25}
\end{eqnarray}%
It characterizes the loss of the energy that the created particles carry
away from the region $S_{\mathrm{int}}$. At the same time, the constant rate
(\ref{m25}) determines the power of the constant effective field $E_{\mathrm{%
pristine}}=\partial _{x}U\left( x\right) $ spent on the pair creation.
Integrating this rate over the time duration of the field from $t_{\mathrm{in%
}}$ to $t_{\mathrm{out}}$, and using the notation%
\begin{equation*}
\Delta T_{\mathrm{cr}}^{00}\left( x\right) =-\int_{t_{\mathrm{in}}}^{t_{%
\mathrm{out}}}\frac{\partial }{\partial t}T_{\mathrm{cr}}^{00}\left(
t,x\right) dt\ ,
\end{equation*}%
we find the total energy density of created pairs per unit of the area $%
V_{\bot }$ as%
\begin{equation}
\int_{x_{\mathrm{L}}}^{x_{\mathrm{R}}}\Delta T_{\mathrm{cr}}^{00}\left(
x\right) dx=\delta U\frac{N^{\mathrm{cr}}}{V_{\perp }}\ .  \label{m26}
\end{equation}

In strong-field QED it is usually assumed that just from the beginning there
exists a classical effective field having a given energy. The system of
particles interacting with this field is closed, that is, the total energy
of the system is conserved\footnote{%
One can, however, imagine an alternative situation when these effective
charges are getting out of the regions $S_{\mathrm{L}}$\ and $S_{\mathrm{R}}$%
\ with the help of the work done by an external storage battery. For
example, dealing with graphene devices, it is natural to assume that the
constant electric strength on the graphene plane is due to the applied fixed
voltage, i.e., we are dealing with an open system of fermions interacting
with a classical electromagnetic field. In that case there would be no
backreaction problem. Note that the evolution of the mean electromagnetic
field in the graphene, taking into account the backreaction of the matter
field to the applied time-dependent external field, was considered in Ref.
\cite{GavGitY12}.}{\large . }It is clear that due to pair creation from the
vacuum, the constant effective field $E_{\mathrm{pristine}}=\partial
_{x}U\left( x\right) $ is losing its energy and should depleted with time.%
\textrm{\ }Thus, the applicability of the constant field approximation,
which is used in the formulation of strong field QED with $x$ step, is
limited by the smallness of the backreaction. The relation (\ref{m26})
allows one to find conditions that provide this smallness, we call these
relations the consistency conditions. These conditions can be obtained from
the requirement that the energy density given by Eq. (\ref{m26}) is
essentially smaller than the energy density of the constant effective field
per unit of the area $V_{\bot }$.

Note that the presence of the matter in the initial state increases the mean
number of created bosons. It is an obvious consequence of the Bose-Einstein
statistics.{\large \ }In the case of fermions, the presence of the matter at
the initial state prevents the pair creation.{\large \ }Assuming that $%
N_{m}^{(+)}\left( \mathrm{in}\right) $\ and $N_{m}^{(-)}\left( \mathrm{in}%
\right) $\ are the mean numbers of particles and antiparticles with quantum
numbers $m$\ at the initial time instant, one obtains that the differential
mean numbers of final particles and antiparticles are
\begin{equation}
N_{m}^{(\zeta )}=\left( 1+N_{m}^{\mathrm{cr}}\right) N_{m}^{(\zeta )}(%
\mathrm{in})+N_{m}^{\mathrm{cr}}\left[ 1+N_{m}^{(-\zeta )}(\mathrm{in})%
\right] \,,  \label{dm30}
\end{equation}%
respectively.{\large \ }The differential mean numbers of particles and
antiparticles created by the external field are given by an increment{\large %
\ } $\Delta N_{m}^{(\zeta )}=N_{m}^{(\zeta )}-N_{m}^{(\zeta )}(\mathrm{in})$%
. One can see that the increments of the numbers of particles and
antiparticles are equal,{\large \ }%
\begin{eqnarray}
&&\Delta N_{m}^{(+)}=\Delta N_{m}^{(-)}=\Delta N_{m}\,,  \notag \\
&&\Delta N_{m}=N_{m}^{\mathrm{cr}}\left[ 1+N_{m}^{(+)}(\mathrm{in}%
)+N_{m}^{(-)}(\mathrm{in})\right] \,.  \label{ex5}
\end{eqnarray}%
In contrast to the previously used methods for studying the production of
bosonic pairs by external fields, our approach allows us to consider the
case of special inhomogeneous external fields supporting the spatial
separation of particles and antiparticles (in the case under consideration,
these are magnons and antimagnons) in the Klein zone.{\large \ }In such a
way, one can see that the equal increments of mean numbers of particles in
the area $S_{\mathrm{R}}$\ and antiparticles in the area $S_{\mathrm{L}}$\
do not depend on the symmetry between the mean numbers of particles and
antiparticles in the initial state.{\large \ }For example, assuming the
absence of the initial antiparticles, $N_{m}^{(-)}(\mathrm{in})=0$, with the
number of initial particles being not zero in the Klein zone, $N_{m}^{(+)}(%
\mathrm{in})\neq 0$,\ one can see that the number of created antiparticles
is growing in comparison with the one created from the vacuum, $\Delta
N_{m}=N_{m}^{\mathrm{cr}}\left[ 1+N_{m}^{(+)}(\mathrm{in})\right] $.
Therefore, the flux of created antiparticles in the area $S_{\mathrm{L}}$ is
growing proportionally to the flux of coming particles from the area{\large %
\ }$S_{\mathrm{R}}$.{\large \ }Such a behavior can be called
statistically-assisted Schwinger effect.

That is why operating with the concept of probability turns out to be
unfruitful in the case when the mean number $N_{m}^{\mathrm{cr}}$ is not
relatively small.\textrm{\ }In our considerations the presence of particles
in the initial state implies{\large \ }that{\large \ }these are ingoing
particles and the mean numbers $N_{m}^{(\zeta )}(\mathrm{in})$\ are
proportional to densities of ingoing fluxes,{\large \ }%
\begin{equation}
\left\langle j_{x}^{\left( \zeta \right) }(\mathrm{in})\right\rangle
_{m}=N_{m}^{(\zeta )}(\mathrm{in})(TV_{\perp })^{-1}.  \label{add6}
\end{equation}%
Densities of outgoing fluxes are:{\large \ }%
\begin{equation}
\left\langle j_{x}^{\left( \zeta \right) }\right\rangle _{m}=N_{m}^{(\zeta
)}(TV_{\perp })^{-1}.  \label{add7}
\end{equation}%
Both ingoing and outgoing magnons are situated in the area\ $S_{\mathrm{R}}$%
{\large \ }while both ingoing and outgoing antimagnons are situated in the
area{\large \ }$S_{\mathrm{L}}${\large . }For example, assuming the absence
of initial antiparticles,{\large \ }$N_{m}^{(-)}(\mathrm{in})=0${\large , }%
the presence of particles in the initial state,{\large \ }$N_{m}^{(+)}(%
\mathrm{in})\neq 0$, leads to the fact that the density of the outgoing
particle flux turns out to be more than the density of incoming particle
flux,
\begin{equation}
\left\langle j_{x}^{\left( +\right) }\right\rangle _{m}\left/ \left\langle
j_{x}^{\left( +\right) }(\mathrm{in})\right\rangle _{m}\right. =1+N_{m}^{%
\mathrm{cr}}\left( 1+1\left/ N_{m}^{(+)}(\mathrm{in})\right. \right) .
\label{add8}
\end{equation}%
Thus, the flux proportional to $N_{m}^{\mathrm{cr}}$ of particles born from
the vacuum is added to the total flux of reflected particles. A similar
picture is observed for antiparticle fluxes in the case when $N_{m}^{(+)}(%
\mathrm{in})=0$ while{\large \ }$N_{m}^{(-)}(\mathrm{in})\neq 0$. In the
areas of $\Omega _{3}$\ adjoining the borders of the ranges{\large \ }$%
\Omega _{2}$ and $\Omega _{4}$, the pair creation is absent,{\large \ }$%
N_{m}^{\mathrm{cr}}\rightarrow 0${\large ,} and the only the total
reflection takes place. However, in general, in the Klein zone, fluxes due
to the total reflection cannot be separated from the fluxes due to the pair
creation, that is one more reason not to use probabilities of the reflection.

\section{Examples of exact solutions with x steps\label{S4}}

In this section, we present a collection of external magnetic fields that
can be used to calculate the characteristics of magnon pair production based
on the exact solutions of Eq. (\ref{d7}). For the sake of convenience, we
discuss examples separately and list pertinent results only. Further details
are placed in Appendix \ref{B}.

\subsection{Differential quantities\label{Ss4a}}

\subsubsection{$L$-constant step\label{Ss4.1}}

The $L$-constant magnetic step is a model of magnetic field inhomogeneity
that grows linearly with $x$ within $S_{\mathrm{int}}$ and is constant
outside of it, $\left. B\left( x\right) \right\vert _{x\leq x_{\mathrm{L}%
}}\neq \left. B\left( x\right) \right\vert _{x\geq x_{\mathrm{R}}}$. We call
this field \textquotedblleft $L$-constant\textquotedblright\ magnetic step
due to its analogy with the \textquotedblleft $L$-constant electric
field\textquotedblright , which is a type of electric field that creates
electron-positron pairs from the vacuum if it is strong enough; see Ref.
\cite{GavGit16b} for a discussion. The field has the following form:%
\begin{equation}
B\left( x\right) =\left\{
\begin{array}{ll}
B^{\prime }L/2\,, & x\in S_{\mathrm{L}}=\left( -\infty ,-L/2\right] \,, \\
-B^{\prime }x\,, & x\in S_{\mathrm{int}}=\left( -L/2,L/2\right) \,, \\
-B^{\prime }L/2 & x\in S_{\mathrm{R}}=\left[ L/2,+\infty \right) \,,%
\end{array}%
\right.  \label{se41.1}
\end{equation}%
where $B^{\prime }>0$, $L>0$, and we set $x_{\mathrm{L}}=-L/2=-x_{\mathrm{R}%
} $ for simplicity.

Beyond the intermediate interval potential energies are constants, $U_{%
\mathrm{L}}=+\mu B^{\prime }L/2$ and $U_{\mathrm{R}}=-\mu B^{\prime }L/2$,
and exact solutions to Eq. (\ref{d7}) are plane waves, classified according
to Eqs. (\ref{d9}). As for the intermediate interval, $S_{\mathrm{int}}$, we
perform a change of variable%
\begin{equation}
\xi \left( x\right) =\frac{\varepsilon +\mu B^{\prime }x}{\sqrt{v_{s}\mu
B^{\prime }}}\,,  \label{se41.2}
\end{equation}%
to rewrite Eq. (\ref{d7}) as%
\begin{equation}
\left( \frac{d^{2}}{d\xi ^{2}}+\xi ^{2}-\lambda \right) \varphi _{m}\left(
\xi \right) =0\,,\ \ \lambda =\frac{\pi _{\bot }^{2}}{v_{s}\mu B^{\prime }}%
\,.  \label{add9}
\end{equation}%
This is Weber's parabolic cylinder differential equation \cite{Erdelyi},
whose independent sets of solutions are $D_{\nu }\left[ \left( 1-i\right)
\xi \right] $, $D_{-\nu -1}\left[ \left( 1+i\right) \xi \right] $ or $D_{\nu
}\left[ -\left( 1-i\right) \xi \right] $, and $D_{-\nu -1}\left[ -\left(
1+i\right) \xi \right] $, where $\nu =-i\lambda /2$.

With the aid of the exact solutions (\ref{se41.3}) and the coefficient (\ref%
{se41.6}) discussed in Appendix \ref{B}, the differential mean numbers of
magnon-antimagnon pairs created from the vacuum by the external field (\ref%
{Na}) has the form:%
\begin{eqnarray}
N_{m}^{\mathrm{cr}}&=&\frac{8e^{-\pi \lambda /2}}{\sqrt{\xi _{1}^{2}-\lambda
}\sqrt{\xi _{2}^{2}-\lambda }}  \notag \\
&\times &\left\vert f_{1}^{\left( -\right) }\left( \xi _{2}\right)
f_{2}^{\left( -\right) }\left( \xi _{1}\right) -f_{2}^{\left( -\right)
}\left( \xi _{2}\right) f_{1}^{\left( -\right) }\left( \xi _{1}\right)
\right\vert ^{-2}\,.  \label{se41.5}
\end{eqnarray}%
Here, $\xi _{1}=\xi \left( x_{\mathrm{L}}\right) $, $\xi _{2}=\xi \left( x_{%
\mathrm{R}}\right) $, and%
\begin{eqnarray}
f_{1}^{\left( \pm \right) }\left( \xi \right) &=&\left( 1\pm \frac{i}{\sqrt{%
\xi ^{2}-\lambda }}\frac{d}{d\xi }\right) D_{-\nu -1}\left[ \pm \left(
1+i\right) \xi \right] \,,  \notag \\
f_{2}^{\left( \pm \right) }\left( \xi \right) &=&\left( 1\pm \frac{i}{\sqrt{%
\xi ^{2}-\lambda }}\frac{d}{d\xi }\right) D_{\nu }\left[ \pm \left(
1-i\right) \xi \right] \,.  \label{se41.8}
\end{eqnarray}

Optimal conditions for the magnon pair production occur when step (\ref%
{se41.1}) is high enough and stretches over a wide region of the space,
characterized by the inequalities%
\begin{equation}
\sqrt{\frac{\left\vert \mu B^{\prime }\right\vert }{v_{s}}}L\gg \max \left\{
1,\frac{\Delta ^{2}}{v_{s}\mu B^{\prime }}\right\} \,.  \label{se41.9}
\end{equation}%
If these conditions are met and $\sqrt{\lambda }$ is fixed, in the sense
that $\sqrt{\lambda }<K_{\perp }$, where $K_{\perp }$ is a reasonably large
number obeying the conditions $\sqrt{\left\vert \mu B^{\prime }\right\vert
/v_{s}}L/2\gg K_{\perp }^{2}\gg \max \left\{ 1,\Delta ^{2}/v_{s}\mu
B^{\prime }\right\} $, then $\left\vert \xi _{1}\right\vert $ and $\xi _{2}$
are large%
\begin{eqnarray}
&&\xi _{2}\geq \sqrt{\frac{\left\vert \mu B^{\prime }\right\vert }{v_{s}}}%
\frac{L}{2}\,,\ \ -\sqrt{\frac{\left\vert \mu B^{\prime }\right\vert }{v_{s}}%
}\frac{L}{2}\leq \xi _{1}\leq -K\,,  \notag \\
&&K_{\perp }^{2}<K\ll \sqrt{\frac{\left\vert \mu B^{\prime }\right\vert }{%
v_{s}}}\frac{L}{2}\,,  \label{se41.10}
\end{eqnarray}%
which means that we can use asymptotic representations of Weber parabolic
cylinder functions (WPCFs) given by Eqs. (1)-(3) in Sec. 8.4 of Ref. \cite%
{Erdelyi}, vol. 2, to show that the mean numbers acquire the form%
\begin{equation}
N_{m}^{\mathrm{cr}}=\exp \left( -\pi \lambda \right) \left[ 1+O\left(
\left\vert \xi _{1}\right\vert ^{-3}\right) +O\left( \xi _{2}^{-3}\right) %
\right] \,.  \label{se41.11}
\end{equation}%
In the limit where the inhomogeneity of the field spreads over the entire $x$%
-axis, i.e., when $L\rightarrow \infty $ (thus $\left\vert \xi
_{1}\right\vert \rightarrow \infty \,,\ \xi _{2}\rightarrow \infty $), we
obtain:%
\begin{equation}
N_{m}^{\mathrm{cr}}\rightarrow N_{m}^{\mathrm{uni}}=\exp \left( -\pi \lambda
\right) .  \label{L22}
\end{equation}

This is a well-known expression that was originally obtained in the context
of electron-positron pair creation from the vacuum by a constant uniform
electric field \cite{Nikis70b}. Its maximum value $\max N_{m}^{\mathrm{uni}%
}=\left. N_{m}^{\mathrm{uni}}\right\vert _{\mathbf{p}_{\perp }=\mathbf{0}}$
becomes pronounced if the derivative $B^{\prime }$ is of the order of the
critical value $B_{\mathrm{c}}^{\prime }=\Delta ^{2}/v_{s}\mu $, which plays
the role of the Schwinger's critical field \cite{Schw51} in the case under
consideration.

Another configuration of the external field worth of discussion is when its
spacial inhomogeneity varies \textquotedblleft abruptly\textquotedblright\
along the $x$-direction. We call this configuration \textquotedblleft
sharply-varying\textquotedblright\ or \textquotedblleft
steep\textquotedblright\ field configuration. A steep $L$-constant magnetic
step is characterized by the set\ of inequalities,%
\begin{equation}
\delta U=\left\vert \mu B^{\prime }\right\vert L=\mathrm{const}>2\Delta ,\ \
\delta UL/v_{s}\ll 1\,,  \label{se41.12}
\end{equation}%
which, in turn, implies in the conditions%
\begin{equation}
\max \left( \left\vert \pi _{0}\left( \mathrm{L}\right) \right\vert \frac{L}{%
v_{s}},\pi _{0}\left( \mathrm{R}\right) \frac{L}{v_{s}}\right) \ll 1\,,
\label{se41.13}
\end{equation}%
as quantum numbers are bounded in the Klein zone $\Omega _{3}$. As a result,
coefficients involving asymptotic momenta are also small since $\left\vert
p^{\mathrm{L/R}}\right\vert <\left\vert \pi _{0}\left( \mathrm{L/R}\right)
\right\vert $. In this case, the argument of the WPCFs is sufficiently small
in $\Omega _{3}$ and we may use their corresponding power-series
representations to demonstrate that
\begin{eqnarray}
&&N_{m}^{\mathrm{cr}}\approx \frac{4\left\vert p^{\mathrm{L}}\right\vert
\left\vert p^{\mathrm{R}}\right\vert }{\left\vert \left\vert p^{\mathrm{L}%
}\right\vert -\left\vert p^{\mathrm{R}}\right\vert +i\sigma \right\vert ^{2}}%
\,,  \notag \\
&&\sigma =\left[ \left\vert p^{\mathrm{L}}\right\vert \left\vert p^{\mathrm{R%
}}\right\vert +\left( i+\lambda \right) \frac{\left\vert \mu B^{\prime
}\right\vert }{v_{s}}\right] L\,.  \label{se41.14}
\end{eqnarray}%
Notice that the limit $L\rightarrow 0$ is admissible provided the difference
$\left\vert \left\vert p^{\mathrm{L}}\right\vert -\left\vert p^{\mathrm{R}%
}\right\vert \right\vert $ is larger compared $\left\vert \sigma \right\vert
$. In particular, if $\left\vert \left\vert p^{\mathrm{L}}\right\vert
-\left\vert p^{\mathrm{R}}\right\vert \right\vert \gg \left\vert \sigma
\right\vert $, then the mean number (\ref{se41.14}) admits form%
\begin{equation}
N_{m}^{\mathrm{cr}}\approx \frac{4k}{\left( 1-k\right) ^{2}}\,,
\label{se41.15}
\end{equation}%
where $k=\left\vert p^{\mathrm{R}}\right\vert /\left\vert p^{\mathrm{L}%
}\right\vert $. This is in agreement with results obtained at $p_{\bot }=0$\
in Refs. \cite{Klein27,Sauter31a,DomCal99,HansRavn81} for the case of the
Klein step formed by an electric field. Additionally, Eq. (\ref{se41.15})
can also be reproducible by other magnetic steps, as shall be seen below.

\subsubsection{Sauter-like step\label{Ss4.2}}

The Sauter-like magnetic step\footnote{%
We name the field (\ref{se42.1})\ \textquotedblleft
Sauter-like\textquotedblright ,\ in reference to F. Sauter \cite{Sauter-pot}%
, who first solved relativistic wave equations for a charged particle with a
potential step of the form $-\alpha E\tanh \left( x/\alpha \right) $.} --- a
more realistic, smoothed version of the $L$-constant magnetic step --- is
another example of magnetic field inhomogeneity for which exact solutions of
Eq. (\ref{d7}) are known. The Sauter (or Sauter-like) electric field is a
popular example of an electric field that may violate the vacuum stability;
see Ref. \cite{GavGi16} for an extensive discussion of the phenomenon in QED
with $x$ steps.

In the present case, the Sauter-like magnetic step has the form:%
\begin{equation}
B\left( x\right) =-B^{\prime }L_{\mathrm{S}}\tanh \left( x/L_{\mathrm{S}%
}\right) \,,\ \ B^{\prime }>0\,,\ \ L_{\mathrm{S}}>0\,.  \label{se42.1}
\end{equation}%
At remote regions $x\rightarrow \mp \infty $, the magnetic field is constant
$B\left( \mp \infty \right) =\pm B^{\prime }L_{\mathrm{S}}$, which means
that $U_{\mathrm{L}}=\mu B^{\prime }L_{\mathrm{S}}=-U_{\mathrm{R}}$.
Therefore, the magnitude of the potential step (\ref{d8}) in this case is $%
\delta U=U_{\mathrm{L}}-U_{\mathrm{R}}=2\mu B^{\prime }L_{\mathrm{S}}$.

Performing the change of variable%
\begin{equation}
\chi \left( x\right) =\frac{1}{2}\left[ 1+\tanh \left( x/L_{\mathrm{S}%
}\right) \right] \,,  \label{se42.2}
\end{equation}%
and seeking for solutions in the form%
\begin{equation}
\varphi _{m}\left( x\right) =\chi ^{-iL_{\mathrm{S}}\left\vert p^{\mathrm{L}%
}\right\vert /2}\left( 1-\chi \right) ^{iL_{\mathrm{S}}\left\vert p^{\mathrm{%
R}}\right\vert /2}f\left( \chi \right) \,,  \label{se42.3}
\end{equation}%
allows us to express Eq. (\ref{d7}) in the same form as the differential
equation for the Gauss hypergeometric function \cite{Erdelyi},%
\begin{equation}
\chi \left( 1-\chi \right) f^{\prime \prime }+\left[ c-\left( a+b+1\right)
\chi \right] f^{\prime }-abf=0\,,  \label{se42.4}
\end{equation}%
provided%
\begin{eqnarray}
&&a=\frac{1}{2}\left[ iL_{\mathrm{S}}\left( \left\vert p^{\mathrm{R}%
}\right\vert -\left\vert p^{\mathrm{L}}\right\vert \right) +1+i\sqrt{\left(
\frac{L_{\mathrm{S}}\delta U}{v_{s}}\right) ^{2}-1}\right] \,,  \notag \\
&&b=\frac{1}{2}\left[ iL_{\mathrm{S}}\left( \left\vert p^{\mathrm{R}%
}\right\vert -\left\vert p^{\mathrm{L}}\right\vert \right) +1-i\sqrt{\left(
\frac{L_{\mathrm{S}}\delta U}{v_{s}}\right) ^{2}-1}\right] \,,  \notag \\
&&\;c=1-iL_{\mathrm{S}}\left\vert p^{\mathrm{L}}\right\vert \,.
\label{se42.5}
\end{eqnarray}%
Using the exact solutions (\ref{b1}) and (\ref{b2}) and the coefficient (\ref%
{b4}) discussed in Appendix \ref{B}, we find that
\begin{widetext}
\begin{equation}
\left\vert g\left( _{+}\left\vert ^{-}\right. \right) \right\vert ^{-2}=%
\frac{\sinh \left( \pi L_{\mathrm{S}}\left\vert p^{\mathrm{R}}\right\vert
\right) \sinh \left( \pi L_{\mathrm{S}}\left\vert p^{\mathrm{L}}\right\vert
\right) }{\sinh ^{2}\left[ \pi L_{\mathrm{S}}\left( \left\vert p^{\mathrm{R}%
}\right\vert -\left\vert p^{\mathrm{L}}\right\vert \right) /2\right] +\cosh
^{2}\left( \frac{\pi }{2}\sqrt{\left( L_{\mathrm{S}}\delta U/v_{s}\right)
^{2}-1}\right) }\,.  \label{se42.7}
\end{equation}
\end{widetext}Result (\ref{se42.7}) holds in the ranges $\Omega _{1}$, $%
\Omega _{5}$, and $\Omega _{3}$. Taking into account relation (\ref{UR1}),
one finds $\left\vert g\left( _{+}\left\vert ^{+}\right. \right) \right\vert
^{2}$. Using these coefficients in Eq. (\ref{RT}), one\emph{\ }can calculate
the relative probabilities of the reflection, $\left\vert R_{m}\right\vert
^{2}=1-\left\vert T_{m}\right\vert ^{2}$, and the transmission,\emph{\ }$%
\left\vert T_{m}\right\vert ^{2}=\left[ 1+\left\vert g\left( _{+}\left\vert
^{-}\right. \right) \right\vert ^{2}\right] ^{-1}$\emph{\ }in the ranges $%
\Omega _{1}$\emph{\ }and\emph{\ }$\Omega _{5}$\emph{. } In the range $\Omega
_{3}$ according to relation (\ref{Na}) the coefficient $\left\vert g\left(
_{+}\left\vert ^{-}\right. \right) \right\vert ^{-2}$ gives the differential
mean numbers of the magnon-antimagnon pairs created\ from the vacuum, $%
N_{m}^{\mathrm{cr}}$. In particular, for any $\pi _{\bot }\neq 0$, one of
the following limits holds true:%
\begin{equation}
\left\vert g\left( _{+}\left\vert ^{-}\right. \right) \right\vert ^{-2}\sim
\left\vert L_{\mathrm{S}}p^{\mathrm{R}}\right\vert \rightarrow 0,\ \
\left\vert g\left( _{+}\left\vert ^{-}\right. \right) \right\vert ^{-2}\sim
\left\vert L_{\mathrm{S}}p^{\mathrm{L}}\right\vert \rightarrow 0.
\label{exs5}
\end{equation}%
This means that in the range $\Omega _{3}$, the mean numbers tend to zero, $%
N_{m}^{\mathrm{cr}}\rightarrow 0$, while in the ranges $\Omega _{1}$ and $%
\Omega _{5}$ the relative probability of the transmission reads $\left\vert
T_{m}\right\vert ^{2}\rightarrow 0$ if $m$ tends to the boundary with \
either the range $\Omega _{2}$ ($\left\vert p^{\mathrm{L}}\right\vert
\rightarrow 0$) or the range $\Omega _{4}$ ($\left\vert p^{\mathrm{R}%
}\right\vert \rightarrow 0$).

In cases where the magnetic step is high enough and stretches over a wide
region of the space, such that $L_{\mathrm{S}}\delta U/v_{s}\gg 1,$ the mean
number of created pairs can be approximated as
\begin{equation}
N_{m}^{\mathrm{cr}}\approx N_{m}^{\mathrm{as}}=e^{-\pi \tau }\,,\ \ \tau =L_{%
\mathrm{S}}\left( 2\left\vert \mu B^{\prime }\right\vert L_{\mathrm{S}%
}/v_{s}-\left\vert p^{\mathrm{R}}\right\vert -\left\vert p^{\mathrm{L}%
}\right\vert \right) \,.  \label{se42.9}
\end{equation}%
When $L_{\mathrm{S}}\delta U/v_{s}\rightarrow \infty $, one obtains $N_{m}^{%
\mathrm{cr}}\rightarrow N_{m}^{\mathrm{uni}}$ where $N_{m}^{\mathrm{uni}}$
is given by Eq. (\ref{L22}).

Sharp-gradient configuration%
\begin{equation}
\delta U=\left\vert \mu B^{\prime }\right\vert L_{\mathrm{S}}=\mathrm{const}%
>2\Delta \,,\ \ \delta UL_{\mathrm{S}}/v_{s}\ll 1\,,  \label{sh1}
\end{equation}%
corresponds to a very sharp field derivative $\partial _{x}U$, highly
concentrated near the origin $x=0$, described by a very \textquotedblleft
steep\textquotedblright\ potential step. This configuration has a special
interest because it corresponds to a regularization of the Klein step
(originally an electric step potential)%
\begin{equation}
U\left( x\right) =\left\{
\begin{array}{l}
U_{\mathrm{L}}\;\mathrm{if}\;\ x<0 \\
U_{\mathrm{R}}\;\mathrm{if}\ x>0%
\end{array}%
\right. ,  \label{exs41}
\end{equation}%
where $U_{\mathrm{R}}$ and $U_{\mathrm{L}}$ are constants, and may be useful
in a discussion of the Klein paradox. In the ranges $\Omega _{1}$ and $%
\Omega _{5}$ the energy $\left\vert \varepsilon \right\vert $ is not
restricted from the above, that is why in what follows we consider only the
subranges, where $\max \left\{ L_{\mathrm{S}}\left\vert p^{\mathrm{L}%
}\right\vert ,L_{\mathrm{S}}\left\vert p^{\mathrm{R}}\right\vert \right\}
\ll 1.$ Note that in these ranges $\left\vert \left\vert p^{\mathrm{L}%
}\right\vert -\left\vert p^{\mathrm{R}}\right\vert \right\vert >\delta U$
then the parameter $k=\left\vert p^{\mathrm{R}}\right\vert \left/ \left\vert
p^{\mathrm{L}}\right\vert \right. $ does not achieve the unit value, $k\neq
1 $.$\ $Then one has:%
\begin{equation}
\left\vert g\left( _{+}\left\vert ^{-}\right. \right) \right\vert
^{-2}\approx \frac{4k}{\left( 1-k\right) ^{2}},  \label{exs49a}
\end{equation}%
and obtains the transmission coefficient as
\begin{equation}
\left\vert T_{m}\right\vert ^{2}\approx \frac{4k}{\left( 1+k\right) ^{2}},
\label{exs49b}
\end{equation}%
that is in agreement with results of the nonrelativistic consideration
obtained in any textbook for one dimensional quantum motion. In the range $%
\Omega _{3}$ for any given $\pi _{\bot }$ the absolute values of $\left\vert
p^{\mathrm{R}}\right\vert $ and $\left\vert p^{\mathrm{L}}\right\vert $ are
restricted from above,%
\begin{equation}
0\leq \left\vert \left\vert p^{\mathrm{L}}\right\vert -\left\vert p^{\mathrm{%
R}}\right\vert \right\vert \leq \sqrt{\delta U\left( \delta U-2\pi _{\bot
}\right) }.  \label{g8}
\end{equation}%
As it follows from Eq.~(\ref{se42.7}), in the range $\Omega _{3}$ the
differential mean numbers of created magnon-antimagnon pairs read:%
\begin{equation}
N_{m}^{\mathrm{cr}}=\left\vert g\left( _{+}\left\vert ^{-}\right. \right)
\right\vert ^{-2}\approx \frac{4\left\vert p^{\mathrm{L}}\right\vert
\left\vert p^{\mathrm{R}}\right\vert }{\left( \frac{\delta U^{2}L_{\mathrm{S}%
}}{2v_{s}^{2}}\right) ^{2}+\left( \left\vert p^{\mathrm{L}}\right\vert
-\left\vert p^{\mathrm{R}}\right\vert \right) ^{2}}.  \label{exs51}
\end{equation}%
They have a maximum at $k=1$ that can be quite large,%
\begin{equation}
\max N_{m}^{\mathrm{cr}}=\frac{4}{\left( L_{\mathrm{S}}\delta U/v_{s}\right)
^{2}}\left[ 1-\left( \frac{2\pi _{\bot }}{\delta U}\right) ^{2}\right] .
\label{51b}
\end{equation}%
The limit $L_{\mathrm{S}}\rightarrow 0$ in Eq. (\ref{exs51}) is possible
only when the difference $\left\vert p^{\mathrm{L}}\right\vert -\left\vert
p^{\mathrm{R}}\right\vert $ is not very small, namely when%
\begin{equation}
\left( \frac{\delta U^{2}L_{\mathrm{S}}}{2v_{s}^{2}}\right) ^{2}\ll \left(
\left\vert p^{\mathrm{L}}\right\vert -\left\vert p^{\mathrm{R}}\right\vert
\right) ^{2}.  \label{add10}
\end{equation}%
Only under the latter condition one can neglect an $L_{\mathrm{S}}$%
-depending term in Eq.~(\ref{exs51}) to obtain the form given by Eq. (\ref%
{se41.15}). Thus, we have an another example of the regularization of the
Klein step.

\subsubsection{Exponential step\label{Ss4.3}}

We present here an example of the magnetic field inhomogeneity whose
analytical form is a piecewise, continuous exponential functions of $x$. In
this case we have a possibility to consider various asymmetric peak
configurations. The electric-analog of this field in QED was considered in
Ref. \cite{GavGitSh17}. The magnetic step has the form:%
\begin{equation}
B\left( x\right) =B^{\prime }\left\{
\begin{array}{ll}
k_{1}^{-1}\left( 1-e^{k_{1}x}\right) \,, & x\in \mathrm{I}=\left( -\infty ,0%
\right] \, \\
k_{2}^{-1}\left( e^{-k_{2}x}-1\right) \,, & x\in \mathrm{II}=\left(
0,+\infty \right) \,%
\end{array}%
\right. ,  \label{se43.1}
\end{equation}%
where $k_{j}$, $j=1,2$, are positive constants that characterizes how steep
or smooth the field decays from $x=-\infty $ to $x=+\infty $. Similarly to
the preceding examples, the exponential magnetic step (\ref{se43.1}) reaches
constant values at remote regions which means that $U_{\mathrm{L}}=\mu
B^{\prime }k_{1}^{-1}$ and $U_{\mathrm{R}}=-\mu B^{\prime }k_{2}^{-1}$. As a
result, the magnitude of the potential step is $\delta U=U_{\mathrm{L}}-U_{%
\mathrm{R}}=\mu B^{\prime }\left( k_{1}^{-1}+k_{2}^{-1}\right) $.

To solve Eq. (\ref{d7}) with this field, we perform the change of variables%
\begin{eqnarray}
\eta _{1} &=&ih_{1}e^{k_{1}x}\,,\ \ h_{1}=\frac{2\mu B^{\prime }}{%
k_{1}^{2}v_{s}}\,,\ \ x\in \mathrm{I}\,,  \notag \\
\eta _{2} &=&ih_{2}e^{-k_{2}x}\,,\ \ h_{2}=\frac{2\mu B^{\prime }}{%
k_{1}^{2}v_{s}}\,,\ \ x\in \mathrm{II}\,,  \label{se43.2}
\end{eqnarray}%
and represent the scalar functions in the form%
\begin{eqnarray}
&&\varphi _{m}\left( x\right) =e^{-\eta _{j}/2}\eta _{j}^{\nu
_{j}}R_{j}\left( \eta _{j}\right) \,,  \notag \\
&&\nu _{1}=\frac{i\left\vert p^{\mathrm{L}}\right\vert }{k_{1}}\,,\ \ \nu
_{2}=\frac{i\left\vert p^{\mathrm{R}}\right\vert }{k_{2}}\,,  \label{se43.3}
\end{eqnarray}%
to learn that the functions $R_{j}\left( \eta _{j}\right) $ obey the
confluent hypergeometric equations%
\begin{equation}
\eta _{j}R_{j}^{\prime \prime }+\left( c_{j}-\eta _{j}\right) R_{j}^{\prime
}-a_{j}R_{j}=0\,,  \label{se43.4}
\end{equation}%
provided%
\begin{equation}
c_{j}=2\nu _{j}+1\,,\ \ a_{1}=\nu _{1}+\frac{1}{2}+\frac{i\pi _{0}\left(
\mathrm{L}\right) }{k_{1}v_{s}}\,,\ \ a_{2}=\nu _{2}+\frac{1}{2}-\frac{i\pi
_{0}\left( \mathrm{R}\right) }{k_{2}v_{s}}\,.  \label{se43.5}
\end{equation}%
Fundamental pairs of solutions to Eq. (\ref{se43.4}) with special asymptotic
properties at remote regions are proportional to confluent hypergeometric
functions $\Phi \left( a_{j},c_{j};\eta _{j}\right) $ and $\eta
_{j}^{1-c_{j}}e^{\eta _{j}}\Phi \left( 1-a_{j},2-c_{j};-\eta _{j}\right) $.

Using exact solutions and their connection via $g$-coefficients discussed in
Appendix \ref{B}, we find:
\begin{widetext}
\begin{equation}
\left\vert g\left( _{+}\left\vert ^{-}\right. \right) \right\vert ^{-2}=%
\frac{4\left\vert p^{\mathrm{L}}\right\vert \left\vert p^{\mathrm{R}%
}\right\vert }{\exp \left[ -\pi \left( k_{1}^{-1}\left\vert p^{\mathrm{L}%
}\right\vert -k_{2}^{-1}\left\vert p^{\mathrm{R}}\right\vert \right) \right]
}\left\vert \left. \left( k_{1}h_{1}y_{1}^{2}\frac{d}{d\eta _{1}}%
y_{2}^{1}+k_{2}h_{2}y_{2}^{1}\frac{d}{d\eta _{2}}y_{1}^{2}\right)
\right\vert _{x=0}\right\vert ^{-2}\,.  \label{se43.6}
\end{equation}
\end{widetext}Taking into account relation (\ref{UR1}) one finds $\left\vert
g\left( _{+}\left\vert ^{+}\right. \right) \right\vert ^{2}$. In the ranges $%
\Omega _{1}$ and $\Omega _{5}$ use of these coefficients in Eq. (\ref{RT})
gives the relative probabilities of the reflection, $\left\vert
R_{m}\right\vert ^{2}$, and the transmission, $\left\vert T_{m}\right\vert
^{2}$. In the range $\Omega _{3}$ according to relation (\ref{Na}) the
coefficient (\ref{se43.6}) gives the differential mean numbers of pairs
created\ from the vacuum, $N_{m}^{\mathrm{cr}}$. In particular, if either $%
\left\vert p^{\mathrm{R}}\right\vert $ or $\left\vert p^{\mathrm{L}%
}\right\vert $ tends to zero for any $\pi _{\bot }\neq 0$, then one of the
following limits holds true:%
\begin{equation}
\left\vert g\left( _{-}|^{+}\right) \right\vert ^{-2}\sim \left\vert p^{%
\mathrm{R}}\right\vert \rightarrow 0,\ \ \left\vert g\left( ^{+}|_{-}\right)
\right\vert ^{-2}\sim \left\vert p^{\mathrm{L}}\right\vert \rightarrow 0.
\label{de.29}
\end{equation}

If the step is high enough and stretches over a wide region of the space, so
that the conditions are satisfied%
\begin{equation}
\min \left( h_{1},h_{2}\right) \gg \max \left\{ 1,\frac{\Delta ^{2}}{%
v_{s}\mu B^{\prime }}\right\} \,,\ \ \frac{k_{1}}{k_{2}}=\mathrm{fixed}\,,
\label{se43.7}
\end{equation}%
then it is possible to show based on the results of Ref. \cite{GavGitSh17}
that the mean numbers of created pairs, given by Eq. (\ref{se43.6}), admit
simpler forms,%
\begin{eqnarray}
&&N_{m}^{\mathrm{cr}}=\left\vert g\left( _{-}|^{+}\right) \right\vert ^{-2}
\notag \\
&\approx &\left\{
\begin{array}{ll}
\exp \left\{ -\frac{2\pi }{k_{1}}\left[ \left\vert \pi _{0}\left( \mathrm{L}%
\right) \right\vert -\left\vert p^{\mathrm{L}}\right\vert \right] \right\}
\,, & 0\leq \varepsilon <U_{\mathrm{L}}-\pi _{\bot }\,, \\
\exp \left\{ -\frac{2\pi }{k_{2}}\left[ \pi _{0}\left( \mathrm{R}\right)
-\left\vert p^{\mathrm{R}}\right\vert \right] \right\} \,, & U_{\mathrm{R}%
}+\pi _{\perp }\leq \varepsilon <0\,.%
\end{array}%
\right.   \label{se43.8}
\end{eqnarray}%
When $h_{1},h_{2}\rightarrow \infty $, one obtains $N_{m}^{\mathrm{cr}%
}\rightarrow N_{m}^{\mathrm{uni}}$.

One can consider an essentially asymmetric configuration with $k_{\mathrm{1}%
}\gg k_{\mathrm{2}}$.

By choosing large parameters $k_{\mathrm{1}}$, $k_{\mathrm{2}}\rightarrow
\infty $ with a fixed ratio $k_{\mathrm{1}}/k_{\mathrm{2}}$, one can obtain
sharp gradient fields that exist only in a small area in a vicinity of the
origin $x=0$. We assume that the corresponding asymptotic potential energies
$U_{\mathrm{R}}$ and $U_{\mathrm{L}}$ define finite magnitudes of the
potential steps $\Delta U_{\mathrm{1}}$ and $\Delta U_{\mathrm{2}}$ for
increasing and decreasing parts,
\begin{equation}
U_{\mathrm{L}}=\Delta U_{\mathrm{1}},\;\;U_{\mathrm{R}}=-\Delta U_{\mathrm{2}%
},  \label{spf.1}
\end{equation}%
respectively, and satisfy the following inequalities:
\begin{equation}
\Delta U_{\mathrm{1}}/k_{\mathrm{1}}\ll 1,\;\;\Delta U_{\mathrm{2}}/k_{%
\mathrm{2}}\ll 1.  \label{spf.2a}
\end{equation}%
This case corresponds to a very sharp peak with a given step magnitude $%
\delta U=\Delta U_{\mathrm{1}}+\Delta U_{\mathrm{2}}$. In the ranges $\Omega
_{1}$ and $\Omega _{5}$ the energy $\left\vert \varepsilon \right\vert $ is
not restricted from the above, that is why in what follows we consider only
the subranges, where%
\begin{equation}
\max \left( \left\vert \pi _{0}\left( \mathrm{L}\right) \right\vert /k_{%
\mathrm{1}},\left\vert \pi _{0}\left( \mathrm{R}\right) \right\vert /k_{%
\mathrm{2}}\right) \ll 1.  \label{spf.2b}
\end{equation}%
In these ranges taking into account that $\left\vert \left\vert p^{\mathrm{L}%
}\right\vert -\left\vert p^{\mathrm{R}}\right\vert \right\vert >\delta U$ we
obtain that the $\left\vert g\left( _{+}\left\vert ^{-}\right. \right)
\right\vert ^{-2}$ and the transmission coefficient can be presented by the
functions (\ref{exs49a}) and (\ref{exs49b}) of the parameter $k=\left\vert
p^{\mathrm{R}}\right\vert \left/ \left\vert p^{\mathrm{L}}\right\vert
\right. $.

In the range $\Omega _{3}$ the difference $\left\vert \left\vert p^{\mathrm{L%
}}\right\vert -\left\vert p^{\mathrm{R}}\right\vert \right\vert $ are
restricted from above by Eq.~(\ref{g8}) and can tend to zero. That is why
the differential mean number of pairs created is
\begin{align}
& N_{m}^{\mathrm{cr}}\approx \frac{4\left\vert p^{\mathrm{L}}\right\vert
\left\vert p^{\mathrm{R}}\right\vert }{\left( \left\vert p^{\mathrm{L}%
}\right\vert -\left\vert p^{\mathrm{R}}\right\vert \right) ^{2}+b^{2}},
\notag \\
& b=\frac{2\Delta U_{\mathrm{1}}}{k_{\mathrm{1}}}\left[ \frac{\Delta U_{%
\mathrm{1}}}{4}+\left\vert \pi _{0}\left( \mathrm{L}\right) \right\vert %
\right] +\frac{2\Delta U_{\mathrm{2}}}{k_{\mathrm{2}}}\left[ \frac{\Delta U_{%
\mathrm{2}}}{4}+\pi _{0}\left( \mathrm{R}\right) \right] .  \label{spf.4a}
\end{align}%
It has a maximum at $k=1$ that can be large, $N_{m}^{\mathrm{cr}%
}=4\left\vert p^{\mathrm{L}}\right\vert \left\vert p^{\mathrm{R}}\right\vert
/b^{2}$. Under the condition $b^{2}\ll \left( \left\vert p^{\mathrm{L}%
}\right\vert -\left\vert p^{\mathrm{R}}\right\vert \right) ^{2}$ one can
verify that the mean number $N_{m}^{\mathrm{cr}}$ is given by Eq. (\ref%
{se41.15}). Thus, we have additional example of the regularization of the
Klein step.

\subsubsection{Inverse-square step\label{Ss4.4}}

As a last example, we present below a model of the magnetic field
inhomogeneity that is also a piecewise and a continuous function of $x$. In
QED, we call the electric field corresponding to the potential step
inverse-square electric field, whose solutions to relativistic wave
equations were found by us recently in Ref.  \cite{AdoGavGit20}. The
magnetic step has the form:%
\begin{equation}
B\left( x\right) =B^{\prime }\left\{
\begin{array}{ll}
\varrho _{1}\left[ 1-\left( 1-x/\varrho _{1}\right) ^{-1}\right] \,, & x\in
\mathrm{I}=\left( -\infty ,0\right] \, \\
\varrho _{2}\left[ \left( 1+x/\varrho _{2}\right) ^{-1}-1\right] \,, & x\in
\mathrm{II}=\left( 0,+\infty \right) \,%
\end{array}%
\right. ,  \label{se44.1}
\end{equation}%
where $\varrho _{j}$, $j=1,2$, are positive constants that characterize how
the field grows/decays along the $x$-axis. The magnitude of the potential
step for this field is $\delta U=U_{\mathrm{L}}-U_{\mathrm{R}}=\mu B^{\prime
}\left( \varrho _{1}+\varrho _{2}\right) $.

By performing the change of variables%
\begin{eqnarray}
z_{1}\left( x\right) &=&2i\left\vert p^{\mathrm{L}}\right\vert \varrho
_{1}\left( 1-x/\varrho _{1}\right) \,,\ \ x\in \mathrm{I}\,,  \notag \\
z_{2}\left( x\right) &=&2i\left\vert p^{\mathrm{R}}\right\vert \varrho
_{2}\left( 1+x/\varrho _{2}\right) \,,\ \ x\in \mathrm{II}\,,  \label{se44.2}
\end{eqnarray}%
we may write the second-order differential equation (\ref{d7}) as%
\begin{equation}
\left( \frac{d^{2}}{dz_{j}^{2}}-\frac{1}{4}+\frac{\kappa _{j}}{z_{j}}+\frac{%
1/4-\mu _{j}^{2}}{z_{j}^{2}}\right) \varphi _{m}\left( x\right) =0\,,
\label{se44.3}
\end{equation}%
whose solutions are proportional to the Whittaker functions $W_{\kappa
_{j},\mu _{j}}\left( z_{j}\right) $ and $W_{-\kappa _{j},\mu _{j}}\left(
e^{-i\pi }z_{j}\right) $, provided the parameters $\kappa _{j}$ and $\mu
_{j} $ have the form:%
\begin{eqnarray}
\kappa _{1} &=&-i\frac{\mu B^{\prime }\varrho _{1}^{2}}{v_{s}}\frac{\pi
_{0}\left( \mathrm{L}\right) /v_{s}}{\left\vert p^{\mathrm{L}}\right\vert }%
\,,\ \ \kappa _{2}=i\frac{\mu B^{\prime }\varrho _{2}^{2}}{v_{s}}\frac{\pi
_{0}\left( \mathrm{R}\right) /v_{s}}{\left\vert p^{\mathrm{R}}\right\vert }%
\,,  \notag \\
\mu _{1} &=&-\sqrt{\frac{1}{4}-\left( \frac{\mu B^{\prime }\varrho _{1}^{2}}{%
v_{s}}\right) ^{2}}\,,\ \ \mu _{2}=\sqrt{\frac{1}{4}-\left(\frac{\mu
B^{\prime }\varrho _{2}^{2}}{v_{s}}\right) ^{2}}\,.  \label{se44.4}
\end{eqnarray}

Using the exact solutions (\ref{b11}), (\ref{b12}) and coefficient (\ref{b15}%
) that connects different solutions as is discussed in Appendix \ref{B}, we
obtain:%
\begin{widetext}
\begin{equation}
N_{m}^{\mathrm{cr}}=\left\vert p^{\mathrm{L}}\right\vert \left\vert p^{%
\mathrm{R}}\right\vert \left\vert \left. \left[ \left\vert p^{\mathrm{L}%
}\right\vert w_{1}^{2}\left( z_{2}\right) \frac{d}{dz_{1}}w_{2}^{1}\left(
z_{1}\right) +w_{2}^{1}\left( z_{1}\right) \left\vert p^{\mathrm{R}%
}\right\vert \frac{d}{dz_{2}}w_{1}^{2}\left( z_{2}\right) \right]
\right\vert _{x=0}\right\vert ^{-2}\,.  \label{se44.5}
\end{equation}
\end{widetext}

Finally, considering cases where the step is sufficiently high and stretches
over a wide region of the space, specified by the conditions%
\begin{equation}
\min \left( U_{\mathrm{L}}\varrho _{1},\left\vert U_{\mathrm{R}}\right\vert
\varrho _{2}\right) \gg \max \left\{ 1,\frac{\Delta ^{2}}{v_{s}\mu B^{\prime
}}\right\} ,\ \ \frac{\varrho _{1}}{\varrho _{2}}=\mathrm{fixed}\,,
\label{se44.6}
\end{equation}%
it is possible to demonstrate based on our previous results \cite%
{AdoGavGit20} that the mean numbers (\ref{se44.5}) admit the asymptotic forms%
\begin{equation}
N_{m}^{\mathrm{cr}}\approx \left\{
\begin{array}{ll}
\exp \left( 2\pi \omega _{1}^{+}\right) \,, & 0\leq \varepsilon <U_{\mathrm{L%
}}-\pi _{\bot }\,, \\
\exp \left( 2\pi \omega _{2}^{-}\right) \,, & U_{\mathrm{R}}+\pi _{\perp
}\leq \varepsilon <0\,.%
\end{array}%
\right.  \label{se44.7}
\end{equation}%
where%
\begin{eqnarray}
\omega _{1}^{\pm } &=&\pm i\left( \kappa _{1}\pm \mu _{1}\right)  \notag \\
&=&\frac{U_{\mathrm{L}}\varrho _{1}}{v_{s}}\left[ \sqrt{1-\left( \frac{2U_{%
\mathrm{L}}\varrho _{1}}{v_{s}}\right) ^{-2}}\pm \frac{\pi _{0}\left(
\mathrm{L}\right) /v_{s}}{\left\vert p^{\mathrm{L}}\right\vert }\right] \,,
\notag \\
\omega _{2}^{\pm } &=&\mp i\left( \kappa _{2}\pm \mu _{2}\right)  \notag \\
&=&\frac{\left\vert U_{\mathrm{R}}\right\vert \varrho _{2}}{v_{s}}\left[
\sqrt{1-\left( \frac{2\left\vert U_{\mathrm{R}}\right\vert \varrho _{2}}{%
v_{s}}\right) ^{-2}}\pm \frac{\pi _{0}\left( \mathrm{R}\right) /v_{s}}{%
\left\vert p^{\mathrm{R}}\right\vert }\right] \,.  \label{se44.8}
\end{eqnarray}%
When $U_{\mathrm{L}}\varrho _{1},\left\vert U_{\mathrm{R}}\right\vert
\varrho _{2}\rightarrow \infty $, we have $N_{m}^{\mathrm{cr}}\rightarrow
N_{m}^{\mathrm{uni}}$ where $N_{m}^{\mathrm{uni}}$. One can see that all of
the above presented examples in the model of smooth-gradient step can be
seen as regularizations of linearly growing magnetic field.

Another field configuration worthy of discussion is the case where the step
is still sufficiently high but the field inhomogeneity is concentrated in a
narrow region of the $x$-axis, characterized by the inequality%
\begin{equation}
\max \left( \frac{U_{\mathrm{L}}\varrho _{1}}{v_{s}},\frac{\left\vert U_{%
\mathrm{R}}\right\vert \varrho _{2}}{v_{s}}\right) \ll 1\,,\ \ \frac{\varrho
_{1}}{\varrho _{2}}\ \mathrm{fixed}\,.  \label{se44.9}
\end{equation}%
In the range $\Omega _{3}$, the most significant contribution to the mean
numbers (\ref{q5}) comes from values for $\left\vert \varepsilon \right\vert
$\ that are sufficiently away from the borders with $\Omega _{2}$\ and $%
\Omega _{4}$, in the sense that the supplementary inequalities%
\begin{equation}
\max \left( \frac{\left\vert \pi _{0}\left( \mathrm{L}\right) \right\vert
\varrho _{1}}{v_{s}},\frac{\pi _{0}\left( \mathrm{R}\right) \varrho _{2}}{%
v_{s}}\right) \ll 1\,,\ \ \frac{\varrho _{1}}{\varrho _{2}}\ \mathrm{fixed}%
\,,  \label{se44.10}
\end{equation}%
are also satisfied. In this case, the argument of the Whittaker functions (%
\ref{b11}) are also small, which means that one can use the connection
formulas given by Eqs. (119) in Ref. \cite{AdoGavGit20} and expand the
Whittaker functions $M_{\kappa _{j},\mu _{j}}\left( z_{j}\right) $\ around
the origin, i.e. $M_{\kappa _{j},\mu _{j}}\left( z_{j}\right) =z^{\mu
_{j}+1/2}\left[ 1-z\kappa _{j}/\left( 2\mu _{j}+1\right) +O\left(
z^{2}\right) \right] $, to show that the differential mean\emph{\ }numbers
of magnon-antimagnon pairs created from the vacuum is approximately given by
the equations%
\begin{eqnarray}
&&N_{n}^{\mathrm{cr}}\approx \frac{4\left\vert p^{\mathrm{L}}\right\vert
\left\vert p^{\mathrm{R}}\right\vert }{\left( \left\vert p^{\mathrm{L}%
}\right\vert -\left\vert p^{\mathrm{R}}\right\vert \right) ^{2}+d^{2}}\,,
\notag \\
&&d=\frac{\pi _{0}\left( \mathrm{L}\right) }{U_{\mathrm{L}}}\left\vert p^{%
\mathrm{L}}\right\vert ^{2}\varrho _{1}+\frac{\pi _{0}\left( \mathrm{R}%
\right) }{U_{\mathrm{R}}}\left\vert p^{\mathrm{R}}\right\vert ^{2}\varrho
_{2}\,.  \label{se44.11}
\end{eqnarray}%
Similar to the previous examples, the limit in which the field is infinitely
steep, given by $\varrho _{1}\rightarrow 0$\ and $\varrho _{2}\rightarrow 0$%
, is admissible as long as the difference $\left\vert \left\vert p^{\mathrm{L%
}}\right\vert -\left\vert p^{\mathrm{R}}\right\vert \right\vert $\ is larger
compared to $\left\vert d\right\vert $. If, in particular, $\left\vert
\left\vert p^{\mathrm{L}}\right\vert -\left\vert p^{\mathrm{R}}\right\vert
\right\vert \gg \left\vert d\right\vert $, then we find the same expression
obtained for the Klein step, given by Eq. (\ref{se41.15}). In addition to
the previous examples, the compatibility between Eq. (\ref{se44.11}) and (%
\ref{se41.15}) shows that the field (\ref{se44.1}) is also a regularization
of the Klein step.

\subsection{Integral quantities\label{Ss4b}}

According to the discussion in preceding section, the total number of the
magnon-antimagnon pairs\ produced from the vacuum by magnetic steps is a sum
of the differential numbers over the quantum numbers within the Klein zone $%
\Omega _{3}$, given by Eq. (\ref{TN}). The created magnons continue to move
in the $x$ direction and form the flux density $j_{x}$, given by Eq. (\ref%
{d19b}), in the area $S_{\mathrm{R}}$, while the antimagnon motion in the
opposite direction form the flux density $-j_{x}$ in the area $S_{\mathrm{L}%
} $. These flux densities can be represented in an integral form as:
\begin{equation}
j_{x}=\frac{1}{\left( 2\pi \right) ^{3}}\int_{U_{\mathrm{R}}+\pi _{\perp
}}^{U_{\mathrm{L}}-\pi _{\bot }}d\varepsilon \int d\mathbf{p}_{\perp }N_{m}^{%
\mathrm{cr}}\,.  \label{se45.1}
\end{equation}%
In general, analytical integration in Eq. (\ref{se45.1}) is impossible.
Nevertheless, due to the analogy between QED with $x$ steps and the present
study, we may use universal expressions in a strong electric field from Ref.
\cite{GGSh19} to conclude that if the magnetic step is sufficiently high but
evolves gradually along the $x$-axis (smooth-gradient step), one can
approximate Eq. (\ref{se45.1}) as
\begin{eqnarray}
&&j_{x}\approx \tilde{j}_{x}=\frac{1}{\left( 2\pi \right) ^{3}}\int_{x_{%
\mathrm{L}}}^{x_{\mathrm{R}}}dx\ U^{\prime }\left( x\right) \int d\mathbf{p}%
_{\bot }N_{n}^{\mathrm{uni}}\left( x\right) ,  \notag \\
&&N_{n}^{\mathrm{uni}}\left( x\right) =\exp \left[ -\pi \frac{\pi _{\bot
}^{2}}{v_{s}U^{\prime }\left( x\right) }\right] ,  \label{np.6}
\end{eqnarray}%
where the prime over the potential denotes differentiation with respect to $%
x $, $U^{\prime }\left( x\right) =\mu dB\left( x\right) /dx$. The quantity $%
N_{n}^{\mathrm{uni}}\left( x\right) $ has a universal form which can be used
to calculate any total characteristic of the pair creation effect.
Integrating the latter expression over $d\mathbf{p}_{\bot }$ one obtain the
final form,
\begin{equation}
j_{x}\approx \tilde{j}_{x}=\frac{1}{(2\pi )^{3}}\int_{x_{\mathrm{L}}}^{x_{%
\mathrm{R}}}dxU^{\prime }\left( x\right) ^{2}\exp \left[ -\pi \frac{\Delta
^{2}}{v_{s}U^{\prime }\left( x\right) }\right] \,.  \label{se45.2}
\end{equation}

For the examples discussed in Secs \ref{Ss4.1}-\ref{Ss4.4}, the total
density of the magnon pair production has the following form:%
\begin{equation}
\tilde{j}_{x}=r^{\mathrm{cr}}\frac{\delta U}{\left\vert \mu B^{\prime
}\right\vert }\left\{
\begin{array}{l}
1\,,\ \text{\textrm{for} }L\text{\textrm{-constant step}} \\
\tilde{\delta}/2\,,\ \text{\textrm{for Sauter-like} \textrm{step}} \\
G\left( 2,\pi \frac{\Delta ^{2}}{v_{s}\left\vert \mu B^{\prime }\right\vert }%
\right) ,\ \text{\textrm{for exponential step}} \\
\frac{1}{2}G\left( \frac{3}{2},\pi \frac{\Delta ^{2}}{v_{s}\left\vert \mu
B^{\prime }\right\vert }\right) ,\ \text{\textrm{for inverse-square step}}%
\end{array}%
\right. ,  \label{se45.3}
\end{equation}%
where $\delta U=U_{\mathrm{L}}-U_{\mathrm{R}}$ is the magnitude of the step
in all the examples,%
\begin{equation}
r^{\mathrm{cr}}=\frac{\left\vert \mu B^{\prime }\right\vert ^{2}}{(2\pi )^{3}%
}\exp \left( -\pi \frac{\Delta ^{2}}{v_{s}\left\vert \mu B^{\prime
}\right\vert }\right) \,,  \label{se45.4}
\end{equation}%
and%
\begin{eqnarray}
\tilde{\delta} &=&\int_{0}^{\infty }dtt^{-1/2}\left( 1+t\right) ^{-5/2}\exp
\left( -\frac{\pi t\Delta ^{2}}{v_{s}\left\vert \mu B^{\prime }\right\vert }%
\right)   \notag \\
&=&\sqrt{\pi }\Psi \left( \frac{1}{2},-2;\pi \frac{\Delta ^{2}}{%
v_{s}\left\vert \mu B^{\prime }\right\vert }\right) \,,  \notag \\
G\left( \alpha ,z\right)  &=&\int_{1}^{\infty }\frac{ds}{s^{\alpha +1}}%
e^{-x\left( s-1\right) }=e^{x}x^{\alpha }\Gamma \left( -\alpha ,x\right) \,.
\label{se45.5}
\end{eqnarray}%
Here $\Psi \left( \alpha ,\beta ;z\right) $ and $\Gamma \left( \beta
,z\right) $ denote a confluent hypergeometric function and an incomplete
Gamma function, respectively.

From the general expression for the vacuum-to-vacuum transition probability,
given by Eq. (\ref{q6}), the universal form of the vacuum-to-vacuum
transition probability for the present case reads:%
\begin{widetext}
\begin{equation}
P_{\mathrm{v}}\approx \exp \left\{ -\frac{V_{\bot }T}{(2\pi )^{3}}%
\sum_{l=1}^{\infty }\left( -1\right) ^{l-1}\int_{x_{\mathrm{L}}}^{x_{\mathrm{%
R}}}dx\frac{U^{\prime }\left( x\right) ^{2}}{l^{2}}\exp \left[ -\pi \frac{%
l\Delta ^{2}}{v_{s}U^{\prime }\left( x\right) }\right] \right\} .
\label{np.11}
\end{equation}
\end{widetext}Representation (\ref{np.11}) coincide with the
vacuum-to-vacuum transition probabilities obtained from the imaginary part
of an effective action in the locally constant field approximation{\Huge \ }
\cite{GiesK17,Karb17}. In this approximation, the effective action $S$ is
expanded about the constant field case, in terms of derivatives of the
background field strength $F_{\mu \nu }$,%
\begin{equation}
S=S^{\left( 0\right) }[F_{\mu \nu }]+S^{\left( 2\right) }[F_{\mu \nu
},\partial _{\mu }F_{\nu \rho }]+...  \label{uni7b}
\end{equation}%
where $S^{\left( 0\right) }$ involves no derivatives of the background field
strength $F_{\mu \nu }$ (that is, $S^{\left( 0\right) }$ is a locally
constant field approximation for $S$ that has a form of the Heisenberg-Euler
action), while the first correction $S^{\left( 2\right) }$ involves two
derivatives of the field strength, and so on; see Ref.~\cite{Dunn04} for a
review. Using representation (\ref{np1a}), one can see that in the locally
constant field approximation the probability $P_{\mathrm{v}}$ is given by
Eq. (\ref{np1a}) where the action $S$ is replaced by $S^{\left( 0\right) }$ .

\section{Conclusion\label{S5}}

In this work we present a Fock space realization of the effective field
model describing low-energy dynamics of the antiferromagnetic magnons.
Mapping the model to the theory of a charged scalar field interacting with
an external constant electric field we apply recently developed approach of
the strong-field QED \ with step potentials to study the Schwinger effect of
the magnon-antimagnon pair production on magnetic field inhomogeneities.
Initial and final one-particle states are constructed from stationary plane
waves satisfied the Klein-Gordon equation. Initial and final vacua are
defined and initial and final states of the Fock space are constructed. Mean
numbers of magnons and antimagnons \ created from the vacuum are expressed
via overlap amplitudes of the stationary plane waves. Observable physical
quantities specifying the vacuum instability are determined. The fluxes of
energy and magnetic moments of created magnons are analyzed. Characteristics
of the vacuum instability obtained for the number of magnetic steps that
allows exact solving the Klein-Gordon equation are presented. In the case of
a smooth-gradient step, universal behavior of the flux density of created
pairs is described and the relation to the imaginary part of a one-loop
effective action $S$\ in a locally constant field approximation is
established. The results are quite general and are not limited to the simple
cubic-type lattice and the G-type antiferromagnet. The presented study
demonstrates a consistent application of strong field QED methods in
magnonics avoiding contradictions and nonexistent paradoxes in the
interpretation of the obtained theoretical results. As a result, one can now
apply many of the results known from strong-field QED in magnonics.

Condensed matter systems provide a possibility for experimental verification
of quantum vacuum effects stimulated by strong fields, in particular, a
laboratory observation the Schwinger effect of the violation of the vacuum
instability due to particle-antiparticle pair production, previously
considered possible only in supercritical fields in astrophysical situations.%
{\large \ }This is due to the fact that in many cases low energy dynamics of
quasiparticle excitations in condensed matter systems can be described by
the Dirac model in which quasiparticles are virtually massless. That is why
external field intensities needed for breaking the vacuum stability are
relatively small and can be observed in the laboratory conditions. Usually
the literature discusses the possibility to observe effects related to the
electron-hole creation from vacuum in the case of Dirac and Weyl semimetals,
such as the graphene and topological insulators, where the low energy
excitations are fermions; see, e.g., Refs. \cite{Adv-QED22,VafVis14,GelTan16}
for the review. Since, the low-energy magnons are bosons with small
effective mass, then for the first time it becomes possible to observe the
Schwinger effect in the case of the Bose statistics, in particular, the
bosonic Klein effect in laboratory conditions.{\large \ }As was already
mentioned above,{\large \ }in the case of the Bose statistics appears a new
mechanism for amplifying the effect of the pair creation, which we call
statistically assisted Schwinger effect. A possible way to detect the
Schwinger mechanism of antiferromagnetic magnons in experiments is discussed
in Ref. \cite{HFMNS21}.{\large \ }For example, one can experimentally
confirm this mechanism by detecting the spin (the magnetic moment) current
using an inverse spin Hall effect.

\begin{acknowledgments}
 The work is supported by Russian Science Foundation (Grant no. 19-12-00042). D.M.G. is grateful to  the Brazilian foundation CNPq for permanent support. The work of T. C. A. in Sec. \ref{S4} and Appendix \ref{B} is funded by XJTLU Research Development Funding, award no. RDF-21-02-056.
\end{acknowledgments}

\appendix

\section{Some details of scalar field quantization in the presence of
critical potential steps\label{Ap}}

The Heisenberg operator of the Klein-Gordon field $\hat{\Phi}\left( X\right)
$ is assigned to the scalar field $\Phi \left( X\right) .$ It is convenient
to consider the canonical pair of the field operator $\hat{\Phi}\left(
X\right) $ and its canonical momentum $\hat{\Pi}\left( X\right) $ as a
column $\hat{\Psi}\left( X\right) ,$

\begin{equation}
\hat{\Psi}\left( X\right) =\left(
\begin{array}{c}
i\hat{\Pi}^{\dag }\left( X\right)  \\
\hat{\Phi}\left( X\right)
\end{array}%
\right) .  \label{add11}
\end{equation}%
The latter satisfies both the Klein-Gordon equation (\ref{d5}) given in the
Hamiltonian form,%
\begin{eqnarray}
&&\left[ i\partial _{0}-U\left( x\right) \right] \hat{\Psi}\left( X\right)
=H^{\mathrm{kin}}\hat{\Psi}\left( X\right) \,,  \notag \\
&&H^{\mathrm{kin}}=\left(
\begin{array}{cc}
0 & -v_{s}^{2}\delta ^{ij}\partial _{i}\partial _{j}+\Delta ^{2} \\
1 & 0%
\end{array}%
\right) ,  \label{a1}
\end{eqnarray}%
and the equal time canonical commutation relations%
\begin{eqnarray}
&&\left. \left[ \hat{\Psi}\left( X\right) ,\hat{\Psi}\left( X^{\prime
}\right) \right] _{-}\right\vert _{t=t^{\prime }}=0,  \notag \\
&&\left. \left[ \hat{\Psi}\left( X\right) ,\hat{\Psi}^{\dag }\left(
X^{\prime }\right) \right] _{-}\right\vert _{t=t^{\prime }}=\delta \left(
\mathbf{r-r}^{\prime }\right) \sigma _{1}\ ,  \label{a2}
\end{eqnarray}%
where $H^{\mathrm{kin}}$ is the one-particle kinetic energy operator, see,
e.g., Refs. \cite{Schwe61,GitTy90}.{\large \ }It follows from Eq.~(\ref{a1})
that
\begin{equation}
i\hat{\Pi}^{\dag }\left( X\right) =\left[ i\partial _{0}-U\left( x\right) %
\right] \hat{\Phi}\left( X\right) ,  \label{add12}
\end{equation}

The Hamiltonian $\widehat{\mathbb{H}}$ of the quantized scalar field has the
form:%
\begin{eqnarray}
&&\widehat{\mathbb{H}}=\widehat{\mathbb{H}}^{\mathrm{kin}}-\frac{1}{v_{s}^{2}%
}\int \hat{\Psi}^{\dagger }\left( X\right) \sigma _{1}U\left( x\right) \hat{%
\Psi}\left( X\right) d\mathbf{r},  \notag \\
&&\widehat{\mathbb{H}}^{\mathrm{kin}}=\int \hat{T}^{00}d\mathbf{r}-\mathbb{H}%
_{0}\mathbf{,\;}  \notag \\
&&\hat{T}^{00}=\frac{1}{v_{s}^{2}}\hat{\Psi}^{\dagger }\left( X\right)
\sigma _{1}\left[ i\partial _{0}-U\left( x\right) \right] \hat{\Psi}\left(
X\right) \mathbf{\ ,}  \label{a3}
\end{eqnarray}%
where $\widehat{\mathbb{H}}^{\mathrm{kin}}$ is a kinetic energy operator,
which, just from the beginning, we write in a renormalized form with a
constant (in general, infinite) term $\mathbb{H}_{0}$\ that corresponds to
the kinetic energy of vacuum fluctuations. In the case under consideration,
in the similar manner as discussing the inner product (\ref{d13}), it is
possible to evaluate integrals (\ref{a3}) for arbitrary field $U\left(
x\right) $, using only the asymptotic behavior (\ref{d9}) of functions in
the regions $S_{\mathrm{L}}$and $S_{\mathrm{R}}$ where particles are free.
Decomposing the field $\hat{\Psi}\left( X\right) $ over the complete set (%
\ref{d6}), and dividing integral (\ref{a3}) in three integrals within the
regions $S_{\mathrm{L}}$, $S_{\mathrm{int}}$, and $S_{\mathrm{R}}$, we
reduce calculating the quantity (\ref{a3}) to calculating its one-particle
matrix elements in the regions $S_{\mathrm{L}}$and $S_{\mathrm{R}}$; see
Sec. IVB.1 and Appendix B in Ref. \cite{GavGi16} for details. One can see
that the matrix elements of the Hamiltonian $\widehat{\mathbb{H}}$ depends
on the total energy of a particle $\varepsilon $. However, probabilities of
the magnon scattering, reflection, and the magnon-antimagnon pair production
depend on the kinetic energy terms $\pi _{0}\left( \mathrm{L/R}\right)
=\varepsilon -U_{\mathrm{L/R}}$, which is a combination of $\varepsilon $
and $U_{\mathrm{L/R}}$. This is due to the fact that not the field $U\left(
x\right) $ itself, acting on the magnons, does the work, but its derivative $%
\partial _{x}U\left( x\right) $. That is why in the field-theoretical
description of magnons embedded into an external inhomogeneous magnetic
field namely eigenvalues of the operator $\widehat{\mathbb{H}}^{\mathrm{kin}}
$ define vacuum states and other state vectors in the Fock space \footnote{%
In QED, where $U_{\mathrm{L/R}}$ are potentials of an electromagnetic field,
the operator $\hat{\mathbb{H}}^{\mathrm{kin}}$ is gauge invariant, which
implies that it is an observable physical quantity in contrast with the
Hamiltonian $\hat{\mathbb{H}}$. In the case under consideration the constant
values $U_{\mathrm{L/R}}$ are physical quantities and are used to tune the
collinear ground state within the regions $S_{\mathrm{L}}$ and $S_{\mathrm{R}%
}$.}

The formal expression of the effective charge (magnetic moment) operator $%
\hat{Q}$ is
\begin{equation}
\hat{Q}=\int \hat{\rho}d\mathbf{r,\;\hat{\rho}=}\frac{\mu }{2v_{s}^{2}}\left[
\hat{\Psi}^{\dagger }\left( X\right) \sigma _{1},\hat{\Psi}\left( X\right) %
\right] _{+},  \label{a4}
\end{equation}%
where $\left[ A,B\right] _{+}=AB+BA$ stands for the anticommutator. The
eigenvalues of the operator $\widehat{\mathbb{H}}^{\mathrm{kin}}$, together
with one-particle mean values of the effective charge, allow one to
distinguish particles and antiparticles.

Before proceeding to the definition of the operators of interest (operators
of fluxes), which usually are not considered in QFT, it is useful to note
that even in the framework of the corresponding classical field theory an
observable $\mathcal{F}$\ can be realized as an inner product of type (\ref%
{d13}) of localizable wave packets $\Phi \left( X\right) $\ and $\hat{F}\Phi
^{\prime }(X)$,%
\begin{equation}
\mathcal{F}\left( \Phi ,\Phi ^{\prime }\right) =\left( \Phi ,\mathcal{\hat{F}%
}\Phi ^{\prime }\right) ,  \label{a5}
\end{equation}%
where $\mathcal{\hat{F}}$\ is a differential operator, whereas $\Phi (X)$\
and $\Phi ^{\prime }(X)$\ are solutions of the Klein-Gordon equation.
Assuming that an observable $\mathcal{F}\left( \Phi ,\Phi ^{\prime }\right) $%
\ is time-independent during the time $T$\ one can represent it in the form
of an average over the period $T$,{\large \ }%
\begin{equation}
\left\langle \mathcal{F}\right\rangle =\frac{1}{T}\int_{-T/2}^{+T/2}\mathcal{%
F}\left( \phi ,\phi ^{\prime }\right) dt.  \label{a6}
\end{equation}%
In general the wave packets $\Phi (X)$\ and $\Phi ^{\prime }(X)$\ can be
decomposed into plane waves $\phi _{m}(X)$\ and $\phi _{m}^{\prime }(X)$,
\begin{equation}
\Phi (X)=\sum_{m}\alpha _{m}\phi _{m}(X),\;\Phi ^{\prime }(X)=\sum_{m}\alpha
_{m}^{\prime }\phi _{m}^{\prime }(X),  \label{add13}
\end{equation}%
where $\phi _{m}(X)$\ and $\phi _{m}^{\prime }(X)$\ are superpositions of
the solutions$_{\;\zeta }\phi _{m}\left( X\right) $\ and $^{\;\zeta }\phi
_{m}(X)$. Taking into account the orthogonality relation (\ref{d10}), one
finds that the corresponding decomposition of $\left\langle \mathcal{F}%
\right\rangle $\ does not contain interference terms,

\begin{equation}
\left\langle \mathcal{F}\right\rangle =\sum_{m}\mathcal{F}\left( \alpha
_{m}\phi _{m},\alpha _{m}^{\prime }\phi _{m}^{\prime }\right) .
\label{add14}
\end{equation}

A physical quantity useful for the further analysis is the average over the
period $T$ of the effective charge current through the surface $x=\mathrm{%
const}$. Since the regions $S_{\mathrm{L}}$and $S_{\mathrm{R}}$ supposed to
be macroscopic and the particles that come there are free, then such a
semiclassical statement of the problem seems to be justified. Moreover, this
is how the problem statement is formulated in the theory of the potential
scattering. The corresponding QFT operator of the effective charge (magnetic
moment) current is proportional to the inner product on this surface given
by Eq. (\ref{scip}),%
\begin{eqnarray}
&&\widehat{\mathbb{J}}=\frac{1}{T}\int \hat{J}_{x}dtd\mathbf{r}_{\bot },
\notag \\
&&\hat{J}_{x}=\frac{\mu i}{2}\left\{ \left[ \hat{\Phi}\left( X\right)
,\partial _{x}\hat{\Phi}^{\dag }\left( X\right) \right] _{+}-\left[ \hat{\Phi%
}^{\dag }\left( X\right) ,\partial _{x}\hat{\Phi}\left( X\right) \right]
_{+}\right\}  \label{a7}
\end{eqnarray}%
Here $\hat{J}_{x}$\ is the longitudinal component of the operator of the
effective current density $\mathbf{\hat{J}}$.\ Note that by virtue of Eq. (%
\ref{d5}) the latter operator and the operator of the effective charge
density $\hat{\rho}$\ satisfy the continuity equation $\nabla \mathbf{\hat{J}%
}+\partial _{t}\hat{\rho}=0$.

In what follows, we consider the energy flux of the scalar field through the
surface $x=\mathrm{const}$. Its QFT operator has the form
\begin{eqnarray}
&&\widehat{\mathbb{F}}\left( x\right) =\frac{1}{T}\int v_{s}\hat{T}^{10}dtd%
\mathbf{r}_{\bot }\ ,  \notag \\
&&\hat{T}^{10}\left( x\right) =-\left[ \partial _{x}\hat{\Phi}^{\dag }\left(
X\right) \right] \hat{\Pi}^{\dag }\left( X\right) -\hat{\Pi}\left( X\right)
\partial _{x}\hat{\Phi}\left( X\right) ,  \label{a8}
\end{eqnarray}%
where $\hat{T}^{10}$ is the component of the operator of the energy momentum
tensor; see Refs. \cite{GavGi16,GavGi20} for more details.

One can decompose the operator $\hat{\Phi}\left( X\right) $\ into solutions
of the either initial or final complete sets (\ref{in-out}) to construct
\textrm{in}- and \textrm{out}-states in an adequate Fock space:%
\begin{eqnarray}
\hat{\Phi}\left( X\right) &=&\sum_{m}\mathcal{M}_{m}^{-1/2}\left[ A_{m}(%
\mathrm{in})\phi _{m}^{\left( \mathrm{in},+\right) }\left( X\right) \right.
\notag \\
&+&\left. B_{m}^{\dagger }(\mathrm{in})\phi _{m}^{\left( \mathrm{in}%
,-\right) }\left( X\right) \right]  \notag \\
&=&\sum_{m}\mathcal{M}_{m}^{-1/2}\left[ A_{m}(\mathrm{out})\phi _{m}^{\left(
\mathrm{out},+\right) }\left( X\right) \right.  \notag \\
&+&\left. B_{m}^{\dagger }(\mathrm{out})\ \phi _{m}^{\left( \mathrm{out}%
,-\right) }\left( X\right) \right]  \label{dec}
\end{eqnarray}
The operator-valued coefficients can be determined with the help of the
inner product (\ref{d13}),%
\begin{eqnarray*}
&&A_{m}(\mathrm{in})=\frac{\mathcal{M}_{m}^{1/2}\left( \phi _{m}^{\left(
\mathrm{in},+\right) },\hat{\Phi}\right) }{\left( \phi _{m}^{\left( \mathrm{%
in},+\right) },\phi _{m}^{\left( \mathrm{in},+\right) }\right) }, \\
&&B_{m}^{\dagger }(\mathrm{in})=\frac{\mathcal{M}_{m}^{1/2}\left( \phi
_{m}^{\left( \mathrm{in},-\right) },\hat{\Phi}\right) }{\left( \phi
_{m}^{\left( \mathrm{in},-\right) },\phi _{m}^{\left( \mathrm{in},-\right)
}\right) }, \\
&&A_{m}(\mathrm{out})=\frac{\mathcal{M}_{m}^{1/2}\left( \phi _{m}^{\left(
\mathrm{out},+\right) },\hat{\Phi}\right) }{\left( \phi _{m}^{\left( \mathrm{%
out},+\right) },\phi _{m}^{\left( \mathrm{out},+\right) }\right) }, \\
&&B_{m}^{\dagger }(\mathrm{out})=\frac{\mathcal{M}_{m}^{1/2}\left( \phi
_{m}^{\left( \mathrm{out},-\right) },\hat{\Phi}\right) }{\left( \phi
_{m}^{\left( \mathrm{out},-\right) },\phi _{m}^{\left( \mathrm{out},-\right)
}\right) },
\end{eqnarray*}%
where $M_{m}${\large \ }are normalization factors given by Eq. (\ref{d14}).%
{\large \ }These operators\ define\ annihilation and creation operators of
initial or final particles in each the range $\Omega _{k}$\ ($m_{k}\in
\Omega _{k}$) as follows:

\begin{eqnarray}
A_{m_{1}}(\mathrm{in}) &=&\ _{+}a_{m_{1}}(\mathrm{in}),\;B_{m_{1}}^{\dagger
}(\mathrm{in})=\ ^{-}a_{m_{1}}(\mathrm{in});  \notag \\
A_{m_{1}}(\mathrm{out}) &=&\ ^{+}a_{m_{1}}(\mathrm{out})\
,\;B_{m_{1}}^{\dagger }(\mathrm{out})=\ _{-}a_{m_{1}}(\mathrm{out})\ ;
\notag \\
A_{m_{2}}(\mathrm{in}) &=&A_{m_{2}}(\mathrm{out})=a_{m_{2}},\;B_{m_{2}}^{%
\dagger }(\mathrm{in})=B_{m_{2}}^{\dagger }(\mathrm{out})=0\,;  \notag \\
A_{m_{3}}(\mathrm{in}) &=&\ _{-}a_{m_{3}}(\mathrm{in}),\;B_{m_{3}}^{\dagger
}(\mathrm{in})=\ ^{-}b_{m_{3}}^{\dagger }(\mathrm{in});  \notag \\
A_{m_{3}}(\mathrm{out}) &=&\ _{+}a_{m_{3}}(\mathrm{out}),\;B_{m_{3}}^{%
\dagger }(\mathrm{out})=\ ^{+}b_{m_{3}}^{\dagger }(\mathrm{out});  \notag \\
A_{m_{4}}(\mathrm{in}) &=&A_{m_{4}}(\mathrm{out})=0,\;B_{m_{4}}^{\dagger }(%
\mathrm{in})=B_{m_{4}}^{\dagger }(\mathrm{out})=b_{m}^{\dagger }\;;  \notag
\\
A_{m_{5}}(\mathrm{in}) &=&\;^{+}b_{m_{5}}^{\dag }(\mathrm{in}%
),\;B_{m_{5}}^{\dagger }(\mathrm{in})=\ _{-}b_{m_{5}}^{\dag }(\mathrm{in});
\notag \\
A_{m_{5}}(\mathrm{out}) &=&_{+}b_{m_{5}}^{\dag }(\mathrm{out}%
),\;B_{m_{5}}^{\dagger }(\mathrm{out})=\ ^{-}b_{m_{5}}^{\dag }(\mathrm{out}),
\label{ab}
\end{eqnarray}

Here $a$\ and $b\ $are annihilation and $a^{\dag }$\ and $b^{\dag }$\ are
creation operators, the operators $a$\ and $a^{\dag }$\ describe magnons and
the operators $b$\ and $b^{\dag }$\ describe antimagnons.{\large \ }The
vacuum vectors $\left\vert 0,\mathrm{in}\right\rangle $\ and $\left\vert 0,%
\mathrm{out}\right\rangle $\ are null vectors for all the annihilation
operators $a$\ and $b$.{\large \ }Operators labeled by the argument "\textrm{%
in}" are interpreted\ as initial particle operators, whereas operators
labeled by the argument "\textrm{out}" are interpreted as final particle
operators.\ Indeed, the commutation relations (\ref{a2}) yield the standard
commutation rules for the introduced creation and annihilation \textrm{in}-
or \textrm{out}-operators. One-particle states of initial (final) magnon and
antimagnon are%
\begin{equation}
a_{m}^{\dagger }(\mathrm{in/out})\left\vert 0,\mathrm{in/out}\right\rangle
,\;b_{m}^{\dagger }(\mathrm{in/out})\left\vert 0,\mathrm{in/out}%
\right\rangle ,  \label{add15}
\end{equation}%
where $a_{m}^{\dagger }(\mathrm{in/out})$ and $b_{m}^{\dagger }(\mathrm{%
in/out})$ are given by Eq. (\ref{ab}) for each the range{\large \ }$\Omega
_{k}${\large . }The unitary transformation (\ref{rel1}) implies a canonical
transformations between the \textrm{in} and \textrm{out}-operators.

Interpretation of the magnons and the antimagnon states which follows from
in Eqs. (\ref{ab}) is consistent with a spectrum analysis of the kinetic
energy operator $\widehat{\mathbb{H}}^{\mathrm{kin}}$ and the effective
charge operator $\hat{Q}${\large .} Inserting decompositions (\ref{dec}) in
Eqs. (\ref{a3}) and (\ref{a4}), we obtain diagonal representations for these
operators in terms of the introduced creation and annihilation operators.
The operators $\widehat{\mathbb{H}}^{\mathrm{kin}}$\ and $\hat{Q}$\ can be
represented as sums of five contributions, each one in the range $\Omega _{k}
$,{\large \ }%
\begin{equation}
{\large \ }\widehat{\mathbb{H}}^{\mathrm{kin}}=\sum_{k=1}^{5}\sum_{m\in
\Omega _{k}}\widehat{\mathbb{H}}_{m},\;\hat{Q}=\sum_{k=1}^{5}\sum_{m\in
\Omega _{k}}\hat{Q}_{m}.  \label{a9}
\end{equation}%
Note that a stationary state%
\begin{equation}
\Psi _{m}\left( X\right) =\left(
\begin{array}{c}
\left[ \varepsilon _{m}-U\left( x\right) \right] \phi _{m}\left( X\right)
\\
\phi _{m}\left( X\right)
\end{array}%
\right) ,  \label{add16}
\end{equation}%
where $\phi _{m}$ is one of the solutions from Eq. (\ref{in-out}),{\large \ }%
satisfies the following eigenvalue problem:%
\begin{equation}
H^{\mathrm{kin}}\Psi _{m}\left( X\right) =\left[ \varepsilon _{m}-U\left(
x\right) \right] \Psi _{m}\left( X\right) .  \label{a10}
\end{equation}%
This implies that the kinetic energy term of the one-particle state reads:%
\begin{eqnarray}
E_{m} &=&\left( \ \phi _{m},\ \phi _{m}\right) ^{-1}\int_{V_{\bot }}d\mathbf{%
r}_{\bot }  \notag \\
&\times &\int\limits_{-K^{\left( \mathrm{L}\right) }}^{K^{\left( \mathrm{R}%
\right) }}\;\Psi _{m}^{\dagger }\left( X\right) \sigma _{1}\left[
\varepsilon _{m}-U\left( x\right) \right] \;\Psi _{m}\left( X\right) dx\ ,
\label{a11}
\end{eqnarray}%
where the quantities $\left( \ \phi _{m},\ \phi _{m}\right) $\ are positive
for $m\in \Omega _{1}\cup \Omega _{2}$\ and are negative for $m\in \Omega
_{4}\cup \Omega _{5}$. In the range $\Omega _{3}$ we have that $\left( \
_{\zeta }\phi _{m},\ _{\zeta }\phi _{m}\right) >0,$ while $\left( \ ^{\zeta
}\phi _{m},\ ^{\zeta }\phi _{m}\right) <0$. Note that the kinetic energy
terms $\ _{\zeta }E_{m}$\ corresponding to the states $\ _{\zeta }\phi _{m}$
and the terms $\ ^{\zeta }E_{m}$ corresponding to the states $\ ^{\zeta
}\phi _{m}$ are different in the general case.{\large \ }The principal value
of integral (\ref{a11}) is determined by integrals over the areas $x\in %
\left[ -K^{\left( \mathrm{L}\right) },x_{\mathrm{L}}\right] $ and $x\in %
\left[ x_{\mathrm{R}},K^{\left( \mathrm{R}\right) }\right] $, where the
field derivative $\partial _{x}U$ is negligible small. Thus, it is possible
to evaluate integrals (\ref{a11}) for any form of the external field, using
only the asymptotic behavior (\ref{d9}) of functions in the regions $S_{%
\mathrm{L}}$and $S_{\mathrm{R}}$ where particles are free; see Sec. IV and
Appendix B in Ref. \cite{GavGi16} for details. Note that $\varepsilon
_{m}-U\left( x\right) =\pi _{0}\left( \mathrm{L/R}\right) $ in the regions $%
S_{\mathrm{L}}/S_{\mathrm{R}}$, respectively.

It can be easily seen from Eq. (\ref{a9}) that in the range $\Omega _{1}\cup
\Omega _{2}$\ a one-particle state is the state of a magnon with the kinetic
energy{\large \ }$E_{m}>0${\large \ }and the magnetic moment $\mu $\ whereas
in the range{\large \ }$\Omega _{4}\cup \Omega _{5}$\ a one-particle state
is the state of an antimagnon with the kinetic energy $-E_{m}>0$\ and the
magnetic moment $-\mu $.

Inserting decompositions (\ref{dec}) in operators (\ref{a7}) and (\ref{a8}),
we obtain a renormalized (with respect to the corresponding vacua) in- and
out-operators of the effective charge (magnetic moment) current and energy
flux flowing\ through the surfaces $x=x_{\mathrm{L}}$ and $x=x_{\mathrm{R}}$%
, respectively,
\begin{eqnarray}
&&\widehat{\mathbb{J}}\left( \mathrm{in}\right) =\widehat{\mathbb{J}}%
-\left\langle 0,\mathrm{in}\left\vert \widehat{\mathbb{J}}\right\vert 0,%
\mathrm{in}\right\rangle \,,  \notag \\
&&\widehat{\mathbb{J}}\left( \mathrm{out}\right) =\widehat{\mathbb{J}}%
-\left\langle 0,\mathrm{out}\left\vert \widehat{\mathbb{J}}\right\vert 0,%
\mathrm{out}\right\rangle \ ,  \notag \\
&&\widehat{\mathbb{F}}\left( x|\mathrm{in}\right) =\widehat{\mathbb{F}}%
\left( x\right) -\left\langle 0,\mathrm{in}\left\vert \widehat{\mathbb{F}}%
\left( x\right) \right\vert 0,\mathrm{in}\right\rangle \,,  \notag \\
&&\widehat{\mathbb{F}}\left( x|\mathrm{out}\right) =\widehat{\mathbb{F}}%
\left( x\right) -\left\langle 0,\mathrm{out}\left\vert \widehat{\mathbb{F}}%
\left( x\right) \right\vert 0,\mathrm{out}\right\rangle \ .  \label{a14}
\end{eqnarray}

The one-particle mean values of the fluxes, the kinetic energy, and the
effective charge through the surfaces\ $x=x_{\mathrm{L}}$\ and\ $x=x_{%
\mathrm{R}}$, given by Eqs. (\ref{a14}),\ are proportional to the inner
product on these surfaces given by Eq. (\ref{d10}),{\large \ }that is, these
fluxes are proportional to the flux densities of\ the particles with given $%
m ${\large . }Of course, in the range $\Omega _{2}\cup \Omega _{4}$\ the
flux densities of\ particles, given by standing waves, are zero.

With account taken of the charge conjugation, it can be seen that the sign
of the flux densities of\ magnons with given $m$\ is equal to $\zeta $\ in
the range $\Omega _{1}$,\ whereas the sign of the flux densities of\
antimagnons with given $m$ \ is equal to $-\zeta $ in the range $\Omega _{5}$%
. In the ranges $\Omega _{1}$ and $\Omega _{5}$ an initial state may be
localized both in the regions $S_{\mathrm{L}}$ and $S_{\mathrm{R}}$. This
follows from the way of choosing initial conditions.{\large \ }Taking into
account directions of motion of magnons and antimagnons in the regions $S_{%
\mathrm{L}}$and $S_{\mathrm{R}}$, we define initial and final states as it
is done in Eqs. (\ref{ab}) and (\ref{in-out}).

In the Klein zone $\Omega _{3}$ the identification of states demands a
special consideration. To this end, we represent explicitly the operators $%
\widehat{\mathbb{H}}_{m}$ and $\hat{Q}_{m}$ as follows:%
\begin{eqnarray}
\widehat{\mathbb{H}}_{m} &=&\ _{+}E_{m}\ \ _{+}a_{m}^{\dagger }(\mathrm{out}%
)\ _{+}a_{m}(\mathrm{out})  \notag \\
&-&\ ^{+}E_{m}\ \ ^{+}b_{m}^{\dagger }(\mathrm{out})\ ^{+}b_{m}(\mathrm{out})
\notag \\
&=&\ _{-}E_{m}\ \ _{-}a_{m}^{\dagger }(\mathrm{in})\ _{-}a_{m}(\mathrm{in})
\notag \\
&-&\ ^{-}E_{m}\ \ \ _{-}b_{m}^{\dagger }(\mathrm{in})\ _{-}b_{m}(\mathrm{in}%
);  \notag \\
\hat{Q}_{m} &=&\mu \left[ \ \ _{+}a_{m}^{\dagger }(\mathrm{out})\ _{+}a_{m}(%
\mathrm{out})\right.  \notag \\
&-&\ \left. \ ^{+}b_{m_{3}}^{\dagger }(\mathrm{out})\ ^{+}b_{m_{3}}(\mathrm{%
out})\right]  \notag \\
&=&\mu \ \left[ \ _{-}a_{m_{3}}^{\dagger }(\mathrm{in})\ _{-}a_{m_{3}}(%
\mathrm{in})\right.  \notag \\
&-&\left. \ _{-}b_{m_{3}}^{\dagger }(\mathrm{in})\ _{-}b_{m_{3}}(\mathrm{in})%
\right] \,,\ \ m\in \Omega _{3}.  \label{a12}
\end{eqnarray}%
Here $\ _{\zeta }E_{m}$\ and $\ ^{\zeta }E_{m}$\ are principal values of
integral (\ref{a11}) for $\ _{\zeta }\Psi _{m}$\ and $^{\zeta }\Psi _{m}$,
respectively. They read:
\begin{eqnarray}
&&\ _{\zeta }E_{m}\mathcal{=}\pi _{0}\left( \mathrm{R}\right) +\frac{\delta U%
}{2}\left\vert g\left( _{+}\left\vert ^{-}\right. \right) \right\vert ^{-2},
\notag \\
&&\;\ ^{\zeta }E_{m}=\pi _{0}\left( \mathrm{L}\right) -\frac{\delta U}{2}%
\left\vert g\left( _{+}\left\vert ^{-}\right. \right) \right\vert ^{-2}.
\label{a13}
\end{eqnarray}%
Thus, we see that in the range $\Omega _{3}$ the one-particle state $\
_{\zeta }\Psi _{m}$\ is the magnon state with the kinetic energy $_{\zeta
}E_{m}>0$\ and the magnetic moment $\mu $\ whereas the one-particle state $\
^{\zeta }\Psi _{m}$\ is the antimagnon state with the kinetic energy $-\
^{\zeta }E_{m}>0$\ and the magnetic moment{\large \ }$-\mu $.

Let us find one-particle mean values of fluxes of the kinetic energy and the
effective charge through the surfaces\ $x=x_{\mathrm{L}}$\ and\ $x=x_{%
\mathrm{R}}$, given by Eqs. (\ref{a14}), in the range $\Omega _{3}$. With
the help of Eq. (\ref{d10}) we obtain:%
\begin{widetext}
\begin{eqnarray}
J_{m}^{a}\left( \mathrm{in}\right) &=&\left\langle 0,\mathrm{in}\left\vert
\;_{-}a_{m}(\mathrm{in})\widehat{\mathbb{J}}\left( \mathrm{in}\right)
\;_{-}a_{m}^{\dagger }(\mathrm{in})\right\vert 0,\mathrm{in}\right\rangle
=-\mu \left( \mathcal{M}_{m}T\right) ^{-1},  \notag \\
J_{m}^{a}\left( \mathrm{out}\right) &=&\left\langle 0,\mathrm{out}\left\vert
\ _{+}a_{m}(\mathrm{out})\widehat{\mathbb{J}}\left( \mathrm{out}\right) \
_{+}a_{m}^{\dagger }(\mathrm{out})\ \right\vert 0,\mathrm{out}\right\rangle
=\mu \left( \mathcal{M}_{m}T\right) ^{-1},  \notag \\
J_{m}^{b}\left( \mathrm{in}\right) &=&\left\langle 0,\mathrm{in}\left\vert
\;^{-}b_{m}(\mathrm{in})\widehat{\mathbb{J}}\left( \mathrm{in}\right)
\;^{-}b_{m}^{\dagger }(\mathrm{in})\right\vert 0,\mathrm{in}\right\rangle
=-\mu \left( \mathcal{M}_{m}T\right) ^{-1},  \notag \\
J_{m}^{b}\left( \mathrm{out}\right) &=&\left\langle 0,\mathrm{out}\left\vert
\ ^{+}b_{m}^{\dagger }(\mathrm{out})\widehat{\mathbb{J}}\left( \mathrm{out}%
\right) \ ^{+}b_{m}^{\dagger }(\mathrm{out})\right\vert 0,\mathrm{out}%
\right\rangle =\mu \left( \mathcal{M}_{m}T\right) ^{-1};  \notag \\
F_{m}^{a}\left( \mathrm{in}\right) &=&\left\langle 0,\mathrm{in}\left\vert
\;_{-}a_{m}(\mathrm{in})\mathbb{\hat{F}}\left( x_{\mathrm{R}},\mathrm{out}%
\right) \;_{-}a_{m}^{\dagger }(\mathrm{in})\right\vert 0,\mathrm{in}%
\right\rangle =-\left( \mathcal{M}_{m}T\right) ^{-1}\pi _{0}\left( \mathrm{R}%
\right) ,  \notag \\
F_{m}^{a}\left( \mathrm{out}\right) &=&\left\langle 0,\mathrm{out}\left\vert
\ _{+}a_{m}(\mathrm{out})\mathbb{\hat{F}}\left( x_{\mathrm{R}},\mathrm{out}%
\right) \ _{+}a_{m}(\mathrm{out})\right\vert 0,\mathrm{out}\right\rangle
=\left( \mathcal{M}_{m}T\right) ^{-1}\pi _{0}\left( \mathrm{R}\right) ,
\notag \\
F_{m}^{b}\left( \mathrm{in}\right) &=&\left\langle 0,\mathrm{in}\left\vert
\;^{-}b_{m}(\mathrm{in})\mathbb{\hat{F}}\left( x_{\mathrm{L}},\mathrm{in}%
\right) \;^{-}b_{m}^{\dagger }(\mathrm{in})\right\vert 0,\mathrm{in}%
\right\rangle =\left( \mathcal{M}_{n}T\right) ^{-1}\left\vert \pi _{0}\left(
\mathrm{L}\right) \right\vert ,  \notag \\
F_{m}^{b}\left( \mathrm{out}\right) &=&\left\langle 0,\mathrm{out}\left\vert
\ ^{+}b_{m}^{\dagger }(\mathrm{out})\mathbb{\hat{F}}\left( x_{\mathrm{L}},%
\mathrm{out}\right) \ ^{+}b_{m}^{\dagger }(\mathrm{out})\right\vert 0,%
\mathrm{out}\right\rangle =-\left( \mathcal{M}_{n}T\right) ^{-1}\left\vert
\pi _{0}\left( \mathrm{L}\right) \right\vert .  \label{a15}
\end{eqnarray}
\end{widetext}Taking into account the space separation of magnons and
antimagnons in the region $S_{\mathrm{R}}$ and $S_{\mathrm{L}}$, one can use
the mean values (\ref{a15}) to distinguish initial and final states in the
range $\Omega _{3}$.

The QFT operators of the effective charge (magnetic moment) current density $%
\hat{J}_{x}$\ and energy flux density $v_{s}\hat{T}^{10}$\ flowing\ through
the surfaces $x=x_{\mathrm{L}}$\ and $x=x_{\mathrm{R}}$, that were
introduced above in the framework of the approximation under consideration,
are time-independent. It is clear that these operators are defined up to
certain $C$-numbers, which may affect explicit forms of the corresponding
vacuum means.{\large \ }This circumstance allows one to relate matrix
elements of these operators with matrix elements of the corresponding exact
strong-field{\large \ }QED\ operators (which are time-dependent in the
presence of the time-dependent field\ $E_{\mathrm{pristine}}$). This can be
done based on the following physical considerations:{\large \ }Let us
consider a relation of the time-independent quantity $J_{x}^{\mathrm{cr}}$,
obtained in the framework of approximation under consideration, to the
matrix elements of the time-dependent longitudinal component $\hat{J}_{x}^{%
\mathrm{true}}\left( t\right) $ of an exact current density operator of the
strong-field QED{\large .\ }According to the general theory the difference $%
\delta J_{x}^{\mathrm{true}}$ of the true final vacuum from the initial one
is due to the contribution of the current density of the created particles
and antiparticles,{\large \ }%
\begin{equation}
\delta J_{x}^{\mathrm{true}}=\left\langle 0,\mathrm{true\;in}\left\vert %
\left[ \hat{J}_{x}^{\mathrm{true}}\left( t_{2}\right) -\hat{J}_{x}^{\mathrm{%
true}}\left( t_{1}\right) \right] \right\vert 0,\mathrm{true\;in}%
\right\rangle ,  \label{q7}
\end{equation}%
where $t_{1}<t_{\mathrm{in}}$\ and $t_{2}>t_{\mathrm{out}}$\ are the time
instants of switching on and off of the field $E_{\mathrm{pristine}}$,
respectively.{\large \ }Assuming that effects of fast switching-on and -off
are small, we can use approximation (\ref{q5}) for the total number of
created pairs. In this case we obtain:{\large \ }%
\begin{equation}
\delta J_{x}^{\mathrm{true}}=J_{x}^{\mathrm{cr}}\left\{ 1+O\left( \left[
\sqrt{v_{s}\left\vert \partial _{x}U\right\vert }T\right] ^{-1}\right)
\right\} .  \label{q9}
\end{equation}%
We see that the quantity $J_{x}^{\mathrm{cr}}$\ can be represented as the
following mean value with respect to the $\left\vert 0,\mathrm{in}%
\right\rangle $\ vacuum:{\large \ }%
\begin{equation}
J_{x}^{\mathrm{cr}}\approx \left\langle 0,\mathrm{in}\left\vert \left[ \hat{J%
}_{x}^{\mathrm{true}}\left( t_{2}\right) -\hat{J}_{x}^{\mathrm{true}}\left(
t_{1}\right) \right] \right\vert 0,\mathrm{in}\right\rangle \ .  \label{q10}
\end{equation}

With the help of this result we may find a relation of the operators $\hat{J}%
_{x}^{\mathrm{true}}\left( t_{2}\right) $\ and $\hat{J}_{x}^{\mathrm{true}%
}\left( t_{1}\right) $\ with the time-independent current density operator $%
\hat{J}_{x}$\ given by Eq. (\ref{a7}). In particular, we see that the
difference $\hat{J}_{x}^{\mathrm{true}}\left( t_{2}\right) -\hat{J}_{x}^{%
\mathrm{true}}\left( t_{1}\right) $\ can be approximated by a $C$-number,%
{\large \ }%
\begin{equation}
\hat{J}_{x}^{\mathrm{true}}\left( t_{2}\right) -\hat{J}_{x}^{\mathrm{true}%
}\left( t_{1}\right) \approx J_{x}^{\mathrm{cr}}.  \label{q11}
\end{equation}%
Then, for example, the current density operator, $\hat{J}_{x}^{\mathrm{true}%
}\left( t_{1}\right) $\ can be approximated by\ the time-independent current
density operator $\hat{J}_{x}${\large , }$\hat{J}_{x}^{\mathrm{true}}\left(
t_{1}\right) \approx \hat{J}_{x}${\large \ . }In this case, we have $\hat{J}%
_{x}^{\mathrm{true}}\left( t_{2}\right) \approx \hat{J}_{x}+J_{x}^{\mathrm{cr%
}}$, therefore the normal form of both $\hat{J}_{x}^{\mathrm{true}}\left(
t_{1}\right) $\ and $\hat{J}_{x}^{\mathrm{true}}\left( t_{2}\right) $\ with
respect to the in- and out-operators of creation and annihilation are the
same.{\large \ }Thus, calculating one-particle mean values of the effective
charge current in Eq. (\ref{a15}) one can use the operator $\hat{J}_{x}$. In
a similar way, one can relate the time-independent operator $\hat{T}%
^{10}\left( x\right) $\ to time-dependent components of the exact operator
EMT,{\large \ }for example,{\large \ }$\hat{T}_{\mathrm{true}}^{10}\left(
t_{1},x\right) \approx \hat{T}^{10}\left( x\right) $\ and{\large \ }$\hat{T}%
_{\mathrm{true}}^{10}\left( t_{2},x\right) \approx \hat{T}^{10}\left(
x\right) +T_{\mathrm{cr}}^{10}(x)${\large . }In addition, calculating
one-particle mean values of fluxes of the kinetic energy in Eq. (\ref{a15})
one can use the operator $\hat{T}^{10}\left( x\right) $.

\section{Examples of exact solutions with x steps\label{B}}

In this appendix, we provide additional information on the computation of
differential quantities that are relevant to the investigation of magnon
pair production stimulated by the external fields mentioned in Sec. \ref{S4}%
. We present results only and refer the reader to Refs. \cite%
{GavGi16,GavGit16b,GavGitSh17,AdoGavGit20} for more comprehensive
discussions.

For the $L$-constant magnetic step (\ref{se41.1}), solutions of Eq. (\ref{d7}%
) to all intervals can be expressed in the form:%
\begin{widetext}
\begin{eqnarray}
\ ^{-}\varphi _{m}\left( x\right) &=&Y\left\{
\begin{array}{l}
\ _{+}C\exp \left[ i\left\vert p^{\mathrm{L}}\right\vert \left( x-x_{\mathrm{%
L}}\right) \right] g\left( _{+}|^{-}\right) -\ _{-}C\exp \left[ -i\left\vert
p^{\mathrm{L}}\right\vert \left( x-x_{\mathrm{L}}\right) \right] g\left(
_{-}|^{-}\right) \,,\ \ x\in S_{\mathrm{L}}\,, \\
\ ^{-}C\left\{ b_{1}D_{\nu }\left[ -\left( 1-i\right) \xi \right]
+b_{2}D_{-\nu -1}\left[ -\left( 1+i\right) \xi \right] \right\} \,,\ \ x\in
S_{\mathrm{int}}\,, \\
\ ^{-}C\exp \left[ -i\left\vert p^{\mathrm{R}}\right\vert \left( x-x_{%
\mathrm{R}}\right) \right] \,,\ \ x\in S_{\mathrm{R}}\,.%
\end{array}%
\right.  \label{se41.3} \\
\ ^{+}\varphi _{m}\left( x\right) &=&Y\left\{
\begin{array}{l}
\ _{+}Cg\left( _{+}|^{+}\right) \exp \left[ i\left\vert p^{\mathrm{L}%
}\right\vert \left( x-x_{\mathrm{L}}\right) \right] -\ _{-}Cg\left(
_{-}|^{+}\right) \exp \left[ -i\left\vert p^{\mathrm{L}}\right\vert \left(
x-x_{\mathrm{L}}\right) \right] \,,\ \ x\in S_{\mathrm{L}}\,, \\
\ ^{+}C\left\{ b_{1}^{\prime }D_{\nu }\left[ \left( 1-i\right) \xi \right]
+b_{2}^{\prime }D_{-\nu -1}\left[ \left( 1+i\right) \xi \right] \right\}
\,,\ \ x\in S_{\mathrm{int}}\,, \\
\ ^{+}C\exp \left[ i\left\vert p^{\mathrm{R}}\right\vert \left( x-x_{\mathrm{%
R}}\right) \right] \,,\ \ x\in S_{\mathrm{R}}\,,%
\end{array}%
\right.  \label{se41.4}
\end{eqnarray}
\end{widetext}where $b_{j}$, $b_{j}^{\prime }$, $g\left( ^{\zeta }|_{\zeta
^{\prime }}\right) $, $g\left( _{\zeta }|^{\zeta ^{\prime }}\right) $ are
constants, which can be obtained via the continuity conditions:%
\begin{eqnarray}
&&\ _{-}^{+}\varphi _{m}\left( x_{\mathrm{L/R}}-0\right) =\ _{-}^{+}\varphi
_{m}\left( x_{\mathrm{L/R}}+0\right) \,,  \notag \\
&&\frac{d}{dx}\ _{-}^{+}\varphi _{m}\left( x_{\mathrm{L/R}}-0\right) =\frac{d%
}{dx}\ _{-}^{+}\varphi _{m}\left( x_{\mathrm{L/R}}+0\right) \,.
\label{add17}
\end{eqnarray}%
By demanding continuity of the above functions and their derivatives at $%
x=x_{\mathrm{R}}$ we find%
\begin{eqnarray}
b_{j} &=&\left( -1\right) ^{j+1}\exp \left[ \frac{i\pi }{2}\left( \nu +\frac{%
1}{2}\right) \right] \sqrt{\frac{\xi _{2}^{2}-\lambda }{2}}f_{j}^{\left(
-\right) }\left( \xi _{2}\right) \,,  \notag \\
b_{j}^{\prime } &=&\left( -1\right) ^{j+1}\exp \left[ \frac{i\pi }{2}\left(
\nu +\frac{1}{2}\right) \right] \sqrt{\frac{\xi _{2}^{2}-\lambda }{2}}%
f_{j}^{\left( +\right) }\left( \xi _{2}\right) \,,  \label{add18}
\end{eqnarray}%
where $\lambda =\pi _{\bot }^{2}/v_{s}\mu B^{\prime }$ and $f_{j}^{\left(
\pm \right) }\left( \xi \right) $ were defined before, see Eqs. (\ref{se41.8}%
). Now, by imposing the continuity of the functions and derivatives at $x=x_{%
\mathrm{L}}$ we obtain the coefficients $g\left( _{+}|^{+}\right) $ and $%
g\left( _{+}|^{-}\right) $:%
\begin{widetext}
\begin{eqnarray}
g\left( _{+}|^{-}\right) &=&e^{\frac{i\pi }{2}\left( \nu +\frac{1}{2}\right)
}\sqrt{\frac{\sqrt{\xi _{1}^{2}-\lambda }\sqrt{\xi _{2}^{2}-\lambda }}{8}}%
\left[ f_{1}^{\left( -\right) }\left( \xi _{2}\right) f_{2}^{\left( -\right)
}\left( \xi _{1}\right) -f_{2}^{\left( -\right) }\left( \xi _{2}\right)
f_{1}^{\left( -\right) }\left( \xi _{1}\right) \right] \,,  \label{se41.6} \\
g\left( _{+}|^{+}\right) &=&e^{\frac{i\pi }{2}\left( \nu +\frac{1}{2}\right)
}\sqrt{\frac{\sqrt{\xi _{1}^{2}-\lambda }\sqrt{\xi _{2}^{2}-\lambda }}{8}}%
\left[ f_{1}^{\left( +\right) }\left( \xi _{2}\right) \tilde{f}_{2}^{\left(
-\right) }\left( \xi _{1}\right) -f_{2}^{\left( +\right) }\left( \xi
_{2}\right) \tilde{f}_{1}^{\left( -\right) }\left( \xi _{1}\right) \right]
\,,  \label{se41.7}
\end{eqnarray}
\end{widetext}where%
\begin{eqnarray}
\tilde{f}_{1}^{\left( \pm \right) }\left( \xi \right) &=&\left( 1\pm \frac{i%
}{\sqrt{\xi ^{2}-\lambda }}\frac{d}{d\xi }\right) D_{-\nu -1}\left[ \mp
\left( 1+i\right) \xi \right] \,,  \notag \\
\tilde{f}_{2}^{\left( \pm \right) }\left( \xi \right) &=&\left( 1\pm \frac{i%
}{\sqrt{\xi ^{2}-\lambda }}\frac{d}{d\xi }\right) D_{\nu }\left[ \mp \left(
1-i\right) \xi \right] \,.  \label{add19}
\end{eqnarray}

For the Sauter-like magnetic step (\ref{se42.1}), solutions of Eq. (\ref{d7}%
) and (\ref{se42.4}) with real asymptotic momenta (\ref{d9}) in remote
regions $x\rightarrow \mp \infty $ read:%
\begin{eqnarray}
\ _{\zeta }\varphi _{m}\left( x\right)  &=&\ _{\zeta }\mathcal{N}\exp \left(
i\zeta \left\vert p^{\mathrm{L}}\right\vert x\right)   \notag \\
&\times &\left[ 1+\exp \left( 2x/L_{\mathrm{S}}\right) \right] ^{-iL_{%
\mathrm{S}}\left( \zeta \left\vert p^{\mathrm{L}}\right\vert +\left\vert p^{%
\mathrm{R}}\right\vert \right) /2}\ _{\zeta }u_{m}\left( x\right) \,,  \notag
\\
\ ^{\zeta }\varphi _{m}\left( x\right)  &=&\ ^{\zeta }\mathcal{N}\exp \left(
i\zeta \left\vert p^{\mathrm{R}}\right\vert x\right)   \notag \\
&\times &\left[ 1+\exp \left( -2x/L_{\mathrm{S}}\right) \right] ^{iL_{%
\mathrm{S}}\left( \left\vert p^{\mathrm{L}}\right\vert +\zeta \left\vert p^{%
\mathrm{R}}\right\vert \right) /2}\ ^{\zeta }u_{m}\left( x\right) \,,
\label{b1}
\end{eqnarray}%
where%
\begin{eqnarray}
\ _{-}u_{m}\left( x\right)  &=&F\left( a,b;c;\chi \right) \,,  \notag \\
_{+}u_{m}\left( x\right)  &=&F\left( a+1-c,b+1-c;2-c;\chi \right) \,,  \notag
\\
\ ^{-}u_{m}\left( x\right)  &=&F\left( a,b;a+b+1-c;1-\chi \right) \,,  \notag
\\
^{+}u_{m}\left( x\right)  &=&F\left( c-a,c-b;c+1-a-b;1-\chi \right) \,,
\label{b2}
\end{eqnarray}%
where $\chi \left( x\right) $ is the change of variable defined in Eq. (\ref%
{se42.2}) and $a$, $b$, $c$ are given by Eqs. (\ref{se42.5}). Using the
above solutions and Kummer's connection formulas \cite{Erdelyi}%
\begin{eqnarray}
&&\left( 1-\chi \right) ^{c-a-b}\ ^{+}u_{m}\left( x\right) =\frac{\Gamma
\left( c+1-a-b\right) \Gamma \left( 1-c\right) }{\Gamma \left( 1-a\right)
\Gamma \left( 1-b\right) }\ _{-}u_{m}\left( x\right)   \notag \\
&&+\frac{\Gamma \left( c+1-a-b\right) \Gamma \left( c-1\right) }{\Gamma
\left( c-a\right) \Gamma \left( c-b\right) }\chi ^{1-c}\ _{+}u_{m}\left(
x\right) \,,  \notag \\
&&\ ^{-}u_{m}\left( x\right) =\frac{\Gamma \left( a+b+1-c\right) \Gamma
\left( 1-c\right) }{\Gamma \left( a+1-c\right) \Gamma \left( b+1-c\right) }\
_{-}u_{m}\left( x\right)   \notag \\
&&+\frac{\Gamma \left( a+b+1-c\right) \Gamma \left( c-1\right) }{\Gamma
\left( a\right) \Gamma \left( b\right) }\chi ^{1-c}\ _{+}u_{m}\left(
x\right) \,,  \label{b3}
\end{eqnarray}%
we conclude that%
\begin{eqnarray}
g\left( _{+}|^{+}\right)  &=&\frac{\ ^{+}\mathcal{N}}{\ _{+}\mathcal{N}}%
\frac{\Gamma \left( c+1-a-b\right) \Gamma \left( c-1\right) }{\Gamma \left(
c-a\right) \Gamma \left( c-b\right) }\,,  \notag \\
g\left( _{+}|^{-}\right)  &=&\frac{\ ^{-}\mathcal{N}}{\ _{+}\mathcal{N}}%
\frac{\Gamma \left( a+b+1-c\right) \Gamma \left( c-1\right) }{\Gamma \left(
a\right) \Gamma \left( b\right) }\,.  \label{b4}
\end{eqnarray}

For the exponential step (\ref{se43.1}), general solutions of Eq. (\ref{d7})
with such a field can be presented as a linear combination of two functions,
$y_{1}^{j}\left( \eta _{j}\right) $ and $y_{2}^{j}\left( \eta _{j}\right) $%
\begin{eqnarray}
y_{1}^{j}\left( \eta _{j}\right) &=&e^{-\eta _{j}/2}\eta _{j}^{\nu _{j}}\Phi
\left( a_{j},c_{j};\eta _{j}\right)  \notag \\
& =&e^{\eta _{j}/2}\eta _{j}^{\nu _{j}}\Phi \left( c_{j}-a_{j},c_{j};-\eta
_{j}\right) \,,  \notag \\
y_{2}^{j}\left( \eta _{j}\right) &=&e^{\eta _{j}/2}\eta _{j}^{-\nu _{j}}\Phi
\left( 1-a_{j},2-c_{j};-\eta _{j}\right)  \notag \\
&=&e^{-\eta _{j}/2}\eta _{j}^{-\nu _{j}}\Phi \left(
a_{j}-c_{j}+1,2-c_{j};\eta _{j}\right) \,,  \label{b5}
\end{eqnarray}%
where $\eta _{j}\left( x\right) $ are defined in Eqs. (\ref{se43.2}) and the
parameters $a_{j}$, $c_{j}$ by Eqs. (\ref{se43.5}). In particular, solutions
with special asymptotic properties in remote regions $x\rightarrow \mp
\infty $ are classified as%
\begin{eqnarray}
\ _{+}\varphi _{m}\left( x\right)&=&\ _{+}\mathcal{N}e^{-i\pi \nu
_{1}/2}y_{1}^{1}\left( \eta _{1}\right) \,,  \notag \\
\ _{-}\varphi _{m}\left( x\right)& =&\ _{-}\mathcal{N}e^{i\pi \nu
_{1}/2}y_{2}^{1}\left( \eta _{1}\right) \,,  \notag \\
\ ^{+}\varphi _{m}\left( x\right) &=&\ ^{+}\mathcal{N}e^{i\pi \nu
_{2}/2}y_{2}^{2}\left( \eta _{2}\right) \,,  \notag \\
\ ^{-}\varphi _{m}\left( x\right)&=&\ ^{-}\mathcal{N}e^{-i\pi \nu
_{2}/2}y_{1}^{2}\left( \eta _{2}\right) \,,  \label{b6}
\end{eqnarray}

With the aid of this classification, we may express solutions to all
intervals in the form%
\begin{widetext}
\begin{eqnarray}
\ ^{+}\varphi _{m}\left( x\right) &=&\left\{
\begin{array}{ll}
\ _{+}\varphi _{m}\left( x\right) g\left( _{+}|^{+}\right) -\ _{-}\varphi
_{m}\left( x\right) g\left( _{-}|^{+}\right) \,, & x\in \mathrm{I} \\
\ ^{+}\mathcal{N}e^{i\pi \nu _{2}/2}y_{2}^{2}\left( \eta _{2}\right) \,, &
x\in \mathrm{II}%
\end{array}%
\right.  \label{b7} \\
\ ^{-}\varphi _{m}\left( x\right) &=&\left\{
\begin{array}{ll}
\ _{+}\varphi _{m}\left( x\right) g\left( _{+}|^{-}\right) -\ _{-}\varphi
_{m}\left( x\right) g\left( _{-}|^{-}\right) \,, & x\in \mathrm{I} \\
\ ^{-}\mathcal{N}e^{-i\pi \nu _{2}/2}y_{1}^{2}\left( \eta _{2}\right) \,, &
x\in \mathrm{II}%
\end{array}%
\right.  \label{b8}
\end{eqnarray}
\end{widetext}where $\mathrm{I}=-\infty <x\leq 0$, and $\mathrm{II}%
=0<x<+\infty $. To calculate the decomposition coefficients $g\left( _{\zeta
}|^{\zeta ^{\prime }}\right) $, $g\left( ^{\zeta }|_{\zeta ^{\prime
}}\right) $, we impose continuity of the functions and their derivatives at $%
x=0$:%
\begin{eqnarray}
\ ^{\zeta }\varphi _{m}\left( x-0\right) &=&\ ^{\zeta }\varphi _{m}\left(
x+0\right) \,,  \notag \\
\frac{d}{dx}\ ^{\zeta }\varphi _{m}\left( x-0\right) &=&\frac{d}{dx}\
^{\zeta }\varphi _{m}\left( x+0\right) \,,  \label{b9}
\end{eqnarray}%
In particular, the coefficients $g\left( _{+}|^{-}\right) $ and $g\left(
_{+}|^{+}\right) $ have the form:%
\begin{widetext}
\begin{eqnarray}
g\left( _{+}|^{+}\right) &=&-\frac{\exp \left[ \frac{i\pi }{2}\left( \nu
_{1}+\nu _{2}\right) \right] }{2\sqrt{\left\vert p^{\mathrm{L}}\right\vert
\left\vert p^{\mathrm{R}}\right\vert }}\left. \left( k_{1}h_{1}y_{2}^{2}%
\frac{d}{d\eta _{1}}y_{2}^{1}+k_{2}h_{2}y_{2}^{1}\frac{d}{d\eta _{2}}%
y_{2}^{2}\right) \right\vert _{x=0}\,,  \notag \\
g\left( _{+}|^{-}\right) &=&-\frac{\exp \left[ \frac{i\pi }{2}\left( \nu
_{1}-\nu _{2}\right) \right] }{2\sqrt{\left\vert p^{\mathrm{L}}\right\vert
\left\vert p^{\mathrm{R}}\right\vert }}\left. \left( k_{1}h_{1}y_{1}^{2}%
\frac{d}{d\eta _{1}}y_{2}^{1}+k_{2}h_{2}y_{2}^{1}\frac{d}{d\eta _{2}}%
y_{1}^{2}\right) \right\vert _{x=0}\,.  \label{b10}
\end{eqnarray}
\end{widetext}

Lastly, exact solutions to Eq. (\ref{d7}) with inverse-square magnetic steps
(\ref{se44.1}) can be represented as a linear combination of Whittaker
functions $W_{\kappa _{j},\mu _{j}}\left( z_{j}\right) $, $W_{-\kappa
_{j},\mu _{j}}\left( e^{-i\pi }z_{j}\right) $%
\begin{eqnarray}
&&w_{1}^{j}\left( z_{j}\right) =e^{-i\pi \kappa _{j}/2}W_{\kappa _{j},\mu
_{j}}\left( z_{j}\right) \,,  \notag \\
&&w_{2}^{j}\left( z_{j}\right) =e^{-i\pi \kappa _{j}/2}W_{-\kappa _{j},\mu
_{j}}\left( e^{-i\pi }z_{j}\right) \,,  \label{b11}
\end{eqnarray}%
where\ the\ variables $z_{j}\left( x\right) $ are given by Eqs. (\ref{se44.2}%
) and the parameters $\kappa _{j}$, $\mu _{j}$\ by Eqs. (\ref{se44.4}).
Their Wronskian determinant $\mathbb{W}$ reads \cite{DLMF}%
\begin{equation}
\mathbb{W}=w_{1}^{j}\left( z_{j}\right) \frac{d}{dz_{j}}w_{2}^{j}\left(
z_{j}\right) -w_{2}^{j}\left( z_{j}\right) \frac{d}{dz_{j}}w_{1}^{j}\left(
z_{j}\right) =1\,.  \label{add20}
\end{equation}%
Sometimes, it is convenient to represent the set of solutions (\ref{b11}) in
terms of confluent hypergeometric functions%
\begin{eqnarray}
w_{1}^{j}\left( z_{j}\right) &=&\exp \left[ -\frac{i\pi }{2}\left( \kappa
_{j}-\mu _{j}-\frac{1}{2}\right) \right]  \notag \\
&\times &e^{-z_{j}/2}\left\vert z_{j}\right\vert ^{c_{j}/2}\Psi \left(
\tilde{a}_{j},\tilde{c}_{j};z_{j}\right) \,,  \notag \\
w_{2}^{j}\left( z_{j}\right) &=&\exp \left[ -\frac{i\pi }{2}\left( \kappa
_{j}+\mu _{j}+\frac{1}{2}\right) \right]  \notag \\
&\times &e^{z_{j}/2}\left\vert z_{j}\right\vert ^{c_{j}/2}\Psi \left( \tilde{%
c}_{j}-\tilde{a}_{j},\tilde{c}_{j};e^{-i\pi }z_{j}\right) \,,  \label{add21}
\end{eqnarray}%
in which $\tilde{a}_{j}=\mu _{j}-\kappa _{j}+1/2$, $\tilde{c}_{j}=1+2\mu
_{j} $.

Based on asymptotic properties of the Whittaker functions with large
argument \cite{DLMF}, solutions with real asymptotic momenta (\ref{d9}) in
remote regions $x\rightarrow \mp \infty $ are classified as follows:%
\begin{eqnarray}
\ _{+}\varphi _{m}\left( x\right) &=&\ _{+}\mathcal{N}w_{1}^{1}\left(
z_{1}\right) \,,\ \ \ _{-}\varphi _{m}\left( x\right) =\ _{-}\mathcal{N}%
w_{2}^{1}\left( z_{1}\right) \,,  \notag \\
\ ^{+}\varphi _{m}\left( x\right) &=&\ ^{+}\mathcal{N}w_{2}^{2}\left(
z_{2}\right) \,,\ \ \ ^{-}\varphi _{m}\left( x\right) =\ ^{-}\mathcal{N}%
w_{1}^{2}\left( z_{2}\right) \,.  \label{b12}
\end{eqnarray}%
Thanks to the above classification, we may represent solutions valid at all $%
x$ in two equivalent forms:%
\begin{widetext}
\begin{eqnarray}
\ ^{+}\varphi _{m}\left( x\right) &=&\left\{
\begin{array}{ll}
\ _{+}\varphi _{m}\left( x\right) g\left( _{+}|^{+}\right) -\ _{-}\varphi
_{m}\left( x\right) g\left( _{-}|^{+}\right) \,, & x\in \mathrm{I} \\
\ ^{+}\mathcal{N}w_{2}^{2}\left( z_{2}\right) \,, & x\in \mathrm{II}%
\end{array}%
\right.  \label{b13} \\
\ ^{-}\varphi _{m}\left( x\right) &=&\left\{
\begin{array}{ll}
\ _{+}\varphi _{m}\left( x\right) g\left( _{+}|^{-}\right) -\ _{-}\varphi
_{m}\left( x\right) g\left( _{-}|^{-}\right) \,, & x\in \mathrm{I} \\
\ ^{-}\mathcal{N}w_{1}^{2}\left( z_{2}\right) \,, & x\in \mathrm{II}%
\end{array}%
\right.  \label{b14}
\end{eqnarray}
\end{widetext}

Demanding continuity of the solutions and their derivatives at $x=0$ (\ref%
{b9}) we discover that $g\left( _{+}|^{+}\right) $ and $g\left(
_{+}|^{-}\right) $ admit the forms:%
\begin{widetext}
\begin{eqnarray}
g\left( _{+}|^{+}\right) &=&\frac{1}{\sqrt{\left\vert p^{\mathrm{L}%
}\right\vert \left\vert p^{\mathrm{R}}\right\vert }}\left. \left[ \left\vert
p^{\mathrm{L}}\right\vert w_{2}^{2}\left( z_{2}\right) \frac{d}{dz_{1}}%
w_{2}^{1}\left( z_{1}\right) +\left\vert p^{\mathrm{R}}\right\vert
w_{2}^{1}\left( z_{1}\right) \frac{d}{dz_{2}}w_{2}^{2}\left( z_{2}\right) %
\right] \right\vert _{x=0}\,,  \notag \\
g\left( _{+}|^{-}\right) &=&\frac{1}{\sqrt{\left\vert p^{\mathrm{L}%
}\right\vert \left\vert p^{\mathrm{R}}\right\vert }}\left. \left[ \left\vert
p^{\mathrm{L}}\right\vert w_{1}^{2}\left( z_{2}\right) \frac{d}{dz_{1}}%
w_{2}^{1}\left( z_{1}\right) +w_{2}^{1}\left( z_{1}\right) \left\vert p^{%
\mathrm{R}}\right\vert \frac{d}{dz_{2}}w_{1}^{2}\left( z_{2}\right) \right]
\right\vert _{x=0}\,.  \label{b15}
\end{eqnarray}
\end{widetext}


\end{document}